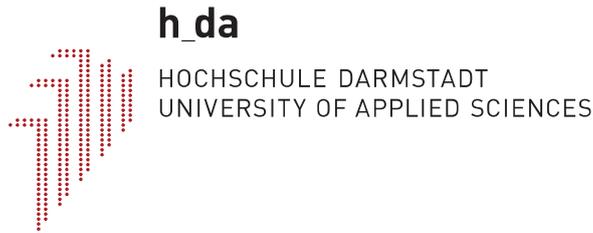

# Hochschule Darmstadt

– Fachbereich Informatik –

## Performanz Evaluation von PQC in TLS 1.3 unter variierenden Netzwerkcharakteristiken

Abschlussarbeit zur Erlangung des akademischen Grades

Master of Science (M.Sc.)

vorgelegt von

**Johanna Henrich**

Matrikelnummer: 748328

Referent : Prof. Dr. Andreas Heinemann
Korreferent : Prof. Dr. Alexander Wiesmaier

# ERKLÄRUNG

Ich versichere hiermit, dass ich die vorliegende Arbeit selbstständig verfasst und keine anderen als die im Literaturverzeichnis angegebenen Quellen benutzt habe.

Alle Stellen, die wörtlich oder sinngemäß aus veröffentlichten oder noch nicht veröffentlichten Quellen entnommen sind, sind als solche kenntlich gemacht.

Die Zeichnungen oder Abbildungen in dieser Arbeit sind von mir selbst erstellt worden oder mit einem entsprechenden Quellennachweis versehen.

Diese Arbeit ist in gleicher oder ähnlicher Form noch bei keiner anderen Prüfungsbehörde eingereicht worden.

*Darmstadt, 08. Februar 2022*

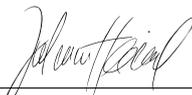

Johanna Henrich

# ABSTRACT

The used cryptographic primitives rely on the computational difficulty of certain mathematical problems. In the last years there has been much research on quantum computers which could be able to efficiently solve these problems in future years. Especially asymmetric primitives, used for authentication and key exchange could be broken. The affected algorithms are actually used within many internet protocols and applications and quantum-safe alternatives are urgently needed. NIST started a process to find and standardize quantum-safe digital signature schemes and key establishment schemes, but the candidates and alternatives come along with specific characteristics and differ from classical proceedings. So, besides analyzing the security of these new algorithms, it is also necessary to evaluate their performance and integrability into existing infrastructures and applications. Especially the integration into TLS protocol, used within about 90 percent of today's internet connections, plays an important role. The current version 1.3 uses the threatened asymmetric primitives for both, digital signatures and key establishment.

In this work, NIST candidates and alternatives for quantum-safe key establishment were evaluated while using them within TLS 1.3. The focus was on analyzing the performance trend while changing certain network parameters like rate or packetloss and examining the suitability of the PQC algorithms under different network scenarios and in the entire application context. To achieve this, the framework of Paquin, Stebila, and Tamvada [196] was extended to emulate various network conditions while frequently establishing a TLS 1.3 connection and measuring handshake duration.

Among our key results, we observe that on the one hand the evaluated candidates Kyber, Saber and NTRU as well as the alternative NTRU Prime achieve very good overall performance and partially beat the classical ECDH. Choosing a higher security level or hybrid versions does not have a significant impact to the handshake times. On the other hand the alternatives FrodoKEM, HQC, SIKE and BIKE show individual disadvantages and the performance is linked to the used security level and variant. This applies in particular to FrodoKEM. SIKE seems to be a worthwhile alternative in specific circumstances, like rates less than 2 Mbps, due to its small key and ciphertext sizes. In general, network conditions should be taken into account while choosing the algorithm and parameter set. Furthermore, it becomes clear that the handshake performance dependents on numerous factors, like TCP mechanisms and MTU, which could compensate the disadvantages of PQC or make them obsolete.

# ZUSAMMENFASSUNG

Klassische kryptografische Verfahren beruhen auf komplexen mathematischen Problemen, deren Lösung aufgrund des exorbitanten Rechenaufwands praktisch unmöglich ist. Quantencomputer, deren Entwicklung vorangetrieben wird, könnten jedoch in einigen Jahren eine Lösung in Polynomialzeit ermöglichen. Insbesondere die asymmetrische Kryptografie, welche die Sicherung von Authentizität und einen Schlüsselaustausch ermöglicht, könnte gebrochen werden. Sie wird in zahlreichen Protokollen und Anwendungen des Internets verwendet, sodass dringend quantensichere Alternativen benötigt werden. Die NIST initiierte einen Prozess zur Standardisierung quantensicherer Algorithmen, wobei die Kandidaten und Alternativen individuelle Charakteristiken aufweisen und sich von den klassischen Verfahren unterscheiden. Dies erfordert eine intensive Evaluierung, welche Performanz und Integrierbarkeit innerhalb bestehender Infrastrukturen berücksichtigt. Unter anderem spielt das Protokoll TLS 1.3 eine wichtige Rolle, da es bei 90 Prozent aller Internetverbindungen eingesetzt wird und asymmetrische Verfahren für Authentifizierung und Schlüsselaustausch nutzt.

In der vorliegenden Arbeit wurden daher die Algorithmen des Standardisierungsprozesses für einen quantensicheren Schlüsselaustausch während des Verbindungsaufbaus von TLS 1.3 überprüft. Der Fokus lag auf dem konkreten Einfluss dezidierter Netzwerkparameter wie Übertragungsrate oder Paketverlust, um Erkenntnisse bezüglich der Eignung der Algorithmen unter entsprechenden Netzwerkbedingungen zu erlangen. Für die Umsetzung wurde ein Framework von Paquin, Stebila und Tamvada [196] erweitert, welches diverse Netzwerkszenarien emuliert und die Handshakedauer für ausgewählte Algorithmen misst.

Es zeigt sich, dass die evaluierten Kandidaten Kyber, Saber und NTRU sowie die Alternative NTRU Prime eine sehr gute Gesamtperformanz besitzen und teils die Handshakedauer des klassischen ECDH unterbieten. Die Wahl eines höheren Sicherheitslevels oder hybrider Varianten bewirkt keinen signifikanten Unterschied. Anders bei Alternativen wie FrodoKEM, SIKE, HQC oder BIKE, welche individuelle Nachteile aufzeigen und deren Performanz abhängig von Sicherheitslevel und hybrider Durchführung stark variiert. Dies gilt besonders für den datenintensiven Algorithmus FrodoKEM. SIKE ist aufgrund der geringen Datenmengen in speziellen Fällen, wie Übertragungsraten unter 2 Mbps, eine lohnenswerte Alternative. Generell sollte bei Wahl des Algorithmus und dessen Variante die vorherrschende Netzwerkcharakteristik berücksichtigt werden. Zudem wird deutlich, dass die Performanz des Handshakes durch weitere Faktoren wie TCP Mechanismen oder MTU beeinflusst wird. Diese könnten mögliche Nachteile durch PQC ausgleichen respektive obsolet erscheinen lassen.

# INHALTSVERZEICHNIS

# ABBILDUNGSVERZEICHNIS

# TABELLENVERZEICHNIS

## ABKÜRZUNGSVERZEICHNIS

API  Anwendungsschnittstelle

IoT  Internet of Things

IP  Internet Protocol

TCP  Transport Control Protocol

UDP  User Datagram Protocol

PQC  Post Quantum Cryptography

HTTP  Hyper Text Transfer Protocol

FTP  File Transfer Protocol

TLS  Transport Layer Security

DoS  DenialofService

NIST  National Institute of Standards and Technology

NetEm  Network Emulator

DES  Data Encryption Standard

AES  Advanced Encryption Standard

KEM  Key Encapsulation Mechanism

DH  Diffie-Hellmann

ITU  International Telecommunication Union

PKI  Public-Key-Infrastruktur

IETF  Internet Engineering Task Force

HTTPS  HyperText Transfer Protocol Secure

SSL  Secure Sockets Layer

BSI  Bundesamt für Sicherheit in der Informationstechnik

DNS  Domain Name System

DNSSEC  Domain Name System Security Extensions

ECDSA  Elliptic Curve Digital Signature Algorithm

DTLS  Datagram TLS

RTT  Round Trip Time

SHA  Secure Hash Algorithm

MAC  Message Authentication Code

HMAC  Hash-based Message Authentication Code

MQV  Menezes–Qu–Vanstone

SVP  Shortest Vector Problem

CVP   Closest Vector Problem

LWE   Learning With Errors

ring-LWE   Ring Learning With Errors

NTRU   N'th Degree Truncated Polynomial Ring

QKD   Quantum Key Distribution

PGP   Pretty Good Privacy

PSK   Pre-Shared Key

HKDF   HMAC-based Key Derivation Function

KDF   Key Derivation Function

DHE   DH Schlüsselaustausch

LAN   Local Area Network

ECDH   Elliptic Curve Diffie Hellman

KEX   Key Exchange

OQS   Open Quantum Safe

SIS   Small Integer Solutions

CA   Certificate Authority

OPTLS   OPTimized oder One-Point-Three TLS

IANA   Internet Assigned Numbers Authority

EdDSA   Edwards-Curve Digital Signature Algorithm

ECDHE   Elliptic Curve Diffie-Hellman Ephemeral

CBC   Cipher Block Chaining

GCC   GNU Compiler Collection

NSA   National Security Agency

CNSS   Committee for National Security Systems

ETSI   European Telecommunications Standards Institute

SSH   Secure Shell

S/MIME   Secure Multipurpose Internet Mail Extensions

PQ-Only   Post Quantum Only

DoS   Denial of Service

IKE   Internet Key Exchange

IKEv2   Internet Key Exchange Version 2

VPN   Virtual Private Network

DSS   Digital Signature Standard

OID   Object Identifier

DSA   Digital Signature Algorithm

FIPS   Federal Information Processing Standard

SP      Special Publication
RAM   Random Access Memory
MTU   Maximum Transmission Unit
MSS   Maximum Segment Size
MLWE  Module Learning With Errors
BIKE   Bit Flipping Key Encapsulation
MSIS   Modular Short Integer Solution
FALCON  Fast-Fourier Lattice-based Compact Signatures over NTRU
RLWE  Ring Learning With Errors
WAN   Wide Area Network
QDISC  Queueing Discipline
EC     Elliptic Curve
CSR    Certificate Signing Request
GSO    Generic Segmentation Offload
TSO    TCP Segmentation Offload
GRO    Generic Receive Offload
RTO    Retransmission Timeout
Mbps   Megabits pro Sekunde
Kbps   Kilobits pro Sekunde
V2V    Vehicle to Vehicle
M2M    Machine to Machine
ECB    Electronic Code Book
CBC    Cipher Block Chaining
CTR    Counter
DHCP   Dynamic Host Configuration Protocol
ECC    Elliptic Curve Cryptography
RT     Round-Trip
UX     User Experience
SQL    Structured Query Language

Teil I

THESIS

# EINLEITUNG

## 1.1 MOTIVATION

Ein Großteil der Internetverbindungen wird mittels kryptografischer Algorithmen abgesichert. Sie werden bei Abruf von E-Mails, Durchführung von Online-Überweisungen, Übertragung von YouTube Videos oder Bezahlung mittels EC-Karte verwendet. Die tägliche Nutzung erfolgt meist völlig unbemerkt. Asymmetrische kryptografische Verfahren ermöglichen unter anderem den Austausch von Schlüsseln für eine vertrauliche Kommunikation oder die Authentifizierung mittels digitaler Signaturen. Dazu zählen Algorithmen wie RSA, Digital Signature Algorithm (DSA), Elliptic Curve Digital Signature Algorithm (ECDSA) und Elliptic Curve Diffie Hellman (ECDH), welche auf bestimmten mathematischen Problemen, beispielsweise aus den Bereichen Faktorisierung oder Diskrete Logarithmen, basieren. Es wird allgemein angenommen, dass diese Probleme mit gängigen Computertechnologien nicht in vertretbarer Zeit lösbar sind. Quantencomputer, welche aktuell Gegenstand der Forschung sind, würden jedoch eine Lösung in Polynomialzeit ermöglichen und könnten schon in wenigen Jahren allgemein zugänglich sein.

Mit Verfügbarkeit von Quantencomputern wären viele IT-Systeme nicht mehr ausreichend geschützt. Easttom [94] stellt fest, dass annähernd alle Bereiche der Netzwerksicherheit auf Aspekte asymmetrischer Kryptografie zurückgreifen würden. Vertraulichkeit und Integrität der Daten sowie Authentizität der Kommunikationsteilnehmer könnten plötzlich nicht mehr gewährleistet werden, sodass Unternehmen, staatliche Instanzen und jeder Einzelne gefährdet werden könnte. Für die rechtzeitige Absicherung muss sowohl die Identifikation zuverlässiger quantensicherer Technologien als auch deren Migration innerhalb der bestehenden Infrastruktur abgeschlossen sein, bevor eine Verfügbarkeit von Quantencomputern gegeben ist:

$$DevelopPQCTime + MigratePQCTime \leq QuantumComputersAvailableTime$$

Ein einfacher Eins-zu-eins Austausch der Algorithmen ist jedoch aufgrund unterschiedlicher Charakteristiken nicht möglich. Die neuen Technologien beinhalten teils lange Signaturen beziehungsweise Chiffrate, führen aufwändige Berechnungen durch oder verwenden große öffentliche oder private Schlüssel. Eine flächendeckende Umstellung könnte mehrere Jahre in Anspruch nehmen, sodass der zuvor zu bewerkstelligende Standardisierungsprozess möglichst bald abgeschlossen werden sollte. Das daraus resultierende neuartige Forschungsgebiet der quantensicheren Kryptografie wird auch als Post Quantum Cryptography (PQC) bezeichnet.

Zur Identifizierung und Standardisierung zuverlässiger PQC Algorithmen zum Austausch symmetrischer Schlüssel sowie zur Authentifizierung startete die National Institute of Standards and Technology (NIST) einen Prozess, welcher zwischen 2022 und 2024 abgeschlossen werden soll[1]. Der Prozess befindet sich aktuell in Runde drei und umfasst eine Liste von sieben potenziellen Kandidaten sowie acht Alternativen[2]. Auch das Bundesamt für Sicherheit in der Informationstechnik (BSI) sieht einen großen Bedarf an quantensicheren kryptografischen Algorithmen in naher Zukunft und beteiligt sich an der Suche sowie Evaluation von PQC Algorithmen [58]. Eine zügige und dennoch qualitativ hochwertige Evaluierung ist von großer Bedeutung, um einen zuverlässigen Schutz gegen Angriffe auf Basis von Quantencomputern zu bieten, bevor diese verfügbar sind.

Dabei müssen insbesondere auch die Herausforderungen bei Integration in bestehenden Infrastrukturen berücksichtigt werden. Unzählige Protokolle, Prozesse und Schemata nutzen kryptografische Primitive und müssen angepasst oder ersetzt werden. Die flächendeckende Umstellung erfordert eine immense Vorlaufzeit und birgt mitunter viele Komplikationen, wie bereits durch die Umstellung von Data Encryption Standard (DES) zu Advanced Encryption Standard (AES) oder Secure Hash Algorithm (SHA)-1 zu SHA-2 ersichtlich wurde [184]. Campagna, LaMacchia und Ott [62] schreiben, dass der Wechsel hin zu quantensicherer Kryptografie deutlich komplexer werde und weitaus mehr Bereiche betreffe als alle anderen, zuvor durchgeführten Migrationen im Kontext des Internets. Eine frühzeitige Evaluation innerhalb des Entwicklungsprozesses, gerade innerhalb der tatsächlichen Wirkumgebung, hat immer einen hohen Stellenwert, denn nur so können Fehler beziehungsweise Probleme frühzeitig erkannt und beseitigt werden.

Eines der betroffenen Protokolle ist Transport Layer Security (TLS), welches einen sicheren Kommunikationskanal zwischen zwei Kommunikationspartner aufbaut und im Jahr 2020 bei 90 Prozent aller Internetverbindungen Verwendung fand [158]. Version 1.3 ist seit 2018 verfügbar und wird als aktueller Standard angesehen. Das Protokoll nutzt während des Verbindungsaufbaus sowohl asymmetrische Algorithmen für den Austausch von Schlüsseln als auch für digitale Signaturen, sodass sich die Frage stellt, welche Kandidaten und Alternativen des NIST Standardisierungsverfahrens integriert werden könnten, ohne dass Sicherheit, Performanz und Nutzbarkeit der Systeme massiv verändert oder eingeschränkt werden. Eine weiterhin performante Durchführung des Verbindungsaufbaus spielt eine tragende Rolle bezüglich der Funktionalität und der User Experience (UX) bei Aufruf einer Webseite. Unzufriedenheit bei Endanwender ist oftmals auf das verlangsamte Laden von Webseiten- beziehungsweise Anwendungsinhalten zurückzuführen und sollte dringlichst vermieden werden [111, 131, 208, 257]. 2 bis 3 Sekunden gelten als maximale Grenze [112, 188]. Der TLS Handshake selbst benötigt nur einen Bruchteil der Gesamtzeit. Der erfolgreiche Verbindungsaufbau

---

[1] https://csrc.nist.gov/Projects/post-quantum-cryptography/workshops-and-timeline, zuletzt aufgerufen 04.07.21

[2] https://csrc.nist.gov/Projects/post-quantum-cryptography/round-3-submissions, zuletzt aufgerufen 04.07.21

ist jedoch Grundvoraussetzung für viele weitere Schritte. Zudem fanden Sy u. a. [243] heraus, dass über 95 Prozent der Webseitenaufrufe mehr als eine TLS Verbindung aufbauen würden, da beispielsweise andere Webinhalte integriert würden. Teils seien 5 bis 25 Handshakes pro Webseitenaufruf durchgeführt worden. Somit fällt die durch PQC verursachte Verzögerung mehrfach ins Gewicht. Daneben kommt TLS auch in anderen Gebieten wie Internet of Things (IoT) zum Einsatz, wobei dort teils besondere Echtzeitanforderungen gelten [205, 248].

Eine besonders frühzeitige Evaluierung ist vor allem im Bezug auf die Schlüsselvereinbarung von Bedeutung, da heute durchgeführte Kommunikationsverläufe aufgezeichnet und bei Verfügbarkeit von Quantencomputern nachträglich entschlüsselt werden könnten. Diese Art von Angriff wird auch als *Harvest-then-Decrypt* Attacke bezeichnet [222]. Bei der Evaluierung sind diverse Konstellationen von Algorithmenparametern, welche unterschiedliche Sicherheitsanforderungen erfüllen und jeweils Einfluss auf die Performanz nehmen, zu berücksichtigen. Auch eine optionale hybride Durchführung als Kombination mit einem klassischen kryptografischen Verfahren sollte betrachtet werden.

Während die reine Integration von PQC innerhalb von TLS und die resultierende Performanz bereits in mehreren Forschungsarbeiten untersucht [46, 77, 141, 193, 240] sowie Entwürfe zur Standardisierung veröffentlicht wurden [147, 221, 238, 261], muss für eine umfassende und aussagekräftige Analyse auch der Anwendungskontext beziehungsweise die vorherrschende Netzwerkcharakteristik berücksichtigt werden. Aufgrund der unterschiedlichen Algorithmeneigenschaften, beispielsweise im Bezug auf Schlüssel- und Chiffratlängen oder die Effizienz der Operationen, sind unterschiedliche Performanzentwicklungen zu erwarten. Es erfolgten bereits erste umfassende Tests mit realen Client-Server-Verbindungen [153, 196], der Einfluss konkreter Netzwerkparameter und -zustände ist jedoch noch weitestgehend unerforscht. Paquin, Stebila und Tamvada [196] entwickelten daher ein Framework für die Emulation von Netzwerkverbindungen, welches die dezidierte Anpassung einzelner Parameter wie Latenz oder Paketverlust ermöglicht und die benötigte Zeit bei Verbindungsaufbau bestimmt. Das Framework bietet jedoch keine flexible Konfiguration und die Autoren evaluierten ausschließlich einige hybride Varianten im Bezug auf variierende Latenz und Paketverlust.

In dieser Arbeit wurde daher das Framework erweitert, um eine flexible Angabe von Algorithmen und deren Konfigurationen sowie von diversen Netzwerkparametern wie Übertragungsrate, Latenz, Jitter oder Paketduplikate zu ermöglichen. Für die verschiedenen Kombinationen wurde anschließend wiederholt ein TLS Verbindungsaufbau durchgeführt und die benötigte Zeit gemessen.

## 1.2 ZIEL DER ARBEIT

Die hier beschriebene Arbeit soll einen Beitrag zur Identifikation und Integration geeigneter quantensicherer kryptografischer Verfahren innerhalb des Protokolls TLS 1.3 leisten. Zunächst soll das von Paquin, Stebila und Tamvada [196] entwickelte Framework so erweitert werden, dass flexibel diverse Algorithmen und Netzwerkkonfigurationen getestet und evaluiert werden können. Anschließende Experimente und Analysen sollen zudem Aufschluss darüber geben, wie sich die Performanz des Verbindungsaufbaus in Abhängigkeit vom gewählten PQC Algorithmus und emulierten Netzwerkbedingungen verhält.

Konkret sollen folgende Fragestellungen anhand der Untersuchungen beleuchtet werden:

1. Welche Auswirkungen hat die Integration der verfügbarer Algorithmen für einen quantensicheren Schlüsselaustausch auf die Performanz des TLS 1.3 Handshakes in Abhängigkeit von:

    (A) dem gewählten Algorithmus,

    (B) den gewählten Algorithmenparameter beziehungsweise dem angestrebten Sicherheitslevel,

    (C) der Entscheidung, ob der Algorithmus allein oder als hybride Variante genutzt wird und

    (D) dem speziell betrachteten Netzwerkparameter beziehungsweise der -konfiguration.

    Dies alles im Hinblick auf die Verwendung unter typischen Netzwerkbedingungen

2. Lassen sich Zusammenhänge oder Widersprüche zwischen der gemessenen Performanz des Handshakes und der Charakteristik des jeweiligen Algorithmus erkennen? Welche Charakteristika sind bei Entwicklung eines Algorithmus in diesem Kontext besonders erstrebenswert? An welcher Stelle können PQC Kandidaten respektive Alternativen verbessert werden, um dem Vergleich zu anderen Algorithmen standzuhalten?

3. Welche Algorithmen und -konfigurationen können in Anbetracht der vorherrschenden Netzwerkqualität ohne eine Einschränkung des Endnutzers verwendet werden, welche sind zu bevorzugen, welche Kompromisse sind gegebenenfalls einzugehen und welche Verfahren sollten gänzlich vermieden werden?

4. Welche Auswirkung auf die Performanz haben die Struktur von TLS oder der darunterliegende Netzwerkaufbau, speziell im Hinblick auf das Verhalten unterschiedlicher PQC Algorithmen und sind eventuelle Anpassungen sinnvoll oder gar notwendig?

Im Bezug auf die zugrundeliegende Arbeit von Paquin, Stebila und Tamvada [196] soll für eine aussagekräftige Analyse die Testmenge erweitert werden, sodass alle Algorithmen der dritten Runde des NIST Verfahrens, welche als direkte Kandidaten oder mögliche Alternativen für eine Standardisierung in Frage kommen, berücksichtigt werden. Zudem werden die Variationen der Netzwerkparameter um die Variablen *Übertragungsrate*, *Paketduplikate*, *Reordering*, *fehlerhafte Pakete* und *Jitter* ergänzt.

## 1.3 AUFBAU

Der Aufbau dieser Arbeit ist wie folgt: In Kapitel 2 werden zunächst die Grundlagen bezüglich aktueller und quantensicherer Kryptografie, TLS sowie Netzwerkaufbau und -emulation näher erläutert. Kapitel 3 schildert anschließend den aktuellen Stand der Wissenschaft bezüglich der Integration und Evaluation von PQC Algorithmen innerhalb von TLS. Kapitel 4 und 5 beinhalten letztlich den Hauptteil. Kapitel 4 schildert den Aufbau des Frameworks und das Vorgehen bei Durchführung der Experimente, während in Kapitel 5 die Ergebnisse aufgeführt und diskutiert werden. Zuletzt wird in Kapitel 6 ein Fazit gezogen.

# EINFÜHRUNG IN DIE GRUNDLAGEN

Dieses Kapitel soll der Einführung in die für diese Arbeit relevanten Themengebiete dienen und beinhaltet die elementaren Grundlagen. Für weiterführende und vollständige Information wird stets auf die angegebene Literatur verwiesen.

## 2.1 KRYPTOGRAFIE

Das Internet, welches als Kommunikationsnetzwerk weltweit Menschen und Maschinen miteinander verbindet, bietet neben einem großen Nutzen in den unterschiedlichsten Bereichen auch ein großes Potential an Bedrohungen. Dabei wird generell unterteilt in passive und aktive Angriffe. Bei passiven Angriffen werden Informationen aus dem Netzwerk von einem unbefugten Dritten mitgelesen, aber nicht verändert. Bei aktiven Angriffen hingegen werden Daten verändert oder neue Daten eingeschleust. Bei einer Denial of Service (DoS) Attacke werden beispielsweise so viele Daten an einen Kommunikationsteilnehmer gesendet, dass eine Überlastung eintritt und der bereitgestellte Service nicht mehr verfügbar ist. Bei SQL-Injection hingegen werden anhand von Sicherheitslücken Structured Query Language (SQL)-Anfragen eingeschleust, welche Einträge in der Datenbank auslesen oder verändern. Generell können passive und insbesondere aktive Angriffe zu massiven ökonomischen wie auch personellen Schäden im gewerblichen und privaten Umfeld führen.

Der Begriff Kryptografie stammt aus dem lateinischen und umfasst die Begriffe „*verstecken*" und „*schreiben*". Ursprünglich wurde er für die Verschlüsselung von Nachrichten verwendet, umfasst heute jedoch generell Verfahren für die Gewährleistung der Sicherheit der Daten. Im Grunde handelt es sich um mathematische Funktionen, mit deren Hilfe viele der oben beschriebenen Angriffe verhindert und die Bedrohungen innerhalb des Internets reduziert werden können. Die Bezeichnung *klassische Kryptografie* wird in dieser Arbeit verwendet, um die aktuell verbreiteten und eingesetzten Verfahren zu beschreiben. Algorithmen aus Bereichen wie Verschlüsselung, kryptografische Hashfunktionen oder Message Authentication Codes (MACs), die Teil einer kryptografischen Anwendung oder eines kryptografischen Systems sind, werden auch als kryptografische Primitive bezeichnet. Viele der aufgeführten Verfahren beruhen auf mathematischen Problemen aus den Bereichen der Faktorisierung oder diskreter Logarithmen, welche als praktisch unlösbar bezeichnet werden. Dies bedeutet, dass für jede bis dato verfügbare Rechenkapazität solche Parameterkonstellationen eines jeweiligen Algorithmus gewählt werden können, dass eine Lösung der entsprechenden Problemstellung nicht möglich ist.

Generell dienen die kryptografische Verfahren somit der Herstellung von Sicherheit. Sicherheit ist jedoch kein fest definiertes Maß und maßgeblich von dem jeweils verfolgten Ziel abhängig. Die Kryptografie verfolgt diverse Schutzziele, wobei drei besonders im Fokus stehen [226]:

1. **Vertraulichkeit:** Eine Nachricht ist nur für eine fest definierte Menge an Teilnehmern im Netzwerk zugänglich. Dies kann beispielsweise anhand von Verschlüsselungsalgorithmen erreicht werden.

2. **Integrität:** Bezieht sich auf die digitalen Daten und sagt aus, dass eine Nachricht auf dem gesamten Weg vom Sender zum Empfänger nicht verändert wurde. Um die Integrität zu überprüfen, müssen Konsistenz, Vollständigkeit, Genauigkeit und Gültigkeit, beispielsweise mithilfe einer Hashfunktion, auf dem gesamten Weg sichergestellt werden [120].

3. **Authentizität:** Generell als Echtheit definiert und beschreibt, dass es sich tatsächlich um eine angegebene Instanz, Kommunikationsteilnehmer oder auch Nachricht, im Netzwerk handelt. Der Prozess der Überprüfung wird auch als Authentisierung bezeichnet und kann beispielsweise anhand digitaler Signaturen erfolgen.

Neben diesen drei Schutzzielen existieren weitere, wie die Anonymität oder die Nichtabstreitbarkeit einer Nachricht.

Die kryptografischen Verfahren, welche für TLS und somit für diese Arbeit relevant sind und der Sicherstellung der oben beschrieben Schutzziele dienen, wurden zur besseren Übersicht in fünf Kategorien unterteilt:

1. Symmetrische Verschlüsselung

2. Asymmetrische Verschlüsselung & Schlüsselvereinbarung

3. Digitale Signaturen

4. Kryptografische Hashfunktionen

5. Digitale Zertifikate

Im Folgenden werden Charakteristik, Nutzen und die bekanntesten Algorithmen kurz beschrieben.

Es sei erwähnt, dass nicht alle Bereiche in diesem Kapitel abgedeckt werden können. Für weiterführende Informationen und ein tiefergehendes Verständnis wird unter anderem folgende Literatur empfohlen:

- Schwenk [228]

- Easttom [93]

- Manz [169]

- Paar und Pelzl [195]

2.1.1 *Symmetrische Verschlüsselung*

Bei symmetrischen kryptografischen Verfahren werden für die Ver- und Entschlüsselung einer Nachricht identische Schlüssel verwendet. Die Algorithmen sind vergleichsweise effektiv und schnell durchführbar [4]. In der Praxis müssen jedoch zunächst entsprechende Schlüssel zwischen den Kommunikationsteilnehmern ausgetauscht werden, ohne dass ein unautorisierter Dritter Diese einsehen kann. Somit gestaltet sich das Management entsprechend aufwändig und es sind oftmals zusätzliche Verfahren für einen sicheren Austausch notwendig. Des Weiteren ermöglichen symmetrische Verfahren keinen sicheren Nachweis bezüglich des Nachrichtenursprungs.

Symmetrische Verfahren können in zwei große Gruppen unterteilt werden:

- **Blockchiffren** zerteilen die Nachricht in Blöcke fester Größe von meist 64 oder 128 Bit, wobei der letzte Block bei Bedarf entsprechend aufgefüllt wird, auch als *Padding* bezeichnet. Anschließend kann eine blockweise Verschlüsselung erfolgen. Für die Ausführung existieren verschiedene Modi wie beispielsweise Electronic Code Book (ECB), Cipher Block Chaining (CBC), Counter (CTR). Die bekanntesten Algorithmen für Blockchiffren sind DES, welches jedoch seit 1999 als unsicher gilt [226] und dessen Nachfolger AES.

  AES wurde ursprünglich von Joan Daemen und Vincent Rijmen unter dem Namen Rijndael entwickelt und im Jahr 2000 vom NIST als Standard veröffentlicht [4]. Es handelt sich um ein iteratives Verfahren, welches auf dem Prinzip von Substitutions-Permutations-Netzwerken beruht. AES weist eine Blocklänge von 128 Bit sowie Schlüssellängen von wahlweise 128, 192 oder 256 Bit auf und ist weltweit verbreitet [4]. Trotz jahrelanger Analyse des Verfahrens und zahlreicher Angriffsversuche, zum Beispiel mithilfe von differenzieller und linearer Kryptoanalyse, ist das Verfahren bis heute sicher [168]. Zahlreiche Hardware- und Software Unterstützungen und ein großer Erfahrungsschatz aufgrund jahrelanger Forschung machen AES zu einem sehr beliebten Algorithmus, welcher anderen Algorithmen wie DES oder 3DES überlegen ist [4]. Für weitere Informationen zu AES und Substitutions-Permutations-Netzwerken siehe auch Manz [168] und [96].

- **Stromchiffren** ver- und entschlüsseln eine Folge von Klartextzeichen nacheinander und in jedem Schritt variierend, ohne die Nachricht zunächst in Blöcke zu unterteilen. So kann Stück für Stück eine zeichenweise Entschlüsselung erfolgen, ohne, dass die Vervollständigung eines Blocks abgewartet oder Padding verwendet werden muss. Für eine sichere Ausführung muss der Schlüssel jedoch die gleiche Länge wie der Klartext haben, was den Austausch und die Speicherung der Schlüssel negativ beeinflusst und einen hohen Ressourcenbedarf sowie ein aufwändigeres Schlüsselmanagement zur Folge hat [40]. Die wohl bekannteste Stromchiffre ist der One-Time Pad, welcher eine XOR-Verknüpfung zwischen einem Klartext und einer zufälligen Bitfolge mit gleicher Länge $n$ durchführt. Das Problem des Schlüsselmanagements kann mithilfe von Pseudozufallsfolgen umgangen werden, welche jedoch gewisse Unsicherheiten bergen. Siehe auch Schwenk [225].

### 2.1.2 Asymmetrische Verschlüsselung & Schlüsselvereinbarung

Bei asymmetrischer Kryptografie werden für die Ver- und Entschlüsselung unterschiedliche Schlüssel verwendet. Die Zeichenfolge zum Verschlüsseln einer Nachricht wird als *öffentlicher* Schlüssel bezeichnet, da dieser ohne Gefahr beliebig geteilt werden kann. Der *private* Schlüssel wird hingegen für die Entschlüsselung benötigt und sollte stets geheim bleiben. Eine alternative Bezeichnung ist daher auch *Public-Key-Verfahren*. Für die Umsetzung werden diverse mathematische Probleme aus Bereichen wie Faktorisierung herangezogen.

Folgende asymmetrische Verfahren sind insbesondere zu erwähnen:

- **RSA** wurde 1978 von seinen Namensgebern Ron Rivest, Adi Shamir und Leonard Adleman entwickelt. Laut Easttom [93] sei RSA der am häufigsten verwendete Algorithmus der asymmetrischen Kryptografie. Er beruht auf dem Phänomen, dass das Multiplizieren zweier großer Zahlen vergleichsweise einfach ist, obwohl es im Gegensatz praktisch unmöglich ist, eine solche große Zahl wieder in Primfaktoren zu zerlegen. Dies wird auch als das Problem der Faktorisierung großer Zahlen bezeichnet. Zur Erzeugung eines Schlüsselpaares werden zwei große, zufällige Primzahlen $p$ und $q$ mit vergleichbarer Größe ausgewählt, wobei $p \neq q$. Anschließend werden diese miteinander multipliziert, sodass $n = p \cdot q$ und die sogenannte Eulersche $\phi$-Funktion kann anhand der Formel $\varphi(n) = (p-1) \cdot (q-1)$ berechnet werden. Weiter wird eine natürliche, zu $\varphi(n)$ teilerfremde Zahl $e$ ausgewählt, wobei $1 < e < \varphi(n)$. Schlussendlich muss noch das multiplikative Inverse zu $e$ bezüglich $\varphi(n)$ bestimmt werden, sodass $e \cdot d \mod \varphi(n) \equiv 1$. Dies kann mithilfe des erweiterten euklidischen Algorithmus erfolgen [183]. $(e, n)$ bilden nun den öffentlichen Schlüssel. Die Parameter $p$, $q$, $\varphi(n)$ und $d$ müssen hingegen geheimgehalten werden, wobei das Paar $(d, n)$ den privaten Schlüssel für die Entschlüsselung bildet. Sei nun $m$ eine zu übertragende Nachricht und $c$ das mit dem öffentlichen Schlüssel verschlüsselte Chiffrat, so gilt:

$$c \equiv m^e \mod n$$
$$m \equiv c^d \mod n$$

  Siehe auch Wendzel [260], Easttom [93] sowie Beutelspacher [39].

  Um die Sicherheit von RSA zu untersuchen wurde 1991 die RSA Factoring Challenge durch das IT-Sicherheitsunternehmen RSA Security[1]

---

1 `https://www.rsa.com`, zuletzt aufgerufen 04.07.21

aufgerufen. Die Aufgabe des öffentlichen Wettbewerbs bestand darin Primfaktorzerlegungen von bestimmten Zahlen zwischen 330 und 2048 Bit zu ermitteln. Obwohl kleinere Zahlen wie RSA-140, RSA-640 oder RSA-768 erfolgreich zerlegt werden konnten [75], [68], [149], wurde der Wettbewerb in 2007 mit dem Fazit beendet, dass das Verfahren auch gegen die aktuell verfügbaren Hochleistungsrechner bei entsprechender Schlüssellänge ausreichenden Schutz bietet [39]. Laut Bundesnetzagentur gelten beispielsweise Schlüssel ab 1976 Bit als geeignet, wobei 2048 Bit oder größer empfohlen werden [263].

- **Diffie-Hellmann** wurde bereits vor RSA veröffentlicht und wird genutzt, um über einen unsicheren Kommunikationskanal einen geheimen symmetrischen Schlüssel auszuhandeln, welcher für die Verschlüsselung aller darauf folgenden Nachrichten verwendet werden kann. Die Umsetzung basiert auf dem diskreten Logarithmus Problem, welches aktuell, ebenso wie das Faktorisierungsproblem, als praktisch unlösbar gilt. Es bezieht sich auf die Suche einer ganzen Zahl $x$ mit der Eigenschaft $g^x mod p \equiv y$, während $p$ eine Primzahl und $g$, $y$ ganze Zahlen sind. Gesucht wird der Logarithmus von $y$ zur Basis $g$, wobei dieser über den modulo einer Primzahl erfolgt, weshalb die Bezeichnung *diskreter* Logarithmus verwendet wird. Die Operationen erfolgen über der Gruppe $\mathbb{Z}_p^\star$. Für tiefergehende Erklärungen siehe Beutelspacher, Neumann und Schwarzpaul [40]. Für den Diffie-Hellmann (DH)-Schlüsselaustausch zwischen zwei Instanzen $A$ und $B$ einigen sich die Teilnehmer zunächst auf eine Primzahl $P$ sowie eine ganze Zahl $g$, wobei es keiner Geheimhaltung bedarf. Im Anschluss wählen beide jeweils eine geheime Zufallszahl $a$ beziehungsweise $b$ und potenzieren $g$ mit dieser, sodass $h_A := g^a \mod p$ und $h_B := g^b \mod p$. Wenn beide die berechneten Werte über eine öffentliche Verbindung austauschen und den erhaltenen Wert jeweils mit der eigenen Zufallszahl potenzieren, erhalten sie das gleiche Ergebnis, denn

$$h_B^a \mod p \equiv g^{ba} \mod p \equiv g^{ab} \mod p \equiv h_A^b \mod p$$

- Daneben existieren **weitere Verfahren**. Das El Gamal Verschlüsselungsverfahren [98] sowie Menezes–Qu–Vanstone (MQV) sind Weiterentwicklungen von DH und beruhen somit ebenfalls auf dem diskreten Logarithmus Problem [94]. Die El Gamal Verschlüsselung greift exakt die gleichen mathematischen Teilschritte wie DH auf, inkludiert jedoch die Verschlüsselung eines Klartextes innerhalb des Prozesses. Zur vereinfachten Durchführung kann der öffentliche Schlüssel des Empfängers publiziert und in einer Datenbank zugänglich gemacht werden, sodass der Austausch des Schlüssel bei der Initiierung einer gesicherten Verbindung entfällt. Alternativ kann der Schlüssel auch innerhalb eines X.509-Zertifikates angegeben werden, siehe Abschnitt 2.1.5. El Gamal kann auch für digitale Signaturen genutzt werden. Bei gleicher Schlüssellänge ergeben sich jedoch im Vergleich zu RSA doppelt so lange

Signaturen, sodass von einer Verwendung in der Praxis in der Regel abgesehen wird [40].

Bei MQV werden zur Vereinbarung eines symmetrischen Schlüssels mehrere Schlüsselpaare genutzt, um die Gefahr durch Known-Key-Attacken zu minimieren, siehe Menezes, Qu und Vanstone [176] sowie Lee, Lim und Kim [157].

#### 2.1.2.1 Elliptische Kurven

Klassischer Weise führen die oben beschriebenen, asymmetrischen Verfahren ihre Operationen auf der Gruppe $\mathbb{Z}_p^\star$ durch. Sie können jedoch relativ leicht auf elliptische Kurven übertragen werden. Diese basieren auf zum diskreten Logarithmus analogen Problemen, sind aber besonders schwer zu lösen, sodass verhältnismäßig kurze Schlüssellängen ausreichen. Praktisch ausgesprochen wird für eine vergleichbare Sicherheit bei einer Gruppe $\mathbb{Z}_p^\star$ eine Primzahl von mindestens 2048 Bit benötigt, während bei elliptischen Kurven Werte in der Größenordnung von 224 Bit genügen. Die Berechnung ist wesentlich effizienter, sodass aktuell viele Verfahren auf elliptische Kurven zurückgreifen [42]. Zudem empfiehlt die NIST die Nutzung von ECDH sowie Elliptic Curve Digital Signature Algorithm [254]. Für weitere Informationen siehe auch Hankerson, Menezes und Vanstone [123] sowie Sadkhan [217].

Die folgende fünf elliptischen Kurven werden von der NIST empfohlen [121]:

| Sicherheitslevel (Bits) | RSA/DH (Bits) | ECDH (Bits) | U.a. verwendete Bezeichnungen |
|---|---|---|---|
| 80 | 1024 | 160 | P-192, nistp192, secp192r1, prime192v1 |
| 112 | 2048 | 224 | P-224, nistp224, secp224r1 |
| 128 | 3072 | 256 | P-256, secp256r1, prime256v1 |
| 192 | 7680 | 384 | P-384, nistp192, secp384r1 |
| 256 | 15360 | S512 | P-521, secp521r1 |

Tabelle 2.1: ECDH Schlüsselstärken im Vergleich zu RSA/DH sowie dem definierten Sicherheitsniveau der NIST, jeweils in Bits, nach [121] bzw. [50]. Das Sicherheitsniveau bezieht sich auf die Bitlänge möglichst gleichwertiger symmetrischer Verschlüsselungsverfahren, siehe auch Abschnitt 2.2.6.1

Die Namen wurden entsprechend der zugrundeliegenden Primzahlen gewählt: *P-192*, *P-224*, *P-256*, *P-384* sowie *P-521*. Daneben existieren weitere Bezeichnungen für die gleichen Werte wie beispielsweise *nistp192* und *nistp224* oder *secp192r1*, *secp224r1*, *secp256r1*, *secp384r1* und *secp521r1*. In OpenSSL werden zudem statt *P-192* beziehungsweise *P-256* die Werte *prime192v1* und *prime256v1* verwendet [50].

#### 2.1.2.2 Key Encapsulation Mechanism

Für den Austausch symmetrischer Schlüssel können auch Key Encapsulation Mechanisms (KEMs) genutzt werden. Im Gegensatz zum klassischen Schlüsselaustausch, beispielsweise mit RSA oder DH, wird eine jeweils abgewandelte Form verwendet. Der Schlüssel wird nicht vom Initiator gewählt und als Chiffrat übertragen, sondern entsteht aus dem öffentlichen Schlüssel des Kommunikationspartners und einer gewählten Zufallszahl. KEMs bestehen aus drei Teilalgorithmen: *KeyGen*, *Encaps* und *Decaps* sowie aus einem zugehörigen Schlüsselraum $K$:

- **KeyGen():** Erzeugung eines Schlüsselpaares $(pk, sk)$, welches aus einem öffentlichen und einem privaten Schlüssel besteht

- **Encaps($pk$):** Erzeugung eines Chiffrats $c$ und eines Schlüssels $k \in K$ mithilfe eines öffentlichen Schlüssels $pk$

- **Decaps($sk, c$):** Erzeugung eines Schlüssels $k \in K$ anhand des entsprechenden privaten Schlüssels $sk$ und des erhaltenen Chiffrats $c$. Bei einem Fehler wird $\perp$ returniert

Siehe auch [45]. Der exemplarische Ablauf einer Schlüsselvereinbarung bezüglich eines symmetrischen Schlüssels $k$ zwischen zwei fiktiven Kommunikationsteilnehmern Alice und Bob ist in Abbildung 2.1 dargestellt und verhält sich wie folgt:

1. Generierung des Schlüsselpaares $(pk, sk)$ bestehend aus dem öffentlichen und privaten Schlüssel von Alice

2. Übermittlung des öffentlichen Schlüssels $pk$ von Alice an Bob

3. Berechnung des Schlüssels $k$ sowie eines Chiffrats $c$ anhand des erhaltenen öffentlichen Schlüssels $pk$ von Alice

4. Übermittlung des Chiffrats $c$ an Alice

5. Berechnung des Schlüssels $k$ anhand des privaten Schlüssels $sk$ von Alice und dem erhaltenen Chiffrat $c$

#### 2.1.2.3 Schlüsselableitung

Key Derivation Functions (KDFs) dienen der Erzeugung eines Schlüssels anhand eines bereits existenten Geheimnisses sowie einem weiteren Parameter wie beispielsweise einer NONCE, welcher nicht geheimgehalten werden muss. KDFs werden unter anderem verwendet, wenn kryptografische Schlüssel zeitlich begrenzt und regelmäßig erneuert werden sollen, um somit die Sicherheit des Systems zu erhöhen. Die Schlüssel werden auch als Sitzungsschlüssel (*Session Key*) oder temporäre Schlüssel (*Ephemeral Key*) bezeichnet. Die Schlüsselableitung erspart hierbei den vergleichsweise teuren erneuten Austausch mithilfe der asymmetrischen Verfahren. Weitere Einsatzgebiete sind die Kombination von Schlüsseln bei hybriden Verfahren oder die Anpassung des Schlüssel an ein gewünschtes Format. Die Ableitung der Schlüssel erfolgt in der Regel durch pseudozufällige Funktionen, welche unter anderem auf kryptografischen Hashfunktionen und Blockchiffren beruhen. HMAC-based Key Derivation Function (HKDF) nutzen beispielsweise hashbasierte MAC und werden bei TLS genutzt, um aus einem mit DH ausgehandelten Wert kryptografische Schlüssel abzuleiten, siehe Krawczyk [150].

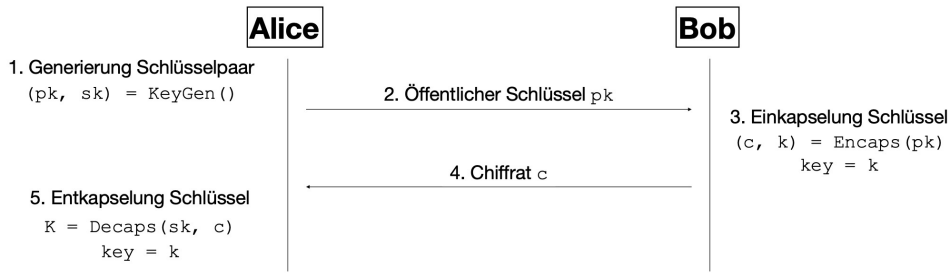

Abbildung 2.1: Ablauf einer KEM Schlüsselvereinbarung zwischen zwei Kommunikationspartnern Alice und Bob

### 2.1.3 Kryptografische Hashfunktion

Kryptografische Hashfunktionen erzeugen auf Grundlage eines Eingabetextes eine bestimmte Ausgabe fester Länge. Sie werden als eine Art eindeutiger Fingerabdruck genutzt, um die Integrität einer Nachricht zu überprüfen, denn sobald die Daten nur minimal verändert werden, ergeben sich mit angrenzender Wahrscheinlichkeit andere Werte. Da es sich hierbei um eine spezielle Art der Hashfunktionen handelt, müssen bestimmte Eigenschaften erfüllt werden. Williams Stalling beschreibt diese im Bezug auf eine Hashfunktion $H$ wie folgt [92]:

1. H kann einen Eingabewert variabler Länge verarbeiten

2. H ergibt eine Ausgabe mit fester Länge

3. Die Berechnung der Ausgabe anhand des Eingabewerts ist relativ einfach und benötigt keine besonderen Ressourcen, sodass H, unabhängig von der Eingabe, praktisch anwendbar ist

4. H ist eine Einwegfunktion. Das heißt, es ist praktisch unmöglich, anhand eines Ausgabewertes die ursprüngliche Eingabe zu berechnen

5. H ist schwach kollisionsresistent. Das heißt, es ist praktisch unmöglich, zwei Eingabewerte $x$ und $y$ zu berechnen, sodass $x \neq y$ und $H(x) = H(y)$

6. H ist stark kollisionsresistent. Das heißt, es ist praktisch unmöglich, ein Paar an Eingabewerte $(x,y)$ zu berechnen, sodass $H(x) = H(y)$

Hashfunktionen können aufgrund der oben beschriebenen Eigenschaften als Prüfsumme oder zur Sicherung von Passwörtern verwendet werden. Sie sind jedoch insbesondere zur Sicherung der Integrität in den Bereichen Kryptografie sowie Computerforensik unabdingbar. Es ist zu beachten, dass es sich bei Prüfsummen nicht per se um kryptografische Hashfunktionen handeln muss. Die Suche nach funktionalen und performanten Hashfunktionen gestaltete sich in der Praxis oftmals schwierig und langwierig [43]. Bekannte Algorithmen sind unter anderem MD5 [251] oder die SHA-Familie [124], wobei für eine sichere Verwendung SHA-224, SHA-256, SHA-386 oder SHA- 512 empfohlen werden. MD5 sollte hingegen nicht mehr verwendet werden, da ungewollt Kollisionen auftreten können, siehe Wang und Yu [259]. Für weitere Informationen zu Hashfunktionen siehe auch Easttom [92] sowie Bellovin und Rescorla [31].

### 2.1.4 Digitale Signaturen

Mit den Verfahren der asymmetrischen Kryptografie kann nicht nur eine Verschlüsselung erfolgen. Umgekehrt kann das Prinzip auch für die Erzeugung und Überprüfung digitaler Signaturen verwendet werden. Ähnlich einer realen Signatur ist nur eine autorisierte Instanz in der Lage, Daten entsprechend zu signieren. Auf der anderen Seite ist jeder in der Lage, diese zu überprüfen. Somit können Authentizität und Nichtabstreitbarkeit bezüglich der Kommunikationsteilnehmer sichergestellt werden. Dabei kann ein asymmetrisches Verfahren immer dann herangezogen werden, wenn folgende Eigenschaft gilt: Seien $E_T(x)$, $D_T(x)$ Funktionen zur Ver- beziehungsweise Entschlüsselung einer Nachricht $x$ und sei $m$ beliebiger Klartext, so ist

$$D_T(E_T(m)) = E_T(D_T(m)) = m.$$

Eine sichere digitale Signatur ist fälschungssicher. Zudem kann der Signierer im Anschluss an die geleistete Signatur nicht leugnen, diese vollzogen zu haben. Aktuell werden für die digitale Signatur vorrangig die Algorithmen RSA sowie ECDSA [136] verwendet.

In der Praxis wird für die Signatur in der Regel zunächst der Hashwert über das Dokument beziehungsweise die Nachricht gebildet. Dies ist erforderlich, da zum einen die Berechnung einer Signatur über größere Datenmengen sehr viel mehr Zeit erfordern würde und zum anderen die Länge der Signatur und somit der zu übermittelnden Daten deutlich reduziert werden kann [43]. Anschließend wird der Hashwert mit dem privaten Schlüssel des Signierers signiert. Umgekehrt kann bei Verifizierung mithilfe des öffentlichen Schlüssels des Signierers der ursprüngliche Hashwert anhand der Signatur berechnet werden. Stimmt dieser Wert mit dem Hash über der erhaltenen Nachricht überein, so ist die Signatur gültig.

Neben Hashwerten können auch MACs beziehungsweise Hash-based Message Authentication Codes (HMACs) genutzt werden. Bei MAC Verfahren wird zur Überprüfung der Integrität zusätzlich ein Schlüssel benötigt, um somit einer unbemerkten Manipulation beziehungsweise Sabotage durch einen Angreifer vorzubeugen. Für die Verschlüsselung wird ein symmetrisches Verfahren genutzt. Bei HMACs wird die Integrität wie gewohnt mittels eines Hashwertes überprüft [250], [78]. Es können jedoch auch alternative Algorithmen wie Blockchiffren verwendet werden [30]. Die Berechnung eines MAC ist per se deutlich effizienter als die einer digitalen Signatur, sie reicht jedoch aufgrund des symmetrischen Schlüssels nicht aus, um die Anforderungen einer Signatur zu erfüllen. Siehe auch Pelzl und Paar [199].

Die aktuellen Algorithmen und Parameter, welche zur Signatur vorgesehen sind, werden in einem offiziellen Standard der NIST spezifiziert, der sogenannte Digital Signature Standard (DSS). Darin beschrieben sind RSA, DSA sowie ECDSA. Die Algorithmen, insbesondere ECDSA, sind schnell und effizient und werden weltweit verwendet [193].

### 2.1.5 Digitale Zertifikate

Digitale Zertifikate sind kryptografische Protokolle, welche der Identifikation der Kommunikationsteilnehmer sowie der zugehörigen öffentlichen Schlüssel dienen. Äquivalent zu amtlichen Dokumenten im realen Leben enthalten sie Informationen zum Aussteller und Inhaber. Durch Angabe des öffentlichen Schlüssels kann die Instanz identifiziert werden und durch eine Signatur des Zertifikatausstellers ist die Integrität des Dokuments gesichert. Ein aktueller und weit verbreiteter Standard für digitale Zertifikate ist der International Telecommunication Union (ITU)-Standard X.509 [49], welcher beispielsweise bei TLS oder Secure Multipurpose Internet Mail Extensions (S/MIME) Verwendung findet. Bei X.509 handelt es sich weniger um ein Protokoll, als mehr um eine Sammlung von Datenformaten und Algorithmen für die Validierung der Identität. Dabei wird der öffentliche Schlüssel durch einen vertrauenswürdigen Dritten signiert und die Überprüfung kann anhand einer sogenannten Public-Key-Infrastruktur (PKI) erfolgen, in welche die Zertifikate eingegliedert sind. Die PKI dient unter anderem der Erzeugung sowie Verteilung digitaler Zertifikate und stellt eine Art Baumstruktur bereit, wobei deren Knoten X.509-Zertifikate abbilden. Die Wurzel des Baumes ist selbstsigniert und allgemein anerkannt. Bei allen anderen Knoten wird die Gültigkeit durch den Elternknoten garantiert, welcher den Herausgeber des Zertifikats repräsentiert. Es handelt sich also um eine Sequenz von Signaturen und man spricht auch von einer Zertifikatskette. Die PKI ermöglicht somit eine sichere Bereitstellung öffentlicher Schlüssel und ist ein zentrales Element bei der Implementierung asymmetrischer Kryptografie [95]. Die Größe eines X.509 Zertifikats variiert entsprechend der gewählten Algorithmen und Parameter sowie der aufgeführten Attribute und der Codierung. In den meisten Fällen beträgt sie zwischen 500 und 2000 Bytes [193], sodass bei der Übertragung der Zertifikatsketten mit meist zwei bis vier Elementen, einige Kilobytes anfallen. Da der Aufbau der Zertifikate anhand der ASN.1 Object Identifier (OID)s festgelegt wird, sind sie hochgradig erweiterbar und bieten eine gewisse Agilität. Durch die Einbindung innerhalb der PKI wird die Komplexität jedoch erhöht. Bei Generierung eines

Zertifikats muss die Signatur des vertrauenswürdigen Dritten mittels eines Certificate Signing Request (CSR) angefragt werden. Ein CSR ist eine Art digitales Formular, welches den öffentlichen Schlüssel des Antragsstellers erhält. Nach erfolgreicher Prüfung an der zuständigen Stelle wird ein gültiges Zertifikat ausgestellt. Für weiterführende Informationen siehe auch [5]. Alternativ können auch Pretty Good Privacy (PGP) Zertifikate genutzt werden, siehe beispielsweise [122, 172]. Sie finden heute vor allem im Bereich der Emailverschlüsselung Verwendung.

Aufgrund der Einbindung der kryptografischen Protokolle in die Kommunikation zweier Teilnehmer ergibt sich eine gewisse Komplexität, sodass neue Angriffsflächen entstehen, welche bedacht und abgesichert werden müssen. Bei Replay-Angriffen werden beispielsweise bereits gesendete und aufgezeichnete Nachrichten nochmals gesendet [244]. Bei Man-in-the-Middle-Angriffe schaltet sich ein Angreifer zwischen zwei Kommunikationspartner und kann Nachrichten belauschen, verändern oder vorenthalten [74].

### 2.1.6 Kryptoagilität

Technologien aus dem Bereich der Informatik entwickeln sich stetig weiter. Der wissenschaftliche Fortschritt ermöglicht neue Innovationen auf dem Markt, deckt aber auch Mängel beziehungsweise Sicherheitslücken auf. Auch Verfahren im Kontext der Kryptografie unterstehen daher einem fortwährenden Verbesserung- und Entwicklungsprozess und gerade hier ist ein zügiger und reibungsloser Ablauf von Bedeutung, um neueste Standards einzuhalten und die Sicherheit der Systeme nicht zu gefährden. Viele der Komponenten werden weltweit in unterschiedlichsten Systemen angewandt, sodass Neuerungen meist nur schleppend umsetzbar sind. Zudem müssen für die Nutzung einer neuen Technologie alle beteiligten Kommunikationspartner ihre Systeme entsprechend umstellen. In der Regel dauert die Umstellung daher mehrere Jahre bis hin zu Jahrzehnten [26] und aufgrund der Heterogenität der betroffenen Infrastruktur können kaum vorhersehbare Probleme auftreten. Negativbeispiele für eine langsame und problematische Umstellungsphase sind unter anderem der Wechsel von 3DES zu AES oder von IPv4 zu IPv6. Für einen möglichst optimalen Ablauf ist eine hohe Kryptoagilität der Systeme maßgebend. Der Begriff Kryptoagilität bezieht sich auf Design der Komponenten und beschreibt, inwiefern dieses auf einen möglichst schnellen und einfachen Wechsel von kryptografischen Primitiven und Algorithmen ausgelegt ist. Problematisch ist eine mangelnde Kryptoagilität vor allem bei Systemen, welche in der Regel über lange Zeiträume unverändert genutzt werden, wie beispielsweise Komponenten in großen Maschinen in Industrieanlagen oder Satelliten im Kontext der Raumfahrt.

## 2.2 NIST POST-QUANTUM CRYPTOGRAPHY STANDARDIZATION

### 2.2.1 *Quantentechnik und Quantencomputer*

Quantencomputer nutzen eine völlig neue Art der Informationsverarbeitung und ermöglichen somit enorme Steigerung der Recheneffizienz.

#### 2.2.1.1 *Technische Grundlagen*

Die Technologie beruht auf den Erkenntnissen der Quantenphysik, welche sich mit dem Verhalten von Quanten befasst. Die Objekte, welche der Größe eines Moleküls entsprechen oder gar noch kleiner sind, weisen ein völlig neuartiges Verhalten auf, welches stark von den Erwartungen aus Perspektive der klassischen Mechanik abweicht. Obwohl diese Phänomene bereits seit über hundert Jahren untersucht werden, siehe Planck [202], konnten bis heute nicht alle Fragestellungen gänzlich geklärt werden. Das beobachtete Verhalten und die daraus resultierenden Theorien und Gesetze lassen sich jedoch für diverse, teils sehr komplexe beziehungsweise langwierige Berechnungen nutzen. Bis dato in endlicher Zeit unberechenbare mathematische Probleme könnten effizient gelöst werden.

Für die Informationsverarbeitung in Quantencomputer werden beispielsweise geladene Atome, sogenannte Ionen, Photonen oder Elektronen innerhalb eines Kreisstroms verwendet. Anlehnend an die Bits in klassischen Computer werden sie als Quanten-Bits oder auch Qubits bezeichnet und die aktuellen Zustände werden anhand der Teilchenzustände beziehungsweise deren Polarisation definiert [189]. Sie werden durch einen Zustandswechsel in einem System mit diskreten Werten erzeugt, aber anders als bei klassischen Bits handelt es sich nicht um zwei diskrete Werte. Werden Qbits in eine sogenannte Superposition gebracht, können sie die Anzahl möglicher Zustände erhöhen. Bei $n$ Qubits können $2^n$ Zustände dargestellt werden, sodass der Zustandsraum während einer Berechnung enorm erweitert wird. Durch die sogenannte Verschränkung sind zudem Verbindungen zwischen den Qubits möglich, welche parallele Berechnungen auf Qubits erlauben. Das Messen der Zustände ist jedoch ein stochastischer Prozess, sodass Berechnungen mehrfach durchgeführt werden sollten. Für weitere Informationen siehe Beutelspacher, Neumann und Schwarzpaul [41].

In der Praxis nutzen Quantencomputer neben den Qubits meist auch zusätzlich klassische Bits, beispielsweise für Fehlerkorrektur oder für die Kontrolle der physikalischen Komponenten [125]. Trotzdem bestehen bei der Umsetzung Herausforderungen, die beachtet werden müssen. So entsteht beispielsweise massive Abwärme, welche die empfindlichen Komponenten beeinflussen kann und den Betrieb im alltäglichen Umfeld problematisch macht. Zudem können Qubits ihre angenommenen Werte nur für geringe Zeit halten und die Möglichkeiten zur ausreichenden Stabilisierung sind beschränkt. Quantencomputer sind daher voraussichtlich weniger ein Ersatz für die klassischen Computer als viel mehr eine Ergänzung, welche über private oder öffentliche Clouddienste in die bestehenden Systeme integriert werden könnte [181].

#### 2.2.1.2 *Entwicklung im Laufe der Jahre*

Ein erster theoretischer Entwurf eines Quantencomputers wurde Anfang der 1980er Jahre von Richard Feynman vorgestellt [106]. Zunächst waren viele Experten der Meinung, dass die Zustände der Quanten zu instabil und die Fehleranfälligkeit zu hoch sei, als dass jemals eine Realisierung möglich sein werde [70]. Nach der Veröffentlichung diverser Verfahren, wie beispielsweise fehlererkennender Codes für Quantentechnologien [206], stieg die Hoffnung jedoch. 1998 präsentierte schließlich eine Partnerschaft von IBM zusammen mit dem Los Alamos Laboratory[2], dem Massachusetts Institute of Technology [3] sowie der University of California at Berkeley [4] die ersten Quantencomputer mit einem beziehungsweise zwei Qubits [133], [115]. Diese beruhen auf Aminosäuren. Die ersten photonenbasierten Ansätze erschienen ab 2004 mit bis zu 12 Qubits [93]. Ab etwa 2012 veröffentlichten auch größere Unternehmen erste Ergebnisse. So veröffentlichten in 2017 und 2018 IBM[5] sowie Google[6] Entwürfe mit 50 beziehungsweise 72 Qubits [262]. Die Systeme sind allerdings bisweilen sehr anfällig für äußere Einflüsse. IBMs Vorschlag hält den Zustand mit 50 Qubits beispielsweise nur für 90 Mikrosekunden, sodass ein Einsatz in der Praxis aktuell noch nicht möglich scheint [94]. Daneben existieren weitere Ansätze, welche nicht universell einsetzbar sind, sonder auf spezielle Optimierungsprobleme ausgerichtet sind [79]. Beispielsweise bietet D-Wave[7] spezielle Recheneinheiten, welche mehrere 1000 Qubits integrieren würden und speziell für bestimmte industrielle Anwendungsfälle geeignet seien [138].

Obwohl die Entwicklung hin zu praxistauglichen Modellen noch nicht abgeschlossen ist, sind die Aussichten verheißungsvoll, denn die Arbeit an den Technologien wird immer weiter voran getrieben. In den letzten Jahren wurden rasante Fortschritte erzielt und viele private und öffentliche Gelder investiert, sodass ein baldiger Durchbruch als wahrscheinlich erscheint [18]. Laut Wittkopp [262] habe beispielsweise die deutsche Bundesregierung den Bereich des Quantencomputings im Jahr 2021 mit Geldern in Höhe von etwa zwei Milliarden Euro unterstützt und laut Easttom [94] rechneten viele Forscher mit einer tatsächlichen Verfügbarkeit in fünf bis zehn Jahren [94]. Die hohen Investitionen lohnen sich, denn die Technologie könnte in diversen Bereichen große Vorteile bringen und neue Perspektiven eröffnen. Ein Einsatz wäre unter anderem

- bei der Analyse von Big Data

---

2 `https://www.lanl.gov`, zuletzt aufgerufen 28.11.21
3 `https://www.mit.edu`, zuletzt aufgerufen 28.11.21
4 `https://www.berkeley.edu`, zuletzt aufgerufen 28.11.21
5 `https://www.ibm.com`, zuletzt aufgerufen 25.11.21
6 `https://www.google.com`, zuletzt aufgerufen 25.11.21
7 `https://www.dwavesys.com`, zuletzt aufgerufen 25.01.22

- bei der Analyse von chemischen und biologischen Substanzen für das Design diverser Materialien und Medikamente

- im Bereich Machine Learning und künstliche Intelligenz

- bei der Simulation komplexer Systeme, beispielsweise im Bereich Weltraumforschung sowie

- im Bereich Kryptografie

denkbar [269]. Im Bereich der Kryptografie können Quanteneffekte für eine neuartige Art des Schlüsselaustausches genutzt werden, erstmals von Bennett und Brassard [33] beschrieben. Das Verfahren wird als Quantum Key Distribution (QKD) bezeichnet und ermöglicht, anders als klassische Kryptografie, die Detektion von passiven Angriffen, da eine Messung innerhalb des quantenmechanischen Systems eine Störung dessen hervorruft. Das Verfahren beruht auf dem Prinzip der Heisenbergschen Unschärferelation, welche unter anderem in [159, 246] beschrieben wird. Die Entwicklung ist jedoch noch nicht abgeschlossen und für die Umsetzung würde eine neuartige Infrastruktur benötigt. Für weitere Informationen siehe [99, 148, 190].

Neben dieser Errungenschaft birgt das Potential auf der anderen Seite die große Gefahr, dass bis dato sichere kryptografische Verfahren in absehbarer Zeit gebrochen werden könnten [235], [196].

### 2.2.1.3 *Grover's Algorithmus & Symmetrische Kryptografie*

Grover entwickelte eine effiziente lineare Suche auf einem Raum mit $n$ Elementen mittels Quantencomputern, welche es ermöglichen könnte, anhand von Klartext-Chiffrat-Paaren symmetrische Schlüssel zu kompromitieren [118, 119]. Das Verfahren würde die Wurzel des Aufwands einer klassischen Suche beanspruchen, das heißt $O(\sqrt{N})$. Im Bezug auf einen Schlüssel bei AES-128 wären somit $2^{64}$ anstatt $2^{128}$ Schritte notwendig. Durch eine Verdopplung der Schlüssellänge könnte die Gefahr jedoch gebannt werden, sodass Grover die Sicherheit der symmetrischen Algorithmen nicht signifikant beeinflussen würde. Ferner konnte gezeigt werden, dass eine exponentielle Beschleunigung des Suchalgorithmus unmöglich ist, sodass die Nutzung der symmetrischen Algorithmen auch unter Quantencomputern empfohlen werden kann. Brassard, Høyer und Tapp [55] nutzen Grovers Ansatz für die Bestimmung von Kollisionen bei Hashfunktionen in $O(\sqrt[3]{N})$. Dieser Bedrohung kann ebenfalls entgegen gewirkt werden, indem der Output der Hashfunktionen entsprechend vergrößert wird [193]. Der potentielle Einfluss der Quantencomputer auf die gängigen symmetrischen Verschlüsselungs- und Hash-Algorithmen ist in Tabelle 2.2 abgebildet. Sie zeigt, dass mit geeigneten, weniger aufwändigeren Maßnahmen weiterhin Sicherheit gewährleistet werden kann. Des Weiteren zeigt Tabelle 2.3 eine Einordnung bezüglich der effektiven Sicherheit ohne und mit Verfügbarkeit von Quantencomputern. Die Angabe der Sicherheitslevel erfolgt in Bit, entsprechend der Beschreibung in Abschnitt 2.2.6.1.

#### 2.2.1.4 Shor's Algorithmus & Asymmetrische Kryptografie

Shor präsentierte in 1994, noch bevor die Nutzbarkeit von Quantencomputern realistisch erschien, eine Möglichkeit um mithilfe der Qubits das in den vorherigen Kapiteln beschriebene Faktorisierungsproblem vergleichsweise effizient zu lösen, siehe [233]. Somit könnte anhand des öffentlichen Schlüssels der Private vergleichsweise schnell und effizient bestimmt werden. Sein Algorithmus nutzt Superposition und Verschränkung der Quanten und ermöglicht eine Berechnung in Polynomialzeit. Das Vorgehen wird durch Beutelspacher, Neumann und Schwarzpaul [41] eingehend erläutert.

Viele der aktuell verwendeten asymmetrischen Verfahren beruhen auf dem Faktorisierungsproblem und würden bei Verfügbarkeit von praktisch einsetzbaren Quantencomputern keine ausreichende Sicherheit mehr bieten. Zudem konnte Shor zeigen, dass auch das Problem des diskreten Logarithmus mit ähnlichem Vorgehen effizient lösbar ist. Auch, wenn für die Berechnung mit Shor's Algorithmus bei den heute verwendeten Schlüsselgrößen und -typen tausende von Qubits mit perfekter Stabilität benötigt würden [125], wären im Zweifel alle gängigen asymmetrischen Algorithmen, sowohl RSA als auch DH oder elliptische Kurven, nicht mehr nutzbar [184]. Die Tabellen 2.2 und 2.3 machen den Einfluss deutlich. Außerdem wurden bereits erste optimierte Ansätze veröffentlicht, die beispielsweise Quantencomputer mit klassischen Technologien verbinden und so eine Berechnung mit weniger Ressourcen ermöglichen könnten [35, 125].

| Krypto Algor. | Typ | Verwendung | QC Algor. | Einfluss durch Quantencomputer |
|---|---|---|---|---|
| AES | Symmetr. | Verschlüsselung | Grover | Größere Schlüssel |
| SHA-2, SHA-3 | —— | Hashfunktion | Grover | Größerer Output |
| RSA | Asymmetr. | Sign & KEX | Shor | UNSICHER |
| Elliptische Kurven (ECDSA, ECDH) | Asymmetr. | Sign & KEX | Shor | UNSICHER |
| Endl. Körper (DSA) | Asymmetr. | Sign & KEX | Shor | UNSICHER |

Tabelle 2.2: Einfluss von Quantencomputer auf die aktuell verwendeten symmetrischen & asymmetrischen kryptografischen Verfahren nach [70]

### 2.2.2 Entwicklung der Post Quanten Kryptografie

Mit PQC entstand ein komplett eigenständiges Forschungsgebiet hinsichtlich der Suche und Analyse quantensicherer Verfahren, welches mit der korrespondierenden PQCrypto[8] sogar eine eigene Konferenz aufweist [70]. Aufgrund der steigenden Bedrohung begann die Suche bereits vor einigen Jahren. Die erste PQCrypto fand im Jahr 2006 statt und wurde durch diverse

---
8 http://pqcrypto.org, zuletzt aufgerufen 07.11.21

| Verschlüsselunsalgor. | Schlüsselgröße | Effektive Sicherheit ohne QC | Effektive Sicherheit mit QC |
|---|---|---|---|
| AES 128 | 128 | 128 | 65 |
| AES 256 | 256 | 256 | 128 |
| RSA 1024 | 1024 | 80 | 0 |
| RSA 2048 | 2048 | 112 | 0 |
| ECC 256 | 256 | 128 | 0 |
| ECC 384 | 384 | 256 | 0 |

Tabelle 2.3: Effektive Sicherheit der gängigen Verschlüsselungsverfahren AES, RSA und Elliptic Curve Cryptography (ECC) nach [171]. Die Angabe der Sichereitheit erfolgt in Bit, entsprechend Abschnitt 2.2.6.1

nationale Förderstrukturen, insbesondere aus Europa und Japan, unterstützt. Damals wurden erste Entwürfe bezüglich gitter-, hash- und codebasierter sowie multivarianter Kryptografie diskutiert [57]. Einige Zeit später wurde die isogeniebasierte Kryptografie ergänzt [70].

Auch Organisationen aus Bereichen der Industrie starteten Forschungsaktivitäten und seit 2013 führte das European Telecommunications Standards Institute (ETSI) mehrere Workshops zu PQC durch. In 2015 wurden erstmals konkrete Pläne hinsichtlich einer Umstellung der regierungseigenen Systeme [191] vom US-amerikanischen Committee for National Security Systems (CNSS)[9] sowie von der US-amerikanischen Sicherheitsbehörde National Security Agency (NSA)[10] öffentlich. Darauf folgend hielt die NIST[11], eine Bundesbehörde der Vereinigten Staaten von Amerika, zunächst einen Workshop zu „Cybersecurity in a Post-Quantum World" und initiierte im Jahr 2016 einen offiziellen Prozess zur Entwicklung neuer, quantensicherer kryptografischer Standards. Die NIST besitzt eine einzigartige Rolle in diesem Kontext, da ihre Standards und Richtlinien aus dem Bereich der IT-Sicherheit von Behörden und anderen Organisationen weltweit anerkannt werden [70]. Auch, wenn Institutionen wie ETSI bereits früh Studien hinsichtlich einer Standardisierung erarbeiteten, rückte die Thematik erst durch das Verfahren der NIST weltweit derart in den Fokus [63].

### 2.2.3 *Motivation & Bestreben des NIST PQC Verfahrens*

Die US-amerikanischen Behörden begründen die Notwendigkeit und Dringlichkeit hinsichtlich der Entwicklung von PQC Standards insbesondere in zwei Punkten [242]:

1. Die Entwicklung von Quantencomputern werde stetig vorangetrieben, sodass die Gefahr bezüglich eines Angriffes auf die digitale Infrastruktur immer realer werde.

---

9 `https://www.cnss.gov/cnss/`, zuletzt aufgerufen 25.01.22
10 `https://www.nsa.gov`, zuletzt aufgerufen 25.01.22
11 `https://www.nist.gov`, zuletzt aufgerufen 25.01.22

2. Der Prozess bezüglich der Implementierung und Migration der neuen asymmetrischen Kryptosysteme im Kontext unterschiedlichster Anwendungsgebiete benötige mehrere Jahre [26], sodass möglichst bald Technologien gefunden werden sollten, welche entsprechenden Schutz bieten würden. Aufgrund abweichender Charakteristika, wie beispielsweise größerer Schlüssellängen, sei ein Eins-zu-eins-Austausch in vielen Fällen nicht möglich und die Umstellung vermutlich komplex.

Da die zur Zeit verwendeten symmetrischen Verschlüsselungsalgorithmen bei Vergrößerung der Schlüssellänge einen ausreichenden Schutz gegen Quantencomputer bieten, werden zunächst ausschließlich Alternativen für folgende Aspekte untersucht:

1. Asymmetrische Verschlüsselung (Public-Key-Verschlüsselung)

2. Schlüsselaustausch

3. Authentifizierung

Sie sollen die offiziellen Standards und Empfehlungen Special Publications (SPs) 800-56A Revision 3 („*Recommendation for Pair-Wise-Key-Establishment Schemes Using Discrete Logarithm Cryptography*") [24], sowie SP 800-56B Revision 2 („*Recommendation for Pair-Wise Key Establishment Using Integer Factorization Cryptography*") [25] und Federal Information Processing Standard (FIPS) 186-4, („*DSS*") [145], ergänzen.

Das PQC-Verfahren stellt eine Art Wettbewerb dar. Forschungsgruppen aus der ganzen Welt können sich mit Vorschlägen beteiligen, ähnlich dem Standardisierungsprozess von AES in den 1990er Jahren. Die Kandidaten werden hinsichtlich ihrer Sicherheit, aber auch im Bezug auf ihre Kosten sowie Flexibilität evaluiert. Eine umfassende Analyse sei von Bedeutung, denn die Migration der PQC Kandidaten gestalte sich laut NIST in vielen Fällen problematischer als bei den klassischen kryptografischen Algorithmen [242]. Diese Annahme basiert auf drei wesentlichen Aspekten:

1. Bei asymmetrischer Verschlüsselung und digitalen Signaturen handle es sich um vergleichsweise komplexe Verfahren mit vielen spezifischen Anforderungen, die berücksichtigt werden müssten.

2. Das Verständnis bezüglich der Quantencomputer und ihrer zukünftigen Fähigkeiten sei noch nicht ausreichend.

3. Die Algorithmen wiesen mitunter stark abweichende Charakteristika wie größerer Schlüssellängen, größere Signaturen oder einen hohen Rechenaufwand auf. Ein direkter Vergleich oder ein Eins-zu-eins-Austausch sei daher in vielen Fällen nicht möglich. Auch wenn die grundlegenden Funktionalitäten sowie ein ausreichendes Level an Sicherheit gegeben seien, könnten mangelnde Performanz oder Skalierbarkeit zu Problemen führen. Gegebenenfalls sei eine Anpassung von Protokollen beziehungsweise der Infrastruktur notwendig [26].

Auch die breite Öffentlichkeit wird dazu aufgerufen, die Algorithmen in unterschiedlichen Kontexten zu integrieren und zu evaluieren, um somit einen weitreichenden Überblick bezüglich der allgemeinen Eignung und Kompatibilität zu gewinnen [242]. Die NIST betont, dass auch eine Standardisierung mehrerer Algorithmen denkbar sei [69].

### 2.2.4 Verlauf & Status des NIST PQC Verfahrens

Der Prozess begann in 2016 mit einer öffentlichen Ausschreibung, welche bis Ende 2017 befristet war. Letztlich wurden 82 Vorschläge eingereicht, wobei 69 die minimalen Akzeptanzkriterien erfüllten.

Die Algorithmen wurden bereits in der ersten Runde hinsichtlich Sicherheit, Kosten und Flexibilität evaluiert. Da die detaillierten Beschreibungen sowie die Referenz-Implementierungen aller eingereichten Algorithmen frei zugänglich sein mussten, konnten sie von Beginn an auch von der Öffentlichkeit analysiert und diskutiert werden. Mitunter wurden signifikante Schwachstellen im Bezug auf Sicherheit und Performanz entdeckt, sodass einige Kandidaten zu einem frühen Zeitpunkt aussortiert werden konnten oder von den Bewerbern zurückgezogen wurden [97]. Des Weiteren finden seit Prozessbeginn PQC Konferenzen sowie Workshops statt. Die Abschlussberichte der *PQC Standardization Conference* fassen jeweils den aktuellen Status des Prozesses zusammen [10, 11, 70] und beschreiben Änderungen oder Ergänzungen der Algorithmen, welche durch die Autoren vorgenommen wurden.

Die erste Runde wurde zu Beginn 2019 beendet und die Auswahl der Kandidaten auf 26 Vorschläge reduziert: 17 für den Schlüsselaustausch und 9 für Signaturverfahren [11]. Es folgte eine noch intensivere Evaluierung, vor allem im Hinblick auf die Implementierung, Migration und das Benchmarking im Kontext der bestehenden Infrastrukturen. Im Verlauf der zweiten Runde wurden schwerwiegendere Sicherheitsmängel bei einigen Kandidaten festgestellt, welche deren weitere Berücksichtigung ausschloss [11].

Der Abschluss der zweiten Runde erfolgte Mitte 2020 und es fand erneut eine Reduktion auf nun mehr sieben Kandidaten, vier KEM und drei Signaturverfahren, statt. Zusätzlich wurden fünf KEMs und drei Signaturverfahren als Alternativen ausgewählt. Die Kandidaten seien laut NIST [11] mit großer Wahrscheinlichkeit für viele der gängigen Anwendungsfälle geeignet und weitestgehend bereit für eine Standardisierung. Die NIST sieht dabei insbesondere die strukturierten, gitterbasierten Verfahren als vielversprechend an und gibt im Statusbericht von Runde drei an, dass für KEMs sowie Signaturen vermutlich jeweils mindestens einer der gitterbasierten Ansätze als Standard gewählt werde. Alternativen werden weiterhin für eine mögliche Standardisierung in Betracht gezogen, jedoch nicht in der nahegelegen Zukunft. Sie weisen verbesserungswürdige Aspekte hinsichtlich Sicherheit beziehungsweise Performanz auf oder es Bedarf einer gründlicheren Erforschung für ein größeres Vertrauen. Die Sicherheit aller Kandidaten und Alternativen basiert auf fünf Familien von Algorithmen: gitterbasierte, codebasierte, multivarianzbasierte, isogeniebasierte und hashbasierte Ansätze. Diese sowie die Kandidaten und Alternativen selbst werden in den Abschnitten 2.2.7 sowie 2.2.8 beschrieben.

Ein erfolgreicher Abschluss des Verfahrens ist zwischen 2022 und 2024 vorgesehen, wobei einige Kandidaten bereits zu Beginn 2022 standardisiert werden könnten. Die NIST behält sich vor, die dritte Runde zu verlängern, sollten bei den Untersuchungen der Finalisten gravierende Sicherheitsmängel aufgedeckt werden. Zudem könnte bei einer vierten Runde eine ergänzende Auswahl getroffen werden [62].

### 2.2.5 Anforderungen an PQC Einsendungen

Die Anforderungen bezüglich einer vollständigen Einreichung sind in [242] beschrieben und wie folgt gegliedert:

- **Deckblatt**: grundlegende Informationen

- **Spezifikation & Dokumentation**: vollständige Beschreibung der Funktionen und Konfigurationen, detaillierte Performanzanalyse, erwartete Sicherheit, bekannte Angriffe, Vorteile und Schwächen sowie, sofern möglich, Abschätzung bezüglich Integrierbarkeit auf unterschiedlicher Hardware und unterschiedlichen Plattformen

- **Digitale & optische Medien**: mit Referenzimplementierung, optimierter Implementierung sowie entsprechender Dokumentationen

- **Erklärung zum geistigen Eigentum sowie weitere öffentliche Erklärung**

Für alle abstimmbaren Parameter des Algorithmus, welche beispielsweise die Sicherheit und Performanz des Verfahrens beeinflussen, sollten innerhalb der Spezifikation konkrete Werte definiert werden. Auch die Auswirkungen auf den Algorithmus bei Auswahl eines bestimmten Parameterwerts sollte beschrieben werden. Die erwartete Sicherheit sollte mithilfe der Security Level angegeben werden, welche in Abschnitt 2.2.6.1 näher erläutert werden. Des Weiteren sollte bei Verfahren zum Austausch symmetrischer Schlüssel mindestens eine Schlüssellänge von 256 Bit unterstützt werden.

Für eine öffentliche Analyse und Evaluierung der Algorithmen müssen diese frei verfügbar und der Öffentlichkeit zugänglich gemacht werden. Zudem sollten alle Beschreibungen und Spezifikationen in englischer Sprache vorliegen. Zudem sollten Implementierungen in ANSI C vorliegen.

### 2.2.6 Evaluationskriterien bezüglich PQC Algorithmen

Der vorgesehene Prozess und die Bewertungskriterien zur Evaluierung der Kandidaten werden, ebenso wie die Anforderungen, in [242] beschrieben. Die NIST führt eigene Tests durch, wünscht aber gleichzeitig eine öffentliche Bewertung für einen umfassenderen Überblick. Alle Beobachtungen und Ergebnisse sollen letztlich bei der Auswahl geeigneter Verfahren gleichermaßen berücksichtigt werden.

Grundsätzlich werden drei Hauptkriterien betrachtet, wobei die Priorisierung entsprechend der hier aufgeführten Reihenfolge erfolgt:

1. **Sicherheit**: Die Verfahren sollen nicht oder nur mit extrem bis unendlich großem Rechenaufwand gebrochen werden können, auch bei Existenz von Quantencomputern. Das heißt konkret, dass ausgetauschte Schlüssel und angefertigte Signaturen nicht durch Dritte eingesehen oder gefälscht beziehungsweise kompromittiert werden können. Die Einordnung der Sicherheit ist meist nicht trivial. Es wird in fünf Abstufungen unterteilt, siehe Abschnitt 2.2.6.1

2. **Kosten**: Folgende Metriken sind aufgrund der unterschiedlichen Anwendungsgebiete zu berücksichtigen:

    a) Länge öffentlicher Schlüssel, Chiffrate oder Signaturen
    
    b) Recheneffizienz bei Operationen mit öffentlichen oder privaten Schlüsseln
    
    c) Recheneffizienz bei Schlüsselerzeugung
    
    d) Kosten bei Implementierung wie Random Access Memory (RAM)
    
    e) Fehler bei Entschlüsselung

   Dabei ist eine geringe Größe der Primitive insbesondere für eine effiziente Übertragung erforderlich. Die Bewertung aller Performanzkriterien sollte kontextbezogen erfolgen. Bei vielen Anwendungen werden beispielsweise nur öffentliche Schlüssel in Kombination mit Chiffraten respektive Signaturen übertragen. Bei Signaturverfahren ist die Performanz bezüglich der Schlüsselgenerierung zudem weniger wichtig.

3. **Charakteristik von Algorithmus & Implementierung**:

    a) Flexibilität: Kompatibilität mit Plattformen und Protokollen, Erweiterbarkeit durch zusätzliche Funktionalitäten, parallelisierbare Implementierungen
    
    b) Einfachheit: einfaches und elegantes Design, gut verständlich und leicht erweiterbar, wenige Abhängigkeiten. Komplexe Implementierungen und Abhängigkeiten bergen hingegen Sicherheitsrisiken
    
    c) Übernahme: keine Behinderung durch Lizenzen etc.

2.2.6.1   *Security Level*

In der Regel erfolgt die Einstufung anhand der Rechenoperationen, welche für die Lösung des mathematischen Problems beziehungsweise die Berechnung des öffentlichen Schlüssels benötigt werden. Die Werte werden in Bits angegeben, wobei eine Sicherheit von $n$ Bits meint, dass $2^n$ Operationen benötigt werden. Für eine präzisere Einstufung erfolgte eine Einteilung in 5 Klassen, welche im Zuge der Definition der Evaluationskriterien in [242] aufgeführt sind. Sie orientieren sich an den gut analysierbaren Referenzverfahren der klassischen Kryptografie: Für einen erfolgreichen Angriff sollten mindestens so viele oder gar mehr Rechenressourcen notwendig sein, als für...

I. Schlüsselsuche auf Blockchiffre mit 128-Bit-Schlüssel (AES128)

II. Kollisionssuche auf 256-Bit-Hashfunktion (SHA256/SHA3-256)

III. Schlüsselsuche auf Blockchiffre mit 192-Bit-Schlüssel (AES192)

IV. Kollisionssuche auf 384-Bit-Hashfunktion (SHA384/SHA3-384)

V. Schlüsselsuche auf Blockchiffre mit 256-Bit-Schlüssel (AES256)

Die Bestimmung der Rechenressourcen ist nicht eindeutig und kann anhand unterschiedlicher Metriken bewertet werden. Dies gilt insbesondere für die vergleichsweise neuartigen PQC Algorithmen respektive deren Parameterkonstellationen, da voraussichtlich viele Angriffe erst nach der breiten Veröffentlichung entdeckt werden und zudem die konkreten Fähigkeiten der zukünftigen Quantencomputer noch unklar sind. Grundsätzlich sollten alle denkbaren Metriken berücksichtigt werden. Sie alle dürfen hinsichtlich der Einhaltung eines Sicherheitslevels nicht unterboten werden. Im Fokus während den Untersuchungen steht jedoch insbesondere die Ausführungszeit respektive die -zyklen.

Das primäre Augenmerk bei Entwicklung und Evaluierung der Algorithmen solle laut NIST zunächst auf den Sicherheitsleveln 1 bis 3 liegen [242]. Sie seien voraussichtlich für die ersten verfügbaren Quantencomputer und bei gewöhnlichem Nutzerverhalten ausreichend. Für besondere Sicherheitsanforderungen wie beispielsweise bei kritischen Infrastrukturen und im Hinblick auf mögliche neue Erkenntnisse im Bereich der Quantentechnik sei jedoch zusätzlich mindestens eine Parameterkonstellationen für Level 4 oder 5 sinnvoll.

Weitere Sicherheitseigenschaften wie Forward Secrecy, Resistenz gegen Seitenkanal- & Multi-Key-Angriffe beeinflussen nicht die Einstufung des Sicherheitslevels. Es ist jedoch von Vorteil, wenn diese Aspekte ohne größeren Aufwand umsetzbar sind.

### 2.2.6.2  *Plattform & Compiler zur Evaluierung*

Die NIST nutzt für die Tests eine festgelegte Referenzplattform. Dabei handelt es sich um eine x64-Implementierung von Intel[12], welche mit Microsoft Windows[13] oder Linux [14] betrieben werden kann und GNU Compiler Collection (GCC) unterstützt.

---

12 https://www.intel.de/content/www/de/de/architecture-and-technology/microarchitecture/intel-64-architecture-general.html, zuletzt aufgerufen 03.01.22
13 https://www.microsoft.com/en-us/windows/, zuletzt aufgerufen 02.01.22
14 https://www.kernel.org, zuletzt aufgerufen 02.01.22

### 2.2.7 Theoretische Ansätze der PQC Verfahren

#### 2.2.7.1 Gitterbasierte Verfahren

Wie bereits der Name verrät, basieren diese Verfahren auf sogenannten Gittern, wobei es sich um eine Art von mathematischen Matrizen beziehungsweise ganzzahlige Linearkombinationen einer Menge von linear unabhängigen Vektoren im $\mathbb{R}^n$ handelt. Mittels gitterbasierter Ansätzen können sowohl Schlüsselaustausch und asymmetrische Verschlüsselung als auch digitale Signaturen erfolgen.

Erste Arbeiten wurden bereits 1996, 1997 sowie 1998 veröffentlicht. Während Ajtai [7] gitterbasierte Systeme erstmals als mögliche Einweg-Funktionen spezifizierte, wurde einige Zeit später ein darauf aufbauendes Public-Key-Kryptosystem eingeführt. Weitere Fortschritte folgten mit [6, 28, 117, 126, 178]. Aktuell beruhen die Verfahren auf zwei grundlegenden mathematischen Problemen, dem Small Integer Solutions (SIS) Problem [180] sowie dem Learning With Errors (LWE) Problem [12, 210, 211]. Es können zudem weitere Probleme wie Shortest Vector Problem (SVP) [8, 71, 146] oder Closest Vector Problem (CVP) [179] abgeleitet werden. Insbesondere auf LWE basiert eine Vielzahl von Verschlüsselungs- und Authentifizierunsalgorithmen. Zudem ergänzt Ring Learning With Errors (ring-LWE) zusätzliche Strukturen und erlaubt somit noch kleinere Schlüssel.

Gitterbasierte Ansätze erwiesen sich insbesondere im Hinblick auf die Performanz sowie die Schlüssellänge als vorteilhaft gegenüber anderen Kandidaten [90, 97]. Sie arbeiten effizient, sind vergleichsweise einfach aufgebaut und hoch parallelisierbar [70], sodass 28 der 69 zu Beginn ausgewählten Kandidaten gitterbasiert waren.

Für weitere Informationen sei auf eine Ausarbeitung von Eberhardt [97] ebenso wie auf eine Ausarbeitung des BSI bezüglich der Sicherheit gitterbasierter kryptografischer Verfahren [58] verwiesen.

#### 2.2.7.2 Hashbasierte Verfahren

Hashbasierte Ansätze werden für digitale Signaturen verwendet und gelten als die am weitesten ausgereiften und best verstandenen Verfahren. Sie wurden bereits in 1979 von Lamport [154] sowie später von Merkle [177] beschrieben und bauen auf die Sicherheit kryptografischer Hashfunktionen. Die Verfahren gelten als besonders resistent bezüglich möglicher Angriffe durch Quantencomputer [240] und sind flexibel anwendbar, da die Durchführung mit jeder beliebigen sicheren Hashfunktion möglich ist.

Bei vielen effizienten Hash-Verfahren besteht der Nachteil, dass der Signierer die Anzahl der zuvor signierten Nachrichten aufzeichnen muss und ein Fehler in den Aufzeichnungen die Sicherheit gefährden würde. Dies erfordert ein zustandsbehaftetes Schemata, welches einen erheblichen Mehraufwand impliziert. Zudem ist die Anzahl der Signaturen für Schlüssel beschränkt und eine Vergrößerung der Anzahl ist nur durch eine gleichzeitige Vergrößerung der Signatur möglich.

### 2.2.7.3  Isogenien auf supersingulären elliptischen Kurven

Isogeniebasierte Ansätze sind in der Kryptografie vergleichsweise jung und werden in weniger Verfahren verwendet. Sie wurden unter anderem in 2011 durch Jao und De Feo [134] sowie 2014 durch De Feo, Jao und Plût [80] beschrieben und basieren auf der Schwierigkeit, Isogenien zwischen super singulären elliptischen Kurven ausfindig zu machen. Isogeniebasierte Verfahren können bei asymmetrischer Verschlüsselung respektive dem Schlüsselaustausch genutzt werden und ermöglichen sehr geringe Schlüsselgrößen von weniger als 500 Bytes. Die Performanz der Algorithmen ist jedoch mäßig [125].

### 2.2.7.4  Multivarianzbasierte Verfahren

Multivariante Kryptografie basiert auf der Schwierigkeit nicht-linearer, meist quadratischer Polynome über endliche Körper. Die Theorie kann insbesondere für digitale Signaturen herangezogen werden und wurde 1988 zunächst von Matsumoto und Imai [170] beschrieben. Da bei Entschlüsselung lediglich linearer Gleichungssysteme gelöst werden müssen, ist diese sehr effizient durchführbar. Die Schlüsselgrößen sind hingegen übermäßig groß. Hoffman [125] gibt Größen zwischen einem Megabyte und elf Kilobytes an und stellt fest, dass eine genauere Analyse der Verfahren sinnvoll sei. Des Weiteren wurde viele der zunächst veröffentlichten multivarianzbasierten Kryptosysteme bereits gebrochen wurden [56].

### 2.2.7.5  Codebasierte Verfahren

Codebasierte Ansätze nutzen fehlererkennende und -korrigierende Codes und basieren auf der Schwierigkeit zur Dekodierung generischer linearer Codes. Die Theorie wird schon lange Zeit genutzt, um mittels zusätzlicher Redundanz Übertragungsfehler zu ermitteln. Mit ihr können aber insbesondere auch Verschlüsselung und Schlüsselaustausch realisiert werden. Die Idee wurde erstmals in 1978 von McEliece [173] veröffentlicht. Diese Arbeit ist auch direkte Grundlage des NIST PQC Kandidaten Classic McEliece [173]. Insgesamt herrscht ein vergleichsweise ausgeprägtes Verständnis bezüglich der Theorie und es wird bereits seit etwa 40 Jahren an möglichen Sicherheitslücken und Angriffsszenarien geforscht. Laut Hoffman [125] sei das Vertrauen in codebasierte Verfahren daher groß. Zudem ist die Ver- und Entschlüsselung sehr performant durchführbar. Auf der anderen Seite sind die öffentlichen Schlüssel im Vergleich sehr groß und weitere, bis dato unentdeckte Sicherheitslücken können nicht ausgeschlossen werden [230].

### 2.2.8  PQC Kandidaten

Der folgende Abschnitt ist entsprechend der Ausschreibung in Key Exchange (KEX) respektive KEM und Signatur Verfahren unterteilt. Zudem findet sich in den Tabellen 2.4, 2.5, 2.6 sowie 2.7 eine jeweilige Übersicht inklusive der Bytegrößen für öffentliche und private Schlüssel sowie zugehörige Chifffrate respektive Signaturen in Abhängigkeit zu den zugeordneten Sicherheitsstufen. Für eine anschaulichere Darstellung zeigt Abbildung 2.2 eine Einordnung der untersuchten KEM Verfahren anhand eines Graphen, wobei die Abszissenachse die Größen der öffentlichen Schlüssel und die Ordinatenachse die Größen der Ciphertexte abbildet. Für die Einschätzung der Effizienz der Algorithmenoperationen wurden die Ergebnisse der Messungen des Open Quantum Safe (OQS) Profiling Projekts[15] herangezogen. Die Messungen bezüglich der Laufzeiten der KEM Algorithmen werden täglich durchgeführt und mittels entsprechender Visualisierung[16] bereitgestellt.

| Name | Ansatz | Public Key Size | Private Key Size | Data Size |
| --- | --- | --- | --- | --- |
| Classic McEliece | Codes | Level 1: 261120 | Level 1: 6492 | Level 1: 128 |
| | | Level 3: 524160 | Level 3: 13608 | Level 3: 188 |
| | | Level 5: 1044992 | Level 5: 13932 | Level 5: 240 |
| | | Level 5: 1047319 | Level 5: 13948 | Level 5: 226 |
| | | Level 5: 1357824 | Level 5: 14120 | Level 5: 240 |
| CRYSTALS-KYBER | Lattice | Level 1: 800 | Level 1: 1632 | Level 1: 768 |
| | | Level 3: 1184 | Level 3: 2400 | Level 3: 1088 |
| | | Level 5: 1568 | Level 5: 3168 | Level 5: 1568 |
| NTRU | Lattice | | | |
| *NTRU-HPS-2048-509* | | Level 1: 699 | Level 1: 935 | Level 1: 699 |
| *NTRU-HPS-2048-677* | | Level 3: 930 | Level 3: 1234 | Level 3: 930 |
| *NTRU-HPS-4096-821* | | Level 5: 1230 | Level 5: 1592 | Level 5: 1230 |
| *NTRU-HRSS-701* | | Level 3: 1138 | Level 3: 1452 | Level 3: 1138 |
| SABER | Lattice | | | |
| *LightSaber* | | Level 1: 672 | Level 1: 1568 (992) | Level 1: 736 |
| *Saber* | | Level 3: 992 | Level 3: 2304 (1344) | Level 3: 1088 |
| *FireSaber* | | Level 5: 1312 | Level 5: 3040 (1760) | Level 5: 1472 |

Tabelle 2.4: KEM Kandidaten des NIST PQC Standardisierungsprozesses in Runde 3 nach *PQC Wiki* (https://pqc-wiki.fau.edu). Größen in Bytes

#### 2.2.8.1  *Schlüsselaustausch*

**Classic McEliece**[17] war das erste asymmetrische Verschlüsselungsfahren auf Basis fehlererkennender- beziehungsweise -korrigierender Codes und ist der älteste Ansatz des PQC Verfahrens. Classic McEliece in Runde drei ist eine Zusammenfassung aus Classic McEliece sowie NTS-KEM und inkludiert einige Verbesserungen bezüglich Performanz sowie Sicherheit [11]. Der Algorithmus bietet eine sehr hohe Geschwindigkeit bei Verschlüsselung sowie eine akzeptable bei Entschlüsselung. Die Ciphertexte sind sehr kurz, die öffentlichen Schlüssel hingegen extrem groß. Dies ist problematisch im Hinblick auf die Migration bei einem Großteil der Internetanwendungen. In einigen speziellen Fällen könnte diese Eigenschaft jedoch auch von Vorteil sein [127].

---

[15] https://github.com/open-quantum-safe/profiling, zuletzt aufgerufen 06.02.22
[16] https://openquantumsafe.org/benchmarking/visualization/speed_kem.html, zuletzt aufgerufen 06.02.22
[17] https://classic.mceliece.org, zuletzt aufgerufen 02.02.22

| Name | Ansatz | Public Key Size | Private Key Size | Data Size |
|---|---|---|---|---|
| CRYSTALS-DILITHIUM | Lattice | Level 1: 1312 | Level1 : 2528 | Level 1: 2420 |
| | Level 3: 4000 | Level 3: 1952 | | Level 3: 3293 |
| | | Level 5: 2592 | Level 5: 4864 | Level 5: 4595 |
| FALCON | Lattice | Level 1: 897 | Level 1: 1281 | Level 1: 690 |
| | | Level 5: 1793 | Level 5: 2305 | Level 5: 1330 |
| Rainbow | Multivar. | | | |
| *Standard* | | Level 1: 161600 | Level 1: 103648 | Level 1: 66 |
| *Rainbow* | | Level 3: 882080 | Level 3: 626048 | Level 3: 164 |
| | | Level 5: 1930600 | Level 5: 1408736 | Level 5: 212 |
| *CZ-Rainbow* | | Level 1: 60192 | Level 1: 103648 | Level 1: 66 |
| | | Level 3: 264608 | Level 3: 626048 | Level 3: 164 |
| | | Level 5: 536136 | Level 5: 1408736 | Level 5: 212 |
| *Compressed* | | Level 1: 60192 | Level 1: 64 | Level 1: 66 |
| *Rainbow* | | Level 3: 264608 | Level 3: 64 | Level 3: 164 |
| | | Level 5: 536136 | Level 5: 64 | Level 5: 212 |

Tabelle 2.5: Signatur Kandidaten des NIST PQC Standardisierungsprozesses in Runde 3 nach *PQC Wiki* (https://pqc-wiki.fau.edu). Größen in Bytes

**NTRU**[18] wurde von Hoffstien, Pipher and Sillverman entwickelt, kann sowohl für die asymmetrische Verschlüsselung als auch für digitale Signaturen herangezogen werden und gilt als das am meisten untersuchte und best verstandene Verfahren im Bereich der (strukturierten) gitterbasierten Kryptografie ([94]). Es besteht aus einer Zusammenführung von NTRUEncrypt und N'th Degree Truncated Polynomial Ring (NTRU)-HRSS-KEM und basiert nicht wie die meisten gitterbasierten Ansätze auf Ring Learning With Errors (RLWE) und Module Learning With Errors (MLWE), sodass es für eine gewisse Diversität in der Menge der finalen Kandidaten sorgt. NTRU bietet eine sehr hohe Sicherheit im Vergleich zu Verfahren der klassischen Kryptografie wie RSA Easttom u. a. [91] und ist im Gesamtbild sehr effizient [11]. Im Vergleich zu den performantesten gitterbasierten Verfahren kann es jedoch nicht mithalten, wobei dies insbesondere im Bezug auf die langsamere Schlüsselgenerierung zutrifft. NTRU inkludiert zwei Versionen mit unterschiedlichen Kosten-Modellen. NTRU-HPS ist weitestgehend gleich mit der originalen NTRU Version [126] und vergleichsweise einfach aufgebaut. Die Spezifikation bietet mehrere Parametersets für unterschiedliche Sicherheitslevel. NTRU-HRSS [128] ist besonders performant [153], bietet aber weniger Möglichkeiten zur Variation.

---

[18] https://ntru.org/, zuletzt aufgerufen 02.02.22

| Name | Ansatz | Public Key Size | Private Key Size | Data Size |
|---|---|---|---|---|
| BIKE | Codes | Level 1: 1541 | Level 1: 5223 | Level 1: 1573 |
|  |  | Level 3: 3083 | Level 3: 10105 | Level 3: 3115 |
| FrodoKEM | Lattice | Level 1: 9616 | Level 1: 19888 | Level 1: 9720 |
| (AES & SHAKE) |  | Level 3: 15632 | Level 3: 31296 | Level 3: 15744 |
|  |  | Level 5: 21520 | Level 5: 43088 | Level 5: 21632 |
| HQC | Codes | Level 1: 2249 | Level 1: 2289 | Level 1: 4481 |
|  |  | Level 3: 4522 | Level 3: 4562 | Level 3: 9026 |
|  |  | Level 5: 7245 | Level 5: 7285 | Level 5: 14469 |
| NTRU PRIME | Lattice |  |  |  |
| *Streamlined* |  | Level 1: 994 | Level 1: 1518 | Level 1: 897 |
| *NTRUPrime* |  | Level 2: 1158 | Level 2: 1763 | Level 2: 1039 |
|  |  | Level 3: 1322 | Level 3: 1999 | Level 3: 1184 |
|  |  | Level 4: 1349 | Level 4: 1652 | Level 4: 1477 |
|  |  | Level 5: 2067 | Level 5: 3059 | Level 5: 1847 |
| *NTRU* |  | Level 1: 897 | Level 1: 1125 | Level 1: 1025 |
| *LPRime* |  | Level 2: 1039 | Level 2: 1294 | Level 2: 1167 |
|  |  | Level 3: 1184 | Level 3: 1463 | Level 3: 1312 |
|  |  | Level 4: 1455 | Level 4: 1773 | Level 4: 1583 |
|  |  | Level 5: 1847 | Level 5: 2231 | Level 5: 1975 |
| SIKE | Isogeny |  |  |  |
|  |  | Level 1: 378 | Level 1: 434 | Level 1: 402 |
|  |  | Level 3: 462 | Level 3: 524 | Level 3: 486 |
|  |  | Level 5: 564 | Level 5: 644 | Level 5: 596 |
| *Compressed* |  | Level 1: 197 | Level 1: 350 | Level 1: 236 |
|  |  | Level 2: 225 | Level 2: 407 | Level 2: 280 |
|  |  | Level 3: 274 | Level 3: 491 | Level 3: 336 |
|  |  | Level 5: 335 | Level 5: 602 | Level 5: 410 |

Tabelle 2.6: KEM Alternativen des NIST PQC Standardisierungsprozesses in Runde 3 nach *PQC Wiki* (https://pqc-wiki.fau.edu). Größen in Bytes

**Kyber**, beschrieben in [20, 53], bietet ein KEM Verfahren sowie ein Verfahren für einen unautorisierten Schlüsselaustausch und greift auf die gleichen Ansätze wie NewHope [14] zurück. Der Algorithmus basiert auf dem MLWE Ansatz [110, 137], welcher einen effizienten Trade-Off zwischen der gewährleisteten Sicherheit und Performanz bietet. Die Performanz im Gesamten betrachtet sei laut NIST exzellent, das MLWE Problem jedoch vergleichsweise jung und wenig erprobt.

| Name | Ansatz | Public Key Size | Private Key Size | Data Size |
|---|---|---|---|---|
| GeMSS | Multivar. | Level 1: 352188 | Level 1: 16 | Level 1: 33 |
|  |  | Level 3: 1237963 | Level 3: 24 | Level 3: 52 |
|  |  | Level 5: 3040700 | Level 5: 32 | Level 5: 72 |
| Picnic | Hash |  |  |  |
| *Picnic-FS* |  | Level 1: 33 | Level 1: 49 | Level 1: 34036 |
|  |  | Level 3: 49 | Level 3: 73 | Level 3: 76776 |
|  |  | Level 5: 65 | Level 5: 97 | Level 5: 132860 |
| *Picnic-UR* |  | Level 1: 33 | Level 1: 49 | Level 1: 53965 |
|  |  | Level 3: 49 | Level 3: 73 | Level 3: 121849 |
|  |  | Level 5: 65 | Level 5: 97 | Level 5: 209510 |
| *Picnic-full* |  | Level 1: 35 | Level 1: 52 | Level 1: 32065 |
|  |  | Level 3: 49 | Level 3: 73 | Level 3: 71183 |
|  |  | Level 5: 65 | Level 5: 97 | Level 5: 126290 |
| *Picnic3* |  | Level 1: 35 | Level 1: 52 | Level 1: 14612 |
|  |  | Level 3: 49 | Level 3: 73 | Level 3: 35028 |
|  |  | Level 5: 65 | Level 5: 97 | Level 5: 61028 |
| SPHINCS+ | Hash |  |  |  |
| *Small* |  | Level 1: 32 | Level 1: 64 | Level 1: 7856 |
|  |  | Level 3: 48 | Level 3: 96 | Level 3: 16224 |
|  |  | Level 5: 64 | Level 5: 128 | Level 5: 29792 |
| *Fast* |  | Level 1: 32 | Level 1: 64 | Level 1: 17088 |
|  |  | Level 3: 48 | Level 3: 96 | Level 3: 35664 |
|  |  | Level 5: 64 | Level 5: 128 | Level 5: 49856 |

Tabelle 2.7: Signatur Alternativen des NIST PQC Standardisierungsprozesses in Runde 3 nach *PQC Wiki* (https://pqc-wiki.fau.edu). Größen in Bytes

Für die Berechnungen werden der Grad $n$ des Polynomrings, eine Primzahl $q$, welche die Struktur des Rings definiert, eine positiven ganze Zahl $\eta$ für die Binomialverteilung sowie eine ganze Zahl $k$ benötigt. Dabei ist $k \cdot n$ die Dimension des zugehörigen LWE Problems. Die Bereitstellung der unterschiedlichen Sicherheitslevel erfolgt durch Variation der Parameter $k$ und $\eta$. Die Spezifikation definiert drei Konfigurationen entsprechend der innerhalb des Standardisierungsprozesses der festgelegten Sicherheitslevels 1, 3 und 5.

**Saber**[19] bietet einen KEM basierend auf MLWE. Das gitterbasierte Verfahren weise laut NIST eine exzellente Performanz auf und öffentliche Schlüssel

---

19 https://www.esat.kuleuven.be/cosic/pqcrypto/saber/, zuletzt aufgerufen 30.12.21

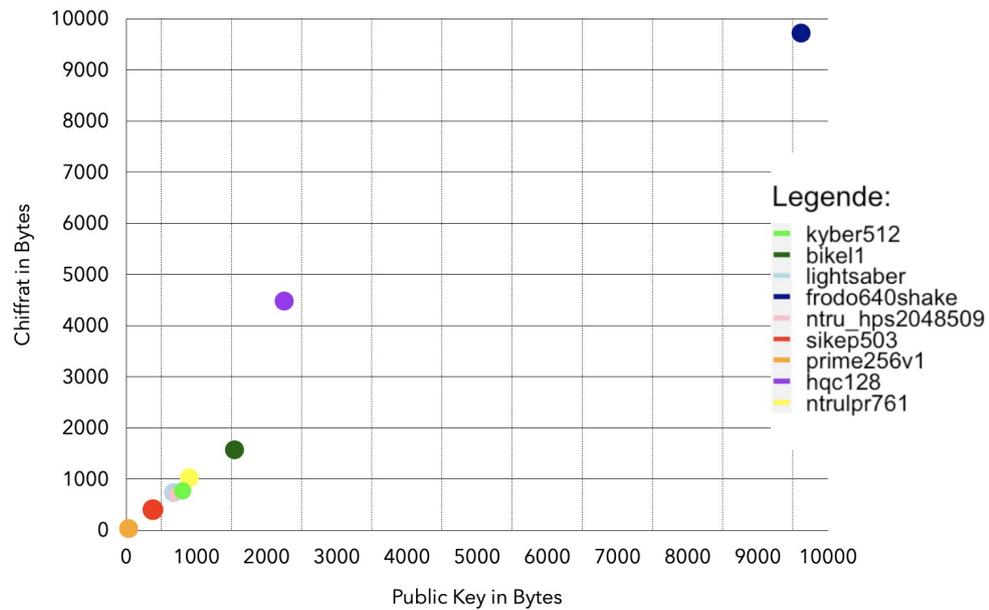

**Abbildung 2.2:** Vergleich der Größen von öffentlichem Schlüssel und Chiffrat in Bytes bei diversen PQC Algorithmen sowie ECDH mit Parameterkonstellationen bezüglich Sicherheitslevel 1. Entnommen aus der Online Spezifikation der C Bibliothek *OQS liboqs* (https://openquantumsafe.org/liboqs/algorithms/, zuletzt aufgerufen 25.01.22)

sowie Chiffrate seien hinreichend klein, sodass der Algorithmus ohne nennenswerte Änderungen in vielen aktuellen Anwendungen integriert werden könne [11]. Saber sei daher ein vielversprechender Kandidat und eine Standardisierung zur Ende der dritten Runde sei denkbar. Trotzdem solle sowohl die Performanz als auch die Sicherheit, insbesondere im Hinblick auf Seitenkanalangriffe, noch intensiver analysiert werden.

#### 2.2.8.2 Authentifizierung

**Crystals Dilithium**[20] ist ein gitterbasiertes Verfahren, aktuell in der dritten Version [21] und basiert auf den Ansätzen MLWE sowie Modular Short Integer Solution (MSIS). Die Implementierung ist vergleichsweise einfach gehalten und das Verhältnis zwischen benötigter Bandbreite und Effizienz der Operationen sehr ausgeglichen, wobei die Schlüsselgrößen ähnlich zu RSA oder DSA sind. Die NIST zieht daher in ihrem Statusbericht das Fazit, dass der Algorithmus für die Anwendung in realen Szenarien sehr geeignet erscheine. Eine ausführliche Beschreibung findet sich auch in [237].

**FALCON**[21] ist ebenfalls ein gitterbasiertes Signaturverfahren und nutzt das SIS Problem. Es erfordert im Vergleich zu allen anderen Signaturverfahren der zweiten Runde die geringste Bandbreite, da der öffentliche Schlüssel sowie die Signatur sehr klein ausfallen. Des Weiteren sind Signatur und Ve-rifikation effizient durchführbar, sodass das Verfahren laut Alagic u. a. [11] vergleichsweise leicht in bestehenden Anwendungen und Protokolle integriert werden könne.

---

20 https://pq-crystals.org/dilithium/, zuletzt aufgerufen 11.01.22
21 https://falcon-sign.info, zuletzt aufgerufen 11.01.22

**Rainbow**[22] basiert auf Multivarianz und wurde bereits 2005 von Ding und Schmidt [86] vorgestellt. Von Vorteil sind die schnelle Signatur und Verifikation sowie die geringe Länge der Signaturen. Sowohl der öffentliche als auch der private Schlüssel sind jedoch mit 500 bis 700 Kilobytes vergleichsweise groß, sodass das Verfahren nicht für den generellen Einsatz innerhalb der bestehenden Infrastrukturen empfohlen werden kann. Insbesondere würden die großen öffentlichen Schlüssel das Volumen der Zertifikatsketten exorbitant steigern. Es existieren jedoch einige Anwendungen, welche nur selten Schlüssel versenden und von den schnellen und kurzen Signaturen profitieren könnten. Zudem erweitert Rainbow die Diversität der Kandidaten aufgrund des multivarianzbasierten Ansatzes.

### 2.2.9 PQC Alternativen

#### 2.2.9.1 Schlüsselaustausch

**SIKE**[23] ist im Vergleich zu allen anderen Kandidaten und Alternativen der zweiten und dritten Runde einzigartig, denn es basiert auf Isogenien elliptischer Kurven. Von großem Vorteil sind die geringen Größen des öffentlichen Schlüssel und Chiffrate. Des Weiteren sind die Parametersets leicht zu skalieren. Nachteilig ist hingegen die schlechte Performanz, welche trotz vieler Verbesserungen im Laufe des Auswahlverfahrens noch immer den meisten Konkurrenten hinterherhinkt. Die NIST sieht SIKE als einen starken Kandidaten für eine Standardisierung in der weiteren Zukunft, da bestimmte Anwendungen die schlechte Performanz für eine geringe Bandbreite in Kauf nehmen könnten. Trotzdem wird zunächst auf weitere Verbesserungen bezüglich der Effizienz gehofft. Auch die Sicherheit müsse laut NIST noch gründlicher erforscht werden.

**HQC**[24] ist ein codebasiertes KEM Verfahren. In der zweiten Runde des NIST Verfahrens konnte gezeigt werden, dass die Fehlerrate bei Entschlüsselung deutlich geringer ausfällt als zunächst vermutet, sodass die Schlüsselgrößen deutlich reduziert werden konnten. Im Vergleich zum direkten Konkurenten Bit Flipping Key Encapsulation (BIKE) besitzen der öffentliche Schlüssel sowie die Chiffrate im Schnitt die 1,6-2-fache beziehungsweise 4-5-fache Größe. Schlüsselgenerierung sowie Entkapselung sind hingegen deutlich schneller. Da gitterbasierte Ansätze jedoch die Performanz deutlich übertreffen und BIKE im Bezug auf die Bandbreite zu bevorzugen ist, wurde HQC zunächst nur als Alternative ausgewählt.

**BIKE**[25] ist ein strukturiertes, codebasiertes KEM Verfahren. Die Performanz sei ausgewogen [11], ähnlich zu gitterbasierten Ansätzen, jedoch mit einer langsameren Entkapselung und Schlüsselgenerierung. Die benötigte Bandbreite ist aufgrund kleinerer öffentlicher Schlüssel und Chiffrate verringert. Auf der anderen Seite benötigt BIKE größere private Schlüssel als viele andere Algorithmen. Für einen Großteil der Anwendungsfälle ist dies irrelevant, bei ressourcenbeschränkten Geräten kann der erhöhte Verarbeitungsaufwand und Speicherbedarf jedoch zu Problemen führen. Die NIST bemängelt im aktuellen Statusbericht, dass die eingereichte Spezifikation 4.1 [17] lediglich Parametersets für die Sicherheitslevel 1 und 3 bietet.

---

22 `https://www.pqcrainbow.org`, zuletzt aufgerufen 11.01.22
23 `https://sike.org`, zuletzt aufgerufen 11.01.22
24 `https://pqc-hqc.org/`, zuletzt aufgerufen 11.01.22
25 `http://bikesuite.org/`, zuletzt aufgerufen 12.01.22

**FrodoKEM**[26] ist in [185] spezifiziert. Der theoretische Ansatz wurde zunächst in [51] beschrieben und basiert zu großen Teilen auf LWE [210, 211], sodass es als gut erforscht und kryptografisch analysiert gilt [11]. Aufgrund der erhöhten Sicherheit müssen kleinere Abstriche bei der Performanz in Kauf genommen werden. Die NIST erachtet die Anwendung vor allem für sinnvoll, wenn ein hohes Vertrauen in die Sicherheit erforderlich ist. Systeme, die auf eine hohe Performanz angewiesen sind, wie beispielsweise TLS Server, könnten hingegen aufgrund der aufwendigen Berechnungen negativ beeinflusst werden. Dennoch ist FrodoKEM performanter als die meisten konservativeren Ansätze, vor allem bezüglich Schlüsselgenerierung und -größe. Zunächst verfügte FrodoKEM nur über Parameterkonstellationen für die Sicherheitsstufen 1 und 3. Für Runde 3 wurde jedoch Level 5 ergänzt. Dabei werden jeweils zwei Varianten angeboten: zum einen die pseudozufällige Generierung einer großen öffentlichen Matrix $A$ mittels AES-128 oder alternativ eine pseudozufällige Generierung mit SHAKE128 aus der SHA-3 Familie. Die SHAKE Varianten weisen in der Regel eine bessere Performanz auf. AES sollte jedoch bevorzugt werden, wenn eine entsprechende Hardwarebeschleunigung vorhanden ist.

**NTRU Prime**[27] ist eine Kombination aus den gitterbasierten KEMs Streamlined NTRU Prime sowie NTRU LPRime und wurde mit der Intention entwickelt, potenzielle Sicherheitslücken bei NTRU zu schließen. Auch, wenn bis dato noch kein konkreter Angriff gegen die betroffenen Strukturen von NTRU gefunden werden konnte. Die Versionen ähneln sich in den meisten Designaspekten, Streamlined orientiert sich jedoch bezüglich der algebraischen Struktur mehr am NTRU-Ansatz, welcher bereits gründlich kryptoanalytisch untersucht wurde. LPRime bezieht sich hingegen auf RLWE, basierend auf der Arbeit von Lyubashevsky, Peikert und Regev [163]. Beide Versionen verfügen seit der dritten Runde über Sicherheitsparameter für alle Sicherheitslevel. Die NIST schreibt jedoch in ihrem Statusbericht, dass die Einstufung noch Fragen aufwerfe und im Vergleich zu anderen Kandidaten ungleich wirke, sodass weitere Forschungsarbeit notwendig sei.

### 2.2.9.2   Authentifizierung

**GeMSS** ist ein multivariantes Signaturverfahren und bietet von allen Kandidaten die kürzesten Signaturen und zudem eine schnelle Verifizierung. Des Weiteren gilt das zugrundeliegende mathematische Problem als stabil und erforscht [200]. Auf der anderen Seite sind öffentliche Schlüssel sehr groß, die Implementierung vergleichsweise komplex und die Signierung langsam. Bei vielen gängigen Anwendungen wie TLS oder Secure Shell (SSH) führe diese Charakteristik laut NIST zu Problemen. Es gebe jedoch auch mögliche Einsatzgebiete. Es werden die Sicherheitslevel 1, 3 und 5 unterstützt, eine Implementierung des Algorithmus ist in OQS respektive liboqs aber nicht verfügbar.

---

[26] `https://frodokem.org/`, zuletzt aufgerufen 12.01.22
[27] `https://ntruprime.cr.yp.to`, zuletzt aufgerufen 12.01.22

**Picnic**[28] ist ein hashbasiertes Signaturverfahren, dessen Sicherheit insbesondere auf der gewählten Hashfunktion sowie auf der LowMC Blockchiffre [13] basiert. Öffentliche Schlüssel sind groß, Signaturen hingegen klein und Signatur sowie Verifikation vergleichsweise langsam. Des Weiteren beschreiben Gellersen, Seker und Eisenbarth [114], dass die einfachste Implementierung ein erhöhtes Potential für Seitenkanalangriffe bietet. Die NIST stellt fest, dass der Ansatz vergleichsweise jung sei und sich stetig weiterentwickle, sodass zwar Potential vorhanden, aber weitere Forschungsarbeit nötig sei.

**SPHINCS+**[29] ist ein zustandloses hashbasiertes Signaturverfahren, welches auf Grundlage des aktuellen Wissensstand als sehr sicher erscheint. Es basiert auf einer hierachischen Struktur von Merkle Hash-Bäumen [29]. Aufgrund der Zustandlosigkeit sind die Signaturen jedoch sehr groß und zudem sind die Operationen vergleichsweise langsam.

OQS respektive liboqs spezifiziert für SPHINCS+ 36 unterschiedliche Parametersets, welche sich aus folgenden Parametern zusammensetzt:

- Hashfunktion: SHAKE-256, SHA-256 und Haraka

- Sicherheitslevel: 1, 3 und 5

- Trade-Off zwischen Signaturgröße (*Small*) und Geschwindigkeit (*Fast*)

- robuste oder einfache Implementierung

In Tabelle 2.7 sind lediglich 6 Variationen aufgeführt, da sich die Größen für die unterschiedlichen Hashfunktionen und Implementierungen nicht signifikant unterscheiden.

### 2.2.10  *Hybrider Modus*

Hybride Ansätze kombinieren einen PQC Algorithmus mit einem Algorithmus der gleichen Sicherheitseinstufung aus der klassischen Kryptografie. Die gleiche Funktionalität wird zweifach und redundant ausgeführt. Es handelt sich um einen konservativen Ansatz, der versucht, die Sicherheit anhand von klassischer und quantensicherer Kryptografie parallel zu gewährleisten. Das Verfahren ist sicher, solange eine der Technologien den Bedrohungen standhält. Diese Vorgehensweise wurde vor PQC in dieser Art noch nicht umgesetzt und birgt einen gewissen Mehraufwand, zunächst bei der Implementierung und später bei den doppelten Ausführungen der gleichen

---

28  https://microsoft.github.io/Picnic/, zuletzt aufgerufen 11.01.22
29  https://sphincs.org, zuletzt aufgerufen 11.01.22

Funktionalität. Trotz allem benennen Crockett, Paquin und Stebila [77] zwei wesentliche Gründe für die Einführung hybrider Verfahren:

1. Beachtung und Einhaltung veralteter Regularien seitens Regierung oder Industrie.

2. Minimierung des Risikos neuer und wenig getesteter Algorithmen trotz Nutzung quantensicherer Kryptografie.

Gerade im Hinblick auf die noch nicht vollständig erforschten PQC Algorithmen könnte dieser Ansatz sinnvoll sein. Er ist sowohl bei Schlüsselaustausch als auch Authentifizierung möglich.

Hybride Verfahren sind kein Teil des NIST Standardisierungsprozesses [242]. Sie sind aktuell vorrangig für den Austausch symmetrischer Schlüssel vorgesehen, denn bei Signaturen zum Nachweis der Authentizität eines Gegenübers genügt eine Absicherung zum Zeitpunkt der Durchführung. Ein Schlüssel kann jedoch nachträglich, das heißt bei Verfügbarkeit leistungsstarker Quantencomputer, kompromittiert und die darauf folgenden Nachrichten ausgespäht werden, sofern der Kommunikationsverlauf aufgezeichnet wurde. Die aktuell diskutierten Ansätze bezüglich eines konkreten hybriden Vorgehens bei Schlüsselaustausch sowie Authentifizierung, auch im Kontext von TLS, werden in Abschnitt 3.1 vorgestellt.

## 2.3 TRANSPORT LAYER SECURITY

TLS ermöglicht die Authentifizierung der Kommunikationsendpunkte im Netzwerk sowie die Errichtung eines sicheren Kommunikationskanals zwischen diesen. Präzise ausgedrückt, werden mithilfe von Mechanismen zur zertifikatbasierten Authentifizierung, Verschlüsselung und Sicherung der Nachrichtenintegrität die Schutzziele Vertraulichkeit, Integrität, Authentizität und Nichtabstreitbarkeit sichergestellt [165]. Dabei werden sowohl Algorithmen aus der symmetrischen als auch asymmetrischen Kryptografie genutzt, sodass TLS auch als hybrides Protokoll bezeichnet wird. Die Durchführung der Prozessschritte erfolgt transparent, weshalb das Protokoll leicht konfigurierbar und universell einsetzbar ist. Ursprünglich wurde das Verfahren von der Internet Engineering Task Force (IETF) als Standardprotokoll für die sichere Durchführung von E-Commerce Diensten im Internet entwickelt [165]. Inzwischen hat sich der Anwendungsraum jedoch auf das gesamte Internet ausgeweitet. Das Protokoll findet in allen HyperText Transfer Protocol Secure (HTTPS) Verbindungen Verwendung [16] und ist mit einem Anteil von über 60 % das meist genutzte Protokoll zur Absicherung von Kommunikationsverbindungen in Netzwerken [235], [236]. Schwenk [227] beschreibt den Standard als äußerst stabil und sehr gepflegt. Zudem sei aufgrund der vielen Neuerungen in Version 1.3 auch eine zukünftige Nutzung empfehlenswert.

2.3.1 *Geschichte*

In den Anfängen des Internets wurden Aspekte bezüglich Sicherheit und Privatheit von Nutzern und Daten nicht in Acht genommen. Grundlegende Protokolle wie Hyper Text Transfer Protocol (HTTP) bieten keine eigenständigen Sicherheitsmechanismen. Um auch sensitive Daten zu übertragen, entwickelte *NetscapeCommunications*[30] das Protokoll Secure Sockets Layer (SSL). Während die erste Version einige schwerwiegende Mängel aufwies, wurde 1995 die zweite Version öffentlich zugänglich gemacht und in den Browser *Netscape Navigator* integriert. 1996 folgte Version 3.0 um weitere Sicherheitslücken zu beseitigen, bevor die IETF schließlich die Standardisierung übernahm und in 1999 die Version SSL 3.1 als Standard unter dem Namen TLS 1.0 [84] veröffentlichte. TLS 1.0 ist nicht mit den SSL Versionen kompatibel. Erst in 2006 folgte die überarbeitete Version 1.1 [83] mit wenigen, aber maßgeblichen Verbesserungen, insbesondere im Bereich des Record Layer. In 2008 erschien TLS 1.2 [85] mit einer Möglichkeit zur authentischen Verschlüsselung und aktualisierten Hashfunktionen. Von 2014 bis 2018 wurde schließlich an der aktuellen Version TLS 1.3 [213] gearbeitet, welche 2018 veröffentlicht wurde und eine Art Revolution darstellte.

TLS 1.3 beinhaltet eine Reihe neuer Ansätze, insbesondere bei Handshake und Schlüsselableitung. Vermeintlich unsichere oder überflüssige Erweiterungen wurden entfernt. Easttom [95] unterteilt die Neuerungen wie folgt:

- Trennung von Algorithmen bezüglich Authentifizierung und Schlüsselaustausch bei Ciphersuiten.

- Ausschluss schwacher oder kaum genutzter elliptischer Kurven.

- Ausschluss der Hashfunktionen MD5 und SHA-224.

- Digitale Signaturen werden grundsätzlich vorausgesetzt.

- Integration von HKDF sowie Semi-Ephemeral DH.

- Optimierung Resumption, siehe 2.3.3.3.

- Verringerung Round Trip Time (RTT) von 2 auf 1.5 sowie Verschlüsselung ab ServerHello, da der Client zu Beginn errät, welche Gruppe der Server bezüglich der elliptischen Kurven auswählen wird. Entsprechende Schlüssel können bereits pro forma gesendet werden und die Schlüsselvereinbarung beginnt im ersten Kommunikationsschritt, wo hingegen vorher im zweiten Schritt der Server begann.

- Erneuter Verbindungsaufbau mit 0 RTT möglich.

- Perfect Forward Secrecy durch vorgeschriebene Nutzung von Ephemeral Keys bei DH Key Agreement.

---

[30] `https://isp.netscape.com`, zuletzt aufgerufen 18.11.21

- Eingestellter Support für diverse unsichere oder überflüssige Features in Bereichen wie Kompression, Neuvereinbarung der Parameter, benutzerdefinierte DH Schlüsselaustausch (DHE) Gruppen oder Change Cipher Spec Protokoll.

- Verbot von SSL oder RC4 auf Kosten der Abwärtskompatibilität.

- Einführung Session Hash.

- Versionsnummer in Record Layer depracted.

- Einführung ChaCha20 Stromchriffre mit Poly1305 MAC.

- Einführung Ed25519 and Ed448 zur digitalen Signatur.

- Einführung x25519 sowie x448 Schlüsselaustausch Protokolle.

Mechanismen wie ClientKeyShare, HKDF sowie Semi-Ephemeral DH, Resumption und Pre-Shared Key (PSK) werden in den folgenden Abschnitten nochmals erläutert. Bei ChaCha handelt es sich um ein symmetrisches Verfahren aus dem Bereich der Stromchiffren, siehe Langley u. a. [155]. Ed25519 und Ed448 sowie x25519 und x448 werden unter anderem in Bernstein [34], Bernstein u. a. [36, 37], Bessalov u. a. [38] und Langley, Hamburg und Turner [156] näher beschrieben.

Aktuell findet TLS bei einem Großteil der Verbindungen innerhalb des Internets Verwendung [191]. Der Prozess bezüglich der Umstellung zu Version 1.3 verläuft jedoch schleppend.

### 2.3.2 Architektur

TLS fungiert als Sicherungsschicht zwischen der Transport- und der Anwendungsebene innerhalb des Internets. Es wird innerhalb der klassischen TCP/IP-Netze in der gängigen Literatur meist oberhalb des Transport Control Protocol (TCP) Protokolls angesiedelt, siehe auch Zhipeng u. a. [272] oder Schwenk [229]. Im Grunde nimmt es die Daten der obersten Schichten aus dem Bereich der Anwendung entgegen, sichert diese entsprechend und gibt sie zur Übermittlung an die Transportschicht weiter, sodass das Paket letztlich den Empfänger erreichen kann. Umgekehrt werden nach Erhalt eines Pakets auch Daten von der Transportschicht entgegengenommen, aufbereitet und an die Anwendung weitergereicht. Der Aufbau ist in Abbildung 2.3 dargestellt. TLS selbst wird dabei nochmals in fünf Unterkomponenten aufgeteilt:

- **TLS Handshake:** wird zu Beginn einer neuen Verbindung genutzt, um kryptografische Modi und Parameter auszuhandeln, Schlüsselmaterial auszutauschen und die Teilnehmer zu authentifizieren. Der Handshake ist die einzige Komponente des TLS Protokolls, welche auf diverse asymmetrische Algorithmen wie RSA oder DH zurückgreift und im Rahmen einer Einführung von PQC Algorithmen entsprechend berücksichtigt werden muss, siehe auch Abschnitt 2.3.3

- **TLS Record:** auch Record Layer, ist für das zentrale Management verantwortlich. Es werden Daten von der Anwendungsschicht als Bytestrom entgegengenommen und in kleinere Pakete aufgeteilt. Bei Bedarf werden diese komprimiert, es wird eine MAC angewandt und sie werden einzeln verschlüsselt, bevor sie schließlich an die Vermittlungsschicht weitergereicht werden. Für die kryptografischen Funktionalitäten werden die im Handshake ausgehandelten Parameter herangezogen, wobei es sich ausschließlich um symmetrische Verfahren handelt

- **TLS Application Data:** Schnittstelle zur oberen Anwendungsebene

- **TLS Alert:** erteilt entsprechende Fehlermeldungen bei fehlerhaften Abläufen. Eine Alert-Nachricht umfasst zwei Bytes, wobei das erste Byte die generelle Schwere des Fehlers klassifiziert und das zweite Byte eine genauere Beschreibung bezüglich der Art des Fehlers liefert

- **TLS ChangeCipherSpec**: wird versendet, nachdem der Handshake weitestgehend erfolgreich durchgeführt wurde und alle Informationen bezüglich der Verschlüsselung ausgetauscht und ausgehandelt wurden. Nach Versand der Nachricht wird die ausgehandelte Verschlüsselung für die weitere Kommunikation aktiviert. Mitunter wird das ChangeCipherSpec Protkoll dem TLS Handshake untergeordnet.

Handshake und Record Layer werden meist als Hauptkomponenten bezeichnet, da sie die grundlegenden Funktionalitäten beinhalten und die Charakteristik von TLS prägen. Sobald die Optionen für die Verschlüsselung ausgehandelt wurden verläuft die gesamte weitere Kommunikation verschlüsselt.

### 2.3.3 TLS Handshake

Das Handshake Protokoll ist das zentrale Protokoll beim Aufbau einer neuen TLS Verbindung. Es dient der Authentifizierung der Kommunikationspartner und dem Austausch der Schlüssel für den Aufbau eines sicheren Kommnikationskanals.

Nach dem Standard von TLS 1.2 wurde der Handshake für den Schlüsselaustausch mithilfe von DH oder RSA konzipiert. Seit TLS 1.3 ist zunächst die Nutzung von ECDH vorgesehen [226]. Die Wahl der Algorithmen und -parameter erfolgt entsprechend der Fähigkeiten der Kommunikationspartner. Von den Konfigurationen, die beide Teilnehmer unterstützen, sollte theoretisch immer die sicherste verwendet werden.

#### 2.3.3.1 TLS Ciphersuites

Bei Ciphersuites innerhalb von TLS handelt es sich um einen String, welcher aus einer Verkettung von Kürzeln besteht. Sie spezifizieren die Protokolle und Parameter, welche während des Handshakes ausgehandelt werden. In TLS 1.2 wurden Zertifikatstypen, HKDFs, MACs, Algorithmen für den Schlüsselaustausch, Authentifizierung und symmetrische Verschlüsselung sowie der

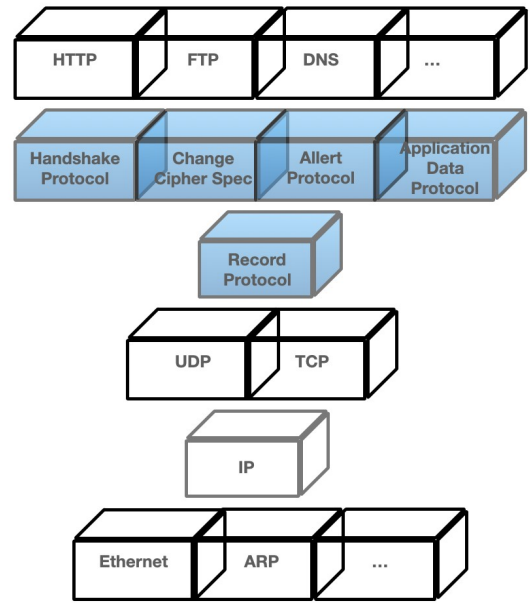

Abbildung 2.3: Eingliederung der TLS Komponenten innerhalb des Protokollstapels des Internets (TCP/IP-Netzwerke) angelehnt an Abbildung 4.1 aus [192]

Betriebsmodi bezüglich der symmetrischen Verschlüsselung innerhalb der Ciphersuite angegeben. Ein gültiges Beispiel wäre

```
TLS_ECDHE_RSA_WITH_AES_128_GCM_SHA256
```

Alle zulässigen Werte sind dabei innerhalb einer zentralen Liste angegeben, bereitgestellt durch die Internet Assigned Numbers Authority (IANA).

Mit Version 1.3 wurden viele unsichere oder veraltete Optionen entfernt. Die Ciphersuiten geben zudem keine Schlüsselaustausch- und Signaturalgorithmen mehr an. Diese Komponenten werden separat ausgehandelt. Aktuell werden fünf Ciphersuiten unterstützt:

```
TLS_AES_128_GCM_SHA256
TLS_AES_256_GCM_SHA384
TLS_CHACHA20_POLY1305_SHA256
TLS_AES_128_CCM_SHA256
TLS_AES_128_CCM_8_SHA256
```

Für den Schlüsselaustausch stehen zusätzlich die Optionen Elliptic Curve Diffie-Hellman Ephemeral (ECDHE), DHE, PSK oder PSK mit einem DH-Verfahren zur Auswahl. Bei Signatur kann zwischen RSA, ECDSA, Edwards-Curve Digital Signature Algorithm (EdDSA) oder symmetrischem PSK gewählt werden.

2.3.3.2 *Ablauf*

1. Zu Beginn sendet der Client ein *ClientHello* an den Server, welches kryptografische Informationen wie die unterstützen Protokolle, Algo-rithmen und Parameter enthält. Für den Austausch des *Ephemeral Keys* wird seit TLS 1.3 ECDH bevorzugt. Mittels einer Liste in der Erweiterung *supported_groups* können die unterstützten Gruppen elliptischer Kurven vermittelt werden. Zudem errät der Client „auf Verdacht", welches Protokoll der Server zur Schlüsselvereinbarung wählen wird und kann einen oder mehrere korrespondierende Schlüsselwerte innerhalb der *keyshare* Erweiterung mit dem Server teilen. Für Signaturverfahren definiert die Erweiterung *signature_algorithms_cert*, welche Signaturen innerhalb der Zertifikate unterstützt werden. *signature_algorithms* gibt an, welche Algorithmen im Protokoll selbst genutzt werden können.

2. Sofern der Client das korrekte Verfahren vorhergesehen hat, kann der Server anhand des mitgeteilten öffentlichen Schlüssels einen Schlüsselwert berechnen und den Sitzungsschlüssel ableiten. Abschließend sendet er ein *ServerHello*, welches die gewählte Ciphersuite, das gewählte Key Agreement Protokoll, sowie den *keyshare* und das Zertifikat des Servers enthält. Ein Teil der zu übertragenden Daten wird durch den Server signiert und mithilfe des Sitzungsschlüssels kann das Zertifikat bereits verschlüsselt werden. Da der Server den Sitzungsschlüssel bereits innehält, sendet er zudem sofort das *Server Finished*.

3. Der Client kann nun ebenfalls den Sitzungsschlüssel mithilfe seines privaten Schlüssels und dem vom Server mitgeteilten Chiffrat berechnen. Er entschlüsselt das Zertifikat und überprüft anschließend dieses sowie die Signatur. Bei erfolgreicher Überprüfung sendet er ein *Client Finished*.

Nach der erfolgreichen Durchführung des Handshakes ist der Server authentifiziert und alle Nachrichten zwischen beiden Parteien können vertraulich ausgetauscht werden. Durch das Erraten des bevorzugten Schlüsselvereinbarungsprotokolls kann in der Regel ein Durchlauf mit 1,5 RTT erreicht werden. Sollte der Client falsch raten, kann der Server mittels eines *HelloRetryRequest* die unterstützte Gruppe übermitteln. Gegebenenfalls fordert der Server zudem eine Authentifizierung des Client, sodass dieser ebenfalls sein Zertifikat übermittelt.

### 2.3.3.3 Resumption & 0-RTT

Bestand bereits eine Verbindung zwischen Client und Server, kann der Neuaufbau beschleunigt werden, indem mittels Session-IDs und Session-Tickets auf die bereits getroffenen Vereinbarungen zurückgegriffen wird. In TLS 1.2 erfolgte dies in einem 1-RTT Handshake. In Version 1.3 werden diese Informationen jedoch „nebenher" in der *early_data* Erweiterung übertragen. Somit kann ein 0-RTT Wiederaufbau erfolgen. Es sollte jedoch beachtet werden, dass beim Versenden der Nachricht des Clients keine Forward Secrecy bezüglich eines kompromittierten Session-Ticket-Keys gewährleistet werden kann. Des Weiteren sind sogenannte Replay-Attacken möglich. Ein Angreifer kann die Nachricht eines Clients erneut senden, ohne dass der Server feststellen kann, ob das Paket bereits gesendet wurde [107]. Server führen daher in der Regel nur 0-RTT-Anfragen aus, welche keinen schreibenden Zugriff erfordern.

#### 2.3.3.4 Perfect Forward Secrecy

Perfect Forward Secrecy bezieht sich auf Protokolle für den Schlüsselaustausch und beschreibt die Eigenschaft, dass eine Kompromittierung der ausgehandelten Schlüssel auch dann nicht möglich ist, wenn die zum Austausch verwendetem Langzeitschlüssel bekannt werden. In der Regel geht dies jedoch mit einem höhere Aufwand zur Generierung von Sitzungsschlüsseln und geringerer Verarbeitungsgeschwindigkeit einher.

### 2.3.4 Implementierungen & OpenSSL

Für TLS ist eine Vielzahl an Implementierungen in annähernd allen Programmiersprachen verfügbar.

OpenSSL ist eine der beliebtesten und weitverbreitetsten Implementierungen [95]. Es ist quelloffen[31] und auf allen gängigen Betriebssystemen nutzbar. Laut Allende u. a. [15] sei OpenSSL inzwischen der defacto Standard für die Anwendung von TLS. Das Projekt wurde in der Sprache C geschrieben und startete bereits in 1998, wobei es auf dem Projekt SSLeay von Eric Young und Tim Hudson aufbaute. Das Tool umfasst diverse Implementierungen von symmetrischen Chiffren, kryptografischen Hashfunktionen, asymmetrischen Algorithmen und Netzwerkprotokollen. Die wichtigsten Bausteine sind:

- **libssl**: Implementierung aller TLS Versionen bis 1.3

- **libcrypto**: Bibliothek mit allen kryptografischen Mechanismen, welche für TLS notwendig sind. libcrypto kann jedoch auch unabhängig von TLS genutzt werden

- **openssl**: Kommandozeilenanwendung, welche unter anderem das Erzeugen von Schlüsselparametern und X.509 Zertifikaten oder die Ver- und Entschlüsselung von Nachrichten ermöglicht.

Im gesamten betrachtet stellt OpenSSL daher eine Vielzahl von Funktionalitäten bereit. Zu den wichtigsten zählen unter anderem:

- Generierung von Pseudo-Zufallszahlen

- Generierung von privaten und öffentlichen Schlüsseln

- Certificate Authority (CA) Management

- Validierung von Zertifikaten

---

[31] `https://github.com/openssl/openssl`, zuletzt aufgerufen 15.12.21

- Management von Krypto-Bibliotheken und Plugins für den Support neuer Algorithmen

- Bereitstellung TLS Clients und Server

Mbed TLS[32] bietet zudem eine TLS Implementierung mit minimiertem Funktionsumfang. Für die Ausführung werden daher 60 Kilobyte Speicher und 64 Kilobyte Arbeitsspeicher benötigt, sodass die Bibliothek auch in eingebetteten System genutzt werden kann.

Weitere bekannte Implementierungen sind BoringSSL[33] und LibreSSL[34].

#### 2.3.4.1 Datagram TLS

Da User Datagram Protocol (UDP) Verbindungen keine Absicherung bezüglich der fehlerfreien Übertragung bieten und insbesondere die Reihenfolge der Pakete nicht garantiert wird, kann das klassische TLS Protokoll in diesem Kontext nicht genutzt werden. Um dem entgegen zu wirken, wurde Datagram TLS (DTLS) entwickelt, welches unterschiedliche Änderungen enthält und daher mit UDP-basierten Anwendungen kompatibel ist. Wichtigste Anforderung ist die zustandslose Entschlüsselung. DTLS und die durchgeführten Anpassungen im Bezug auf TLS 1.2 sind in RFC 6347 festgehalten [214]. Eine Version bezüglich TLS 1.3 wird in einem Entwurf der IETF spezifiziert [215] und ist aktuell noch nicht abgeschlossen. Für weiterführende Informationen zu DTLS siehe [182, 245].

### 2.4 NETZWERKE UND DEREN SIMULATION

Computernetzwerke ermöglichen Verbindungen zwischen zwei oder mehr Kommunikationspartnern und dienen dem Austausch von unterschiedlichsten Informationen. Sie sind Grundbaustein des Internets und auch das in dieser Arbeit fokussierte Protokoll TLS ist Teil dieser Struktur. Die Evaluierung der Performanz sollte daher aufgrund der bestehenden Abhängigkeiten und Restriktionen innerhalb der Netze nicht losgelöst betrachtet werden. Für eine praxisnahe Evaluierung sollten Experimente innerhalb realer Netze oder mittels geeigneter Simulation respektive Emulation durchgeführt werden.

Performanz und Qualität von Netzwerkverbindungen respektive der dort vorgenommenen Datenübertragungen kann durch diverse Parameter quantifiziert und beschrieben werden. Dies ist unter anderem notwendig, um eine gegebene Verbindung bewerten und demgemäß emulieren zu können. In diesem Abschnitt findet sich daher, neben einer Einführung in die Netzwerksimulation und das verwendete Werkzeug **netem**, eine Einführung in die wesentlichen Einflussfaktoren der Netzwerkcharakteristik.

---

[32] https://www.trustedfirmware.org/projects/mbed-tls/, zuletzt aufgerufen 19.12.21
[33] https://boringssl.googlesource.com/boringssl/, zuletzt aufgerufen 15.12.21
[34] https://www.libressl.org, zuletzt aufgerufen 15.12.21

2.4.1   *Netzwerkstruktur & -aufbau*

Die Datenvermittlung innerhalb eines Netzwerkes kann grundsätzlich in zwei Kernaspekte unterteilt werden, welche die Charakteristik und insbesondere die Performanz der Übertragung bestimmen: Zum einen die direkte Übertragung der Signale auf dem jeweils verfügbaren Medium und zum anderen die Verarbeitung der Daten, inklusive Koordination und Verwaltung, an den End- und eventuellen Zwischenknoten innerhalb der Verbindungen.

*Datenübertragung*

Für die Datenübertragung können unter anderem elektrische Signale, elektromagnetische Wellen, Schall oder optische Signale wie Lichtleiter oder Laser genutzt werden. Die Signale werden mithilfe entsprechender Medien übermittelt. Bei kabellosen Technologien erfolgt die Kommunikation beispielsweise über elektromagnetische Wellen. Hoch frequentierte, rechenintensive Server sind hingegen meist an Glasfasernetze angeschlossen, welche optische Signale übermitteln. Die Auswahl der Technologien erfolgt je nach Anwendungsfall und Verfügbarkeit. Bei Vehicle to Vehicle (V2V)- oder Machine to Machine (M2M)-Kommunikation im Bereich IoT werden oft direkte, drahtlose Verbindungen genutzt, um einen flexiblen und performanten Austausch zu ermöglichen [231]. Meist kommen jedoch mehrere Technologien zum Einsatz und der Datenaustausch verläuft über mehrere Subnetze und Zwischenknoten hinweg. Ruft beispielsweise ein Nutzer mit seinem Laptop innerhalb des Heimnetzwerks im Browser eine Webseite auf, so können die Informationen zunächst mithilfe elektromagnetischer Wellen zum Router übertragen werden. Je nach Art des Routers leitet dieser die Daten anschließend weiter an einen Verteiler, welcher im günstigen Fall an ein Glasfasernetz angeschlossen sein könnte und die Daten somit über optische Signale übermittelt, bis sie, möglicherweise über weitere Zwischenschritte, letztlich den Server erreichen.

Jede Technologie hat andere Übertragungseigenschaften, welche sich auf die Netzwerkcharakteristik auswirken. Kabelverbindungen bieten beispielsweise in der Regel sehr gute Voraussetzungen, um möglichst viele Daten in kurzer Zeit zu übertragen. Sofern die Leitungen entsprechend präpariert sind, bleibt die verfügbare Bandbreite in etwa konstant. Funkübertragungen sind wesentlich störanfälliger, da die Signale über die räumliche Umgebung hinweg übertragen werden und diese, entsprechend der Bewegungen des Senders, Empfängers und der umgebenden Hindernisse, stark variieren können [212].

*Datenverarbeitung, Koordination & Vermittlung*

Neben der reinen Übertragung der Informationen über ein oder mehrere Medien wird die Verbindung und insbesondere ihre Qualität durch die Verarbeitung der Bits respektive Datenpakete beeinflusst [19]. Durch die Verarbeitungsschritte an den Knotenpunkten wird eine möglichst effiziente, feh-lerfreie und sichere Zustellung an einen bestimmten Empfänger ermöglicht. Hierzu werden unterschiedliche Prozesse und Protokolle herangezogen. Die für die Kommunikation mit TLS meist verwendeten Protokolle werden im Folgenden kurz beschrieben. Zum besseren Verständnis werden die Schritte innerhalb des international anerkannten ISO/OSI-Schichtenmodells [209] eingeordnet und beschrieben. Das Modell unterteilt die Prozesse in insgesamt sieben Schichten, welche aufeinander aufbauen, wobei sich der Abstraktionsgrad der Funktionen schichtweise erhöht.

1. **Physical Layer**: ist zuständig für die grundlegende Übermittlung der einzelnen Bitwerte und inkludiert die Aktivierung, Deaktivierung und Aufrechterhaltung der Verbindung. Für die Interpretation der Signale werden entsprechende Codierungstechniken eingesetzt.

   Wie beschrieben, differieren die verwendeten Technologien, und innerhalb einer Verbindung können mehrere Übertragungstechniken zum Einsatz kommen. Die verwendeten Übertragungsmedien beeinflussen die Qualität der Netzwerkverbindung massiv, wobei kabellose Übertragung beispielsweise meist störanfälliger ist als kabelgebundene Varianten [19]. Eine entscheidende Rolle spielt zudem die Anzahl der Nutzer, welche sich das Medium teilen müssen.

2. **Data Link Layer**: (Sicherungsschicht) sorgt für eine möglichst zuverlässige Übertragung der Daten. Übertragungsfehler werden detektiert und, sofern möglich, korrigiert. Zudem erfolgt innerhalb dieser Schicht die Regelung von konkurrierendem Zugriff durch eine angemessene Flusskontrolle. Protokolle von Übertragungs- und Sicherungsschicht sind meist eng miteinander verbunden.

3. **Network Layer**: (Vermittlungsschicht) ist für die Weiterleitung von Datenpaketen beziehungsweise die Verknüpfung von einer Verbindung zwischen zwei Kommunikationspartnern verantwortlich. In dieser Arbeit liegt der Fokus auf Internet Protocol (IP) [204], welches weltweit innerhalb des Internets eingesetzt wird [226]. Es dient insbesondere der logischen Adressierung zur Identifikation der Kommunikationsteilnehmer sowie der Wegfindung (Routing) innerhalb des Netzwerks. Des Weiteren werden die Pakete fragmentiert, um die Kommunikation über verschiedene physikalische Netze hinweg zu ermöglicht. Die Pakete umfassen jeweils einen sogenannten Header, welcher Metadaten über das Paket bereit hält, sowie einen Body, welcher die eigentlichen Nutzdaten enthält. Die maximale Länge von IP-Datenpaketen ist auf 65.535 Bytes beschränkt, wobei 20 bis 60 Bytes für den Header benötigt werden.

   IPv6 [81] ist Nachfolger der ersten Version, welche auch als IPv4 bezeichnet wird, und hat insbesondere zum Ziel, bekannte Einschränkungen des Vorgängers zu beseitigen. So wurden Adressen von 32 Bit auf 128 Bit erweitert, um den weltweit gestiegenen Bedarf zu decken. Weiter sind eine Autokonfiguration der IPv6-Adressen, ein schnelleres

Routing und größere Datenpakete von bis zu 4 Gigabyte möglich. Mit IPsec [32] werden zudem zusätzliche Mechanismen zur Sicherung von Vertraulichkeit, Authentizität und Integrität angeboten. Der Umstieg von IPv4 auf IPv6 begann bereits im Dezember 1998, ist jedoch noch nicht abgeschlossen. Die Versionen werden aktuell parallel betrieben.

4. **Transport Layer**: dient insbesondere der Segmentierung der Daten, der reibungslosen Übertragung und der Flusskontrolle. Sie ermöglicht eine gesicherte und direkte Ende-zu-Ende-Kommunikation zwischen zwei Anwendungsprozessen, welche durch IP nicht gegeben ist. Die Identifikation der beteiligten Prozesse erfolgt über sogenannte Portnummern. Die Protokolle der Transportschicht arbeiten unabhängig von darunterliegenden Schichten und dem gewählten Übertragungsmedium. Des Weiteren ist die Netzwerkübertragung transparent und tieferliegende Protokolle sind oberen Schichten verborgen. Oftmals wird bei Internetverbindungen ausschließlich auf TCP in Kombination mit IP zurückgegriffen [72]. Mithilfe von TCP können Punkt-zu-Punkt-Verbindungen zwischen zwei Teilnehmern mit symmetrischen Regeln aufgebaut werden. Das Protokoll ist zuverlässig, da ein gewisser Schutz gegen Verlust oder Verfälschung der gesendeten Daten gewährleistet wird, paketvermittelnd und zudem verbindungsorientiert, mit explizitem Verbindungsauf- und -abbau zwischen den Kommunikationspartnern [249]. Des Weiteren ist TCP stromorientiert, das heißt, die Daten kommen in der gesendeten Reihenfolge beim Empfänger an und die Übertragung erfolgt gepuffert. Somit kann auch eine Strom- und Überlastungskontrolle gewährleistet werden. Beispielsweise bestätigt der Empfänger grundsätzlich die erhaltenen Pakete mittels einer *ACK*-Nachricht. Bei ausbleibender Bestätigung, aufgrund eines verlorenen oder fehlerhaften Segments, wird ein Paket erneut gesendet. Die Zeit bis zur wiederholten Sendung wird, je nach Netzwerkzustand, variiert. Das Vorgehen zur Berechnung des Retransmission Timeout (RTO) ist in RFC 6198 [219] definiert. Um den Durchsatz zu maximieren, werden Datenpakete zudem teils parallel gesendet. Die Anzahl wird, entsprechend der abgeschätzten Netzwerkkapazität, über das sogenannte *SlidingWindow* festgelegt, siehe auch [167]. Das Sendefenster wird schrittweise vergrößert und bei Verlust eines Pakets wiederum verringert. So wird versucht, eine maximale Auslastung zu erzeugen. Nagle's Algorithmus sorgt dafür, dass kein zu großer Overhead aufgrund von kleinen Datenmengen mit großem Header entsteht. Das genaue Verhalten ist unter anderem in RFC 896 und RFC 1122 beschrieben [54, 73]. Je nach Variante wird auch das sogenannte *Fast-Retransmit* durchgeführt, sofern aufgrund eines fehlenden Pakets wiederholt *ACKs* für die gleiche Sequenznummer eingehen.

Die minimale Größe eines TCP-Segments beträgt 556 Bytes. Dabei werden TCP-Segmente, ebenso wie IP-Pakete, in Header und Body unterteilt. Da der Header 20 Bytes benötigt, bleiben mindestens 536 Bytes für Nutzdaten. Die tatsächliche Größe der Segmente wird zwischen den Kommunikationspartnern ausgehandelt.

Neben TCP kann auch das verbindungslose Protokoll UDP verwendet werden, welches jedoch deutlich weniger Kontrollfunktionen bietet [252]. Vorteile sind hingegen der verringerte Overhead, der entfallende Aufbau und Abbau der Verbindung und die daraus resultierenden schnelleren Prozessabläufe. Die konkreten Auswirkungen auf die Verbindungsqualität bei Wahl von TCP oder UDP sind jedoch noch immer nicht vollständig untersucht und stark von den vorherrschenden Netzwerkmetriken und Szenarien abhängig [1]. Aktuell findet UDP vorwiegend bei fehlertoleranten Anwendungen wie Dynamic Host Configuration Protocol (DHCP), Domain Name System (DNS) oder IP-Telefonie sowie bei Verbindungen mit geringer Latenz Verwendung. TCP hingegen wird meist in Verbindung mit HTTP, Telnet oder File Transfer Protocol (FTP) genutzt.

Zudem führte Google 2012 das Transportprotokoll QUIC ein, welches seit 2021 offizieller IETF Standard ist [132]. Es baut auf UDP auf und soll als Alternative zu TCP eine performante sowie verbindungsorientierte Übertragung von Anwendungsdaten ermöglichen.

5. **Session Layer**: (Sitzungsschicht) dient dem organisierten und synchronisierten Datenaustausch zwischen den Prozessen zweier Systeme, insbesondere bei Unterbrechung der Verbindung.

6. **Presentation Layer**: (Darstellungsschicht) versucht systemabhängige Formate zu abstrahieren, um einen fehlerfreien Austausch zu ermöglichen. Auch Verschlüsselung und Kompression der Daten können innerhalb dieser Schicht umgesetzt werden.

7. **Application Layer**: (Anwendungsschicht) stellt die Schnittstelle mit dem Nutzer und ermöglicht die Ein- und Ausgabe von Daten

TLS, welches der Vertraulichkeit, Authentisierung und Integritätssicherung dient, ist zwischen Anwendungs- und Transportschicht anzusiedeln, siehe auch Abschnitt 2.3.

### 2.4.2 *Netzwerkmetriken*

#### 2.4.2.1 *Datenrate*

Die Datenrate bezeichnet grundsätzlich die Menge an Daten, welche innerhalb einer bestimmten Zeitspanne übertragen werden kann. Der *Datendurchsatz* berücksichtigt hingegen nur die übertragenen Nutzdaten. Die Angabe der Datenrate erfolgt in übertragenen Bits pro Sekunde. In der Praxis finden aufgrund der gängigen Größen oftmals Kilobits pro Sekunde (Kbps) beziehungsweise Megabits pro Sekunde (Mbps) Verwendung.

Zusammen mit der Latenzzeit ist die Übertragungsrate maßgebend für die Leistungsfähigkeit des Verbindungskanals. Ihre Größe ist größtenteils abhängig vom zugrundeliegenden Verfahren. Entsprechend wird in der Praxis nach Technologien gesucht, welche möglichst schnell möglichst viele Daten übertragen können.

Bei wellenartigen Signalen ist die Bandbreite wesentlich für eine effektive Übertragung. Sie beschreibt die Differenz der Grenzfrequenzen innerhalb derer die Signale übertragen werden und wird daher in Hertz angegeben. Je größer die Spanne desto mehr Daten können innerhalb eines bestimmten Zeitabschnitts übertragen werden, weshalb die Begriffe Übertragungsrate und Bandbreite fälschlicherweise oft als Synonym verwendet werden.

### 2.4.2.2 *Latenz*

Die zeitliche Verzögerung ist eng mit der Übertragungsrate verbunden und maßgeblich für die Leistungsfähigkeit der Verbindung. Zwei Kernfaktoren beeinflussen die Latenz: Zum einen die Dauer, welche für die direkte Übertragung der Signale auf dem jeweiligen Medium benötigt wird und zum anderen die Verarbeitung der Daten an End- und eventuellen Zwischenknoten innerhalb der Netzwerkverbindung.

Generell ist die minimale Latenz beschränkt aufgrund der physikalisch festgelegten Geschwindigkeit, welche die Übertragung eines Signals benötigt in Abhängigkeit zu der Länge der Übertragungsstrecke. Bei optischen Signalen ist beispielsweise die Lichtgeschwindigkeit eine natürliche Schranke, die nicht durchbrochen werden kann. Das heißt bei einer RTT von 20 Millisekunden kann entsprechend ein Radius von etwa 3000 Kilometern im Vakuum abgedeckt werden. Durch die Leitung über entsprechende Faserkabel wird die Übertragungsgeschwindigkeit zudem um einen Faktor von etwa 1.52 reduziert und der Radius verkleinert sich auf etwa 2027 Kilometer. Es wird deutlich, dass die Entfernung der Kommunikationspartner maßgeblich die Latenz beeinflusst. Dies wird auch im Experiment von Paquin, Stebila und Tamvada [196] deutlich. Während eine Übertragung über wenige 100 Kilometer etwa 5 bis 30 Millisekunden benötigt, ist die Übertragung über mehrere tausend Kilometer deutlich verlangsamt.

Neben der reinen Übertragung der Informationen über eine gewisse Instanz wird zusätzlich eine Verzögerung durch die Verarbeitung von Routing- und Randinformationen [19] verursacht. Durch die zusätzlichen Verarbeitungsschritte wird eine möglichst effiziente, fehlerfreie und sichere Zustellung an einen gewünschten Empfänger ermöglicht. Hier muss ein guter Ausgleich gefunden werden. Jeder Mechanismus zur Sicherung beansprucht zusätzliche Verarbeitungszeit. Zu wenige Sicherungsmaßnahmen führen zu unbehandelten Störungen und Fehlern, welche die Übertragungszeit ebenfalls negativ beeinflussen.

#### 2.4.2.3 *Jitter*

Jitter beschreibt die Varianz der zeitlichen Verzögerung bei Übermittlung der Daten. Da die äußeren Einflüsse auf die Latenz in der Regel nicht konstant sind und zudem die einzelnen auf der Vermittlungsschicht verarbeiteten Pakete nicht zwangsläufig die gleichen Wegstrecken zurücklegen, differiert die Latenz in der Praxis. Eine Emulation von Jitter entspricht somit eher dem Verhalten eines typischen Datennetzes. Generell sind Funkübertragungen störanfälliger als abgeschirmte Kabelkanäle, sodass die Schwankungen der Latenzzeit ausgeprägter sind.

*Packetloss*

Paketverlust bezieht sich auf Netzwerke, welche die Daten mittels Paketvermittelnder Dienste übertragen. Das heißt die einzelnen Bits werden durch den Sender zu Paketen zusammengefasst und gemeinsam mit entsprechenden Metainformationen übermittelt. Paketverlust beschreibt den Unterschied zwischen allen gesendeten und den während der Sendung verlorenen Paketen. Der Wert kann auch als Wahrscheinlichkeit $p$ des Verlusts für ein einzelnes Paket angegeben werden. Wie bereits mehrfach erläutert, können zwei Bereiche bei der Datenübertragung betrachtet werden und für den Verlust von Daten verantwortlich sein. Zum einen sind auf physikalischer Ebene Fehler möglich, welche den Verlust bedingen. Beispielsweise, wenn Funksignale abgeschirmt werden oder es zu einer Überhitzung von Kabeln kommt. Zum anderen können Verarbeitungsschritte bei Zwischen- und Endknoten einen Verlust hervorrufen. Unter anderem ignorieren gewisse Dienste wie IP oder TCP fehlerhafte Pakete respektive Segmente und für die Endanwendungen ist dies äquivalent zu einem Verlust.

### 2.4.3 *Netzwerkemulation und -werkzeuge*

Ein emuliertes System fertigt in Teilaspekten ein Abbild eines realen System an, wobei es die gleichen Daten innehält, vergleichbare Programme ausführt und möglichst gleiche Ergebnisse bezüglich bestimmter Faktoren produziert. Netzwerkemulatoren sind daher Programme, welche das Verhalten eines Netzwerks abbilden und Funktionen sowie Konfigurationen ersetzen. Dabei können sie Einfluss auf die einzelnen Netzwerkschichten nehmen oder ihre Ein- und Ausgaben verändern, um das zu untersuchende Szenario zu reproduzieren. Im Gegensatz zur Simulation werden nur Teilaspekte nachgebildet und mitunter ist weiterhin ein Einsatz von Netzwerkhardware erforderlich, sodass ein geringerer Abstraktionsgrad möglich ist. Siehe auch Repp [212].

#### 2.4.3.1 *Network Namespaces*

Mithilfe von *ip-netns* können verschiedene Verbindungen beziehungsweise Teilnehmer emuliert werden. *netns* steht für Network Namespace und ist ein vergleichsweise neues Framework des Linux Kernels[35]. Laut Paquin, Stebila und Tamvada [196] sei ein Network Namespace eine „unabhängige, isolierte Kopie des *network stacks*" und somit eine Art Abstraktion, welche an bestimmte Prozesse gebunden ist. Einem *namespace* können eigene Routingtabellen, Netzwerkadressen und - konfigurationen zugeordnet werden. Mittels *veths* (Virtual Ethernets), können virtuelle Schnittstellen zugewiesen werden, welche eine Verbindung der emulierten Netze ermöglicht. Die Verlinkung erfolgt paarweise. Um Befehle innerhalb eines *namespace* durchzuführen, wird in der Regel folgende Syntax verwendet:

```
ip netns exec <network namespace> <command concerning the namespace>
```

Wird jeder Instanz des Experiments ein eigener *namespace* zugeordnet, so kann die Kommunikation zwischen realen Netzen emuliert werden. Eine ausführliche Beschreibung der Anwendung und ihrer Instruktionen befindet sich auf den entsprechenden Linux Manual Pages für *ip-netns*[36] und *veths*[37].

2.4.3.2 *Netem*

*tc-netem* ist ebenfalls Teil der Linux Kernel Features und steht für *Network Emulator*. Das Framework ermöglicht eine Steuerung des ausgehenden Datenverkehrs innerhalb eines Netzes und somit die Emulations von Wide Area Network (WAN) Charakteristiken. Konkret ausgedrückt, können Verbindungseigenschaften wie Latenz, Paketverlust oder verminderte Übertragungsraten emuliert werden. Folgende Attribute sind für diese Arbeit von Bedeutung:

- *delay*: Hinzufügen einer gewünschten Verzögerung in Millisekunden. Die Präzision ist beschränkt durch die Taktung des Kernels in Hertz. Bei 100 Hertz ist somit eine Steigerung um jeweils 10 Millisekunden möglich, sodass der tatsächliche Wert in der Regel abweicht [139].

- *jitter*: Ergänzung des Delays durch eine Art Zufallskomponente für jedes gesendete Paket, wobei die Angabe in Millisekunden erfolgt und die statistische Standardabweichung pro Datenpaket beschreibt. Wie auch bei Latenz ist der Jitter auf realer Hardware nicht exakt umzusetzen und meist etwas geringer [139]

- *loss*: Emulation von Paketverlust durch Angabe der Wahrscheinlichkeit in Prozent, sodass jedes Paket mit einer gewissen Wahrscheinlichkeit nicht an den Empfänger übermittelt wird

- *corrupt*: Bei einer ausgewählten Prozentzahl von Paketen wird jeweils an einer zufälligen Stelle ein Fehler eingefügt

- *duplicate*: Der angegebene Prozentsatz der Pakete wird doppelt angefügt

---

35 https://www.kernel.org, zuletzt aufgerufen 20.01.22
36 https://man7.org/linux/man-pages/man8/ip-netns.8.html, zuletzt aufgerufen 21.12.21
37 https://man7.org/linux/man-pages/man4/veth.4.html, zuletzt aufgerufen 21.12.21

- *reorder*: Durch eine verzögerte Übermittlung ausgewählter Pakete wird deren Reihenfolge beeinflusst. Bei 10 Millisekunden Latenz und einer Angabe von 25 Prozent werden die Pakete beispielsweise mit einer Wahrscheinlichkeit von 0.25 direkt gesendet. Alle anderen haben eine Verzögerung von 10 Millisekunden

- *rate*: Steuerung der Übertragungsrate, wobei die Angabe in unterschiedlichen Einheiten wie Kilobits oder Megabits möglich ist. Die interne Emulation erfolgt anhand der Paketgrößen und entsprechender Latenz

- *limit*: Maximale Anzahl der Pakete, die an die Queueing Discipline (QDISC) angehängt werden können, siehe unten

Somit kann beispielsweise eine geografische Distanz zwischen Server und Client mittels entsprechender Latenz oder eine verminderte Qualität über Paketverluste emuliert werden.

Die Steuerung von Network Emulator (NetEm) erfolgt über das Linux Kernel Feature *tc*, welches generell der Kontrolle des Nachrichtenverkehrs dient. Wesentliches Element ist QDISC, eine Art Warteschlange, welche jeweils einem Interface zugeordnet ist und in welche die Pakete vor der Verarbeitung eingereiht werden.

Ein gültiger Befehlsaufruf für NetEm wäre:

```
ip netns exec cli_ns tc qdisc change dev cli_ve root netem loss 2%
    corrupt 0.25% reorder 0% delay 2.684ms rate 500mbit
```

Eine ausführliche Beschreibung findet sich auf den entsprechenden Linux Manual Pages für NetEm[38] und *tc*[39].

### 2.4.3.3 Alternative Softwarewerkzeuge

Neben NetEm kann eine Emulation unter anderem mittels den Softwarewerkzeugen Mininet[40] oder NetMirage[41] erfolgen. Sie sind komplexer, umfassen mehr Funktionalitäten und ermöglichen somit eine ausgeweitete und mitunter realistischere Emulation, wohingegen Nachvollziehbarkeit und dezidierte Analyse erschwert werden.

---

38 https://man7.org/linux/man-pages/man8/tc-netem.8.html, zuletzt aufgerufen 21.12.21
39 https://man7.org/linux/man-pages/man8/tc-netem.8.html, zuletzt aufgerufen 21.12.21
40 http://mininet.org, zuletzt aufgerufen 10.10.21
41 https://github.com/cryspuwaterloo/netmirage, zuletzt aufgerufen 10.10.21

# 3

# VERWANDTE ARBEITEN & AKTUELLER STAND

Dieses Kapitel beschreibt Arbeiten, welche relevante Ergebnisse im Kontext der Integration von PQC innerhalb von TLS und der Evaluierung der Performanz beigetragen haben. Dies ermöglicht, zumindest in Teilen, einen Überblick bezüglich des aktuellen Wissensstandes und noch notwendigen Schritten. Da der Fokus innerhalb dieser Arbeit aus besagten Gründen auf TLS Version 1.3 liegt, werden Arbeiten bezüglich TLS 1.2 nur angerissen, jedoch nicht näher ausgeführt.

Das Kapitel gliedert sich wie folgt: In Abschnitt 3.1 werden Arbeiten zu Integration von PQC Algorithmen, insbesondere im Kontext von TLS betrachtet. Unter anderem werden verschiedene Entwürfe für mögliche Standards und verfügbare Implementierungen beleuchtet. In Abschnitt 3.2 folgen Arbeiten, welche die Performanz der implementierten TLS Integrationen und der verwendeten PQC Algorithmen untersuchen. Anschließend sind in Abschnitt 3.3 Arbeiten aufgeführt, welche Integrierbarkeit und Performanz der PQC Algorithmen in anderen Protokollen und Kontexten analysieren und zuletzt wird in Abschnitt 3.4 ein Fazit gezogen.

## 3.1 IMPLEMENTIERUNG & MIGRATION QUANTENSICHERER TLS VERSIONEN

### 3.1.1 Architektonische Grundkonzepte

Von einer Umstellung hinzu einer quantensicheren Kryptografie sind aktuell nur asymmetrische Primitive direkt betroffen. Im Bezug auf TLS müssen daher Komponenten zur Vereinbarung symmetrischer Schlüssel sowie zur Authentifizierung der Endpunkte angepasst werden. Die vertrauliche Übertragung der Anwendungsdaten mittels symmetrischer Verschlüsselung ist nicht betroffen. Der Fokus bei PQC Migration liegt somit auf dem TLS Handshake. TLS 1.3 bietet grundsätzlich eine bessere Kryptoagilität als die Vorgängerversion, da anstelle monolithischer Ciphersuites eine separierte Vereinbarung von Algorithmen und Parametern erfolgt und somit eine Einbindung neuer Verfahren leichter umsetzbar ist.

Crockett, Paquin und Stebila [77] beschreiben im Jahr 2019 unterschiedliche Ansätze zur PQC Integration bei TLS 1.2 und 1.3 sowie innerhalb des kryptografischen Netzwerkprotokolls SSH, siehe [162]. Sie beleuchten insbesondere Möglichkeiten bezüglich eines hybriden Vorgehens und geben einen sehr guten Überblick bezüglich der aktuell diskutierten Grundkonzepte, der benötigten Komponenten und der speziellen Anforderungen.

An ein praktikables, System, welches die Nutzung hybrider Verfahren ermöglicht, stellen die Autoren folgende Anforderungen:

- Kommunikation zwischen hybrid-fähigen und nicht hybrid-fähigen Clients und Servern weiterhin möglich

- Keine signifikante Performanzverschlechterung

- Keine substanzielle Latenzvergrößerung

- Keine zusätzliche RTT

- Kein Austausch redundanter Informationen

Integrierbarkeit und Performanz bei hybridem Vorgehen würden zudem durch die folgenden grundlegenden Designentscheidungen beeinflusst:

- Wie können eine hybride Variante, sowie die dort verwendeten Algorithmen und Parameter ausgehandelt werden

- Wie viele kryptografische Algorithmen können kombiniert werden und ist die Anzahl fest oder variabel

- Wie werden kryptografische Daten wie Schlüssel, Chiffrate oder Signaturen übermittelt

- Wie können die Werte der Algorithmen verknüpft werden

Für die Zusammenführung der Algorithmenwerte existierten unterschiedliche Ansätze, wie XOR-Verknüpfung oder KDF. Die Autoren betonen, es müsse beachtet werden, dass die Sicherheit aller Werte unabhängig voneinander garantiert werden könne und sich die zu übermittelnde Datenmenge im Vergleich zum klassischen Vorgehen nicht signifikant erhöhe.

### 3.1.2 Erste Entwürfe & angehende Standards

Im Kontext der Migration von PQC innerhalb von TLS 1.3 wurden bereits mehrere konkrete Architekturentwürfe sowie Vorschläge für neue Standards, insbesondere sogenannte „Internet Drafts" [105], veröffentlicht. Internet Drafts sind Entwürfe für neue Standards, welche über die IETF veröffentlicht und stetig weiterentwickelt werden. Der folgende Abschnitt ist entsprechend der Kategorisierung des NIST PQC Verfahrens in Schlüsselvereinbarung und digitale Signatur unterteilt.

#### 3.1.2.1 Schlüsselvereinbarung

2019 veröffentlichten Stebila und Gueron [239] einen ersten Entwurf bezüglich eines Standards für den hybriden Schlüsselaustausch zur Erweiterung des TLS 1.3 Standards [213]. Im Laufe der Zeit wurden einige Änderungen vorgenommen und aktuell befindet sich der Entwurf in der dritten Version [238], welche 2022 ausläuft. Für die Aushandlung der Algorithmen und Parameter sieht der Entwurf neue *group* Werte innerhalb der *supported_groups* Erweiterung vor, welche zunächst für die Wahl der Gruppen bezüglich ECDH vorgesehen war. Im Hinblick auf die hybriden Varianten wird für alle verfügbaren Kombinationen jeweils ein neuer Eintrag benötigt und die Anzahl der Algorithmen innerhalb einer hybriden Kombination wurde auf zwei beschränkt. Die Standardisierung der Werte könnte durch die IANA erfolgen, aktuell sind jedoch noch keine einheitlichen Identifikatoren verfügbar. Die gegebenenfalls notwendige Verbindung der zwei öffentlichen Schlüssel erfolgt mittels Konkatenation. Das bedeutet, die Werte werden zunächst kombiniert und als ein Wert innerhalb der *KeyShareClientHello*- beziehungsweise *KeyShareServerHello* Erweiterung übermittelt. Während der Client eine Liste mit *key_shares* sendet, welche entsprechend der Priorisierung der Algorithmen geordnet ist, antwortet der Server lediglich mit dem ausgewählten Algorithmus und übermittelt seinen korrespondierenden Wert. Die Anforderung zur unabhängigen Generierung der Schlüsselwerte aus [213] werden gelockert, sodass ein Algorithmus in mehreren *named_groups* auftreten und der *key_exchange* Wert jeweils mehrfach aufgeführt werden kann. Ebenso können auch die Werte für das *shared_secret* konkateniert werden, sodass das Resultat anschließend lediglich anstelle des original ECDH Werts innerhalb des TLS 1.3 *key_schedule* verwendet werden kann. Abbildung 3.1 veranschaulicht den Nachrichtenverlauf zwischen Client und Server bei Durchführung eines Handshakes unter Verwendung einer hybriden KEM Variante. Aspekte, welche die Schlüsselvereinbarung betreffen, sind blau markiert. Die notwendigen, beschriebenen Veränderungen sind mit einem + gekennzeichnet. Zudem wird anhand der geschweiften Klammern ersichtlich, welche Nachrichten bereits verschlüsselt werden können, noch bevor der Client die *Finished* Nachricht übermittelt und der Handshake abgeschlossen ist.

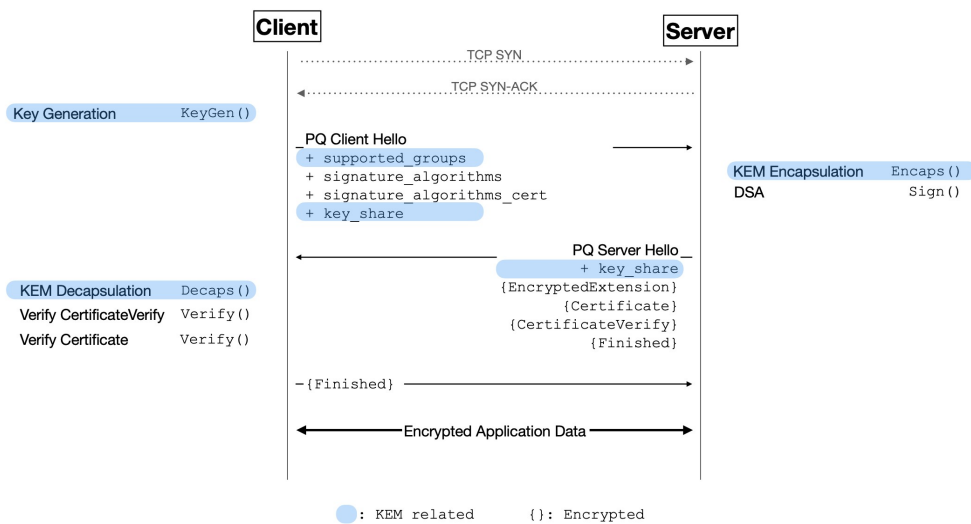

Abbildung 3.1: Skizzierter Ablauf TLS 1.3 Handshake mittels hybridem KEM Verfahren entsprechend des Entwurfs von Stebila, Fluhrer und Gueron [238] (Orientiert an Abbildung 1 aus [198])

Es sollte beachtet werden, dass die *key_exchange* Werte innerhalb des *key_share* auf $2^{16} - 1$ Bytes beschränkt sind. Einige Algorithmen, wie beispielsweise der NIST Kandidat Classic McEliece's, weisen jedoch größere

Schlüssellängen auf. Des Weiteren kann es aufgrund mehrfach gesendeter *key_shares* zu Redundanzen kommen, da jede hybride Kombination ihre eigene Kopie der öffentlichen Schlüssel sendet.

Ähnliche Entwürfe sind in [147, 221, 261] zu finden. Schanck und Stebila [221] schlagen beispielsweise für das Aushandeln hybrider Komponenten eine weitere *supported_groups* Erweiterung mit einer zweiten Liste an Algorithmen vor. Die für das hybride Verfahren herangezogenen Algorithmen können somit unabhängig voneinander ausgehandelt werden. Kiefer und Kwiatkowski [147] benennen zudem weitere Herangehensweisen für die Definition neuer *supported_groups* Identifikatoren. Die Ansätze wurden jedoch in den gängigen Implementierungen kaum berücksichtigt oder in [238] aufgenommen. Seit 2017 beziehungsweise 2018 fanden keine Anpassungen oder Weiterentwicklungen statt.

Hinsichtlich der Kombination der öffentlichen Schlüssel sind neben Konkatenation ebenfalls andere Vorgehensweisen denkbar. Ma und Zhandry [164], Schanck und Stebila [221] sowie Campagna und Petcher [64] beschreiben diverse KDFs und andere komplexe Funktionen. Ott, Peikert u. a. [193] führt aus, dass die geeignete Kombination ein wichtiger Faktor im Bezug auf Sicherheit und Performanz sei, welcher noch zu wenig erforscht sei.

Da sich TLS 1.2 und 1.3 deutlich voneinander unterscheiden und nicht miteinander kompatibel sind, existieren für 1.2 separate Ansätze. Schanck, Whyte und Zhang [223] sowie Campagna und Crockett [61] definieren IETF Entwürfe bezüglich modularer Ciphersuiten beziehungsweise eines einheitlichen Vorgehens für den quantensicheren und hybriden TLS 1.2 Handshake. Bos u. a. [52], Stebila und Mosca [240] sowie Bos u. a. [51] integrierten beispielsweise Gitter- beziehungsweise Ring-LWE-basierte KEM-Algorithmen auf der Basis von OpenSSL, indem sie neue Ciphersuiten für den hybriden Schlüsselaustausch kreierten. Zeier, Wiesmaier und Heinemann [267] geben einen Erfahrungsbericht bezüglich der Integration von McEliece und SPHINCS+ mittels der TLS Implementierung der Krypto Bibliothek Bouncy Castle[1]. Sie bemängeln eine fehlende einheitliche Definition von Ciphersuiten für neue Algorithmen. Des Weiteren sei für die Verwendung einiger Algorithmen wie McEliece aufgrund der zu großen Datenmengen eine Anpassung der darunterliegenden Software notwendig gewesen.

### 3.1.2.2 *Digitale Signatur*

Im Vergleich zur Verfügbarkeit von Forschungsarbeiten und Internet Drafts bezüglich des quantensicheren oder hybriden Schlüsselaustauschs, sind die Fortschritte bezüglich der Authentifizierung innerhalb von TLS bisher geringer. Crockett, Paquin und Stebila [77] sehen diese Tatsache darin begründet, dass zum einen die Berechtigungsnachsweise innerhalb einer verteilten Infrastruktur bereitgehalten und verfügbar gemacht werden müssten, was eine deutlich komplexere und aufwändigere Umstellung impliziere. Zum anderen sei der Wechsel hin zu einer quantensicheren Authentifizierung weitaus weniger dringlich, da die Authentizität des Kommunikationspartners nur zum tatsächlichen Zeitpunkt der Durchführung nachgewiesen werden müsse. Bei gespeichertem Kommunikationsverlauf könne der ausgetauschte Schlüssel hingegen auch noch Jahre später kompromittiert und die gesendeten Daten entschlüsselt werden. Da die Sicherheit bei Authentifizierung nur zum Anwendungszeitpunkt gewährleistet werden müsse, sei zudem ein weniger konservatives Vorgehen erforderlich und der Bedarf an hybriden Modellen eher gering.

---

[1] https://www.bouncycastle.org, zuletzt aufgerufen 28.12.21

Eine mögliche Vorgehensweise für die quantensichere Authentifizierung in TLS 1.3, auch im Hinblick auf hybride Varianten, beschreiben Crockett, Paquin und Stebila [77]. Ähnlich wie bei *supported_groups* bezüglich des Schlüsselaustauschs könne die Angabe der unterstützten Signaturverfahren mittels entsprechenden Identifikatoren für die jeweiligen Algorithmen oder Kombinationen erfolgen. Betroffen seien die *signature_algorithm_cert* Erweiterung, welche die Algorithmen für Signaturen innerhalb der Zertifikate angibt sowie das Feld *signature_algorithms*, welches die Verfahren definiert, die im Protokoll selbst unterstützt werden. Die Übertragung der öffentlichen Schlüssel zur Überprüfung der jeweiligen Signatur erfolgt aktuell innerhalb eines X.509-Zertifikats und Signaturen werden innerhalb der $CertificateVerify$ Erweiterung übertragen. Bei hybrider Vorgehensweise müsse daher abgewogen werden, ob die Übertragung mehrerer Schlüssel in mehreren Zertifikaten erfolgen solle oder ob Zertifikate um mehrere Schlüssel erweitert werden könnten. Die herkömmliche $CertificateVerify$ Erweiterung könne nicht ohne Weiteres durch eine zweite Signatur ergänzt werden. Möglich sei eine Konkatenation der Werte oder eine gänzliche Änderung innerhalb der Protokolllogik, beispielsweise durch zusätzliche Erweiterung. Unter anderem beschreiben Bindel u. a. [46], Ott, Peikert u. a. [193] sowie Kampanakis u. a. [141] Möglichkeiten für die Integration mehrere Schlüssel und Signaturen innerhalb eines Zertifikats mittels *non-critical* Erweiterungen. Die Erweiterungen beinhalten jeweils den zusätzlichen Signaturalgorithmus, den dafür benötigten Schlüssel oder die hiermit durchgeführte Signatur. Zudem machen sie Vorschläge bezüglich der Kompatibilität zu bestehenden Standards. Bindel u. a. [44] präsentieren eine entsprechende Java Implementierung von Zertifikaten, welche die Umsetzung hybrider Authentifizierung bei Verwendung des X.509-Standards ermöglicht. Der IETF Entwurf von Ounsworth und Pala [194] definiert zudem Ansätze hinsichtlich zusammengesetzter Zertifikate.

Fan u. a. [102] untersuchten die PQC Integrationen innerhalb der PKI. Dabei lag der Fokus auf der Erteilung, dem Entzug und dem Management von Zertifikaten mittels der CAs. Für die Untersuchungen wurde eine modifizierte Version von OpenSSL verwendet und neben dem Einfluss auf TLS wurden weitere Protokolle untersucht. Generell sei eine Verwendung aller untersuchten NIST PQC Verfahren ohne eine Änderung der internen Logik möglich.

Schwabe, Stebila und Wiggers [224] beschreiben einen alternativen TLS 1.3 Handshake, welcher anstelle von Signaturen auf eine KEMs-basierte Au-thentifizierung zurückgreift. KEM Verfahren würden somit sowohl für den Austausch des vorläufigen Schlüssels als auch für die Authentifizierung des Servers verwendet. Der Beleg der Identität bezüglich des Servers erfolge anhand der Dechiffrierung eines mit seinem öffentlichen Schlüssel verschlüsselten sogenannten „Challenge Value". Für die Authentifizierung würden jedoch teilweise Signaturen über die längerfristigen KEM Schlüssel (von einer CAs) benötigt. Die Autoren führten Experimente mit den quantensicheren KEM Verfahren Kyber, SIKE sowie NTRU durch. Für die weiterhin notwendigen Signaturen wurden unter anderem Dilithium und Falcon getestet. Eine weitere Möglichkeit zur KEM-basierten Authentifizierung bietet die modifizierte TLS Version OPTimized oder One-Point-Three TLS (OPTLS) [151]. Ursprünglich für DH konzipiert, könnten auch dort PQC Algorithmen genutzt werden. Die Konzepte wurden bis dato nicht in einen TLS Standard aufgenommen, da die Verfahren noch nicht ausreichend erforscht wurden und viele Systeme die Technologie nicht auf Anhieb unterstützen könnten [152]. Zudem wäre gegebenenfalls die severseitige Persistierung eines weiteren Secrets notwendig, welche zusätzliche Sicherheitsrisiken birgt. Trotzdem könnte der Ansatz, gerade aufgrund der bevorstehenden Neuerungen durch Quantencomputer, einen interessanten Lösungsweg bieten. Schwabe, Stebila und Wiggers [224] beschreiben, dass aufgrund des ausgesparten Signaturverfahrens geringere Datenmengen anfallen würden und die Bandbreite beziehungsweise die Übertragungsdauer, ebenso wie die CPU-Last somit reduziert werden könnten. Dies sei von besonderer Bedeutung, da einige PQC Signaturen komplexere Verfahren mit großen zu übertragenden Datenmengen aufweisen würden.

### 3.1.3   Open Quantum Save & OpenSSL

#### 3.1.3.1   Struktur

OQS ist ein öffentliches Projekt und dient der Unterstützung bei Entwicklung, Migration und Evaluierung quantensicherer Algorithmen. Es basiert auf einer Arbeit von Stebila und Mosca [240], ist auf Github[2] frei verfügbar und wird durch zahlreiche Institutionen und Unternehmen wie die Universität Waterloo oder Amazon unterstützt. OQS setzt sich aus zwei Hauptkomponenten zusammen:

- **liboqs**: C Bibliothek mit PQC Algorithmen als einheitliche Schnittstelle für KEMs sowie digitale Signaturen. Alle Kandidaten und Alternativen der dritten Runde des NIST Verfahrens, bis auf GeMSS, sind verfügbar und die Bibliothek ist mit den gängigen Betriebssystemen und Prozessorarchitekturen kompatibel.

- **Prototypen**: Integrationen der PQC Verfahren in Protokolle und Anwendungen wie TLS, SSH oder S/MIME [220] (Protokoll für Absicherung des Email-Verkehrs)

---

2 `https://github.com/open-quantum-safe`, zuletzt aufgerufen 15.01.22

Für diese Arbeit von besonderem Interesse ist die modifizierte Variante der TLS Implementierung OpenSSL, welche sowohl für TLS 1.2 als auch für 1.3 verfügbar ist und den überwiegenden Teil der Kandidaten des NIST PQC Verfahrens unterstützt. Sie ermöglicht die Nutzung klassischer, quantensicherer sowie hybrider Verfahren. Zudem sind die Schnittstellenfunktionalitäten äquivalent zu denen der Standardversion, sodass eine Nutzung innerhalb bekannter Anwendungen wie Apache httpd[3], curl[4] oder NGNIX[5] möglich ist.

Es ist zu beachten, dass aufgrund der enormen Schlüsselgröße eine einfache Integration des PQC KEM Kandidaten Classic McEliece nicht möglich war. Die Größenbeschränkung der genutzten Erweiterungen innerhalb der TLS Handshake-Nachrichten können nicht eingehalten werden, siehe auch Abschnitt 3.1.2.1. Eine Implementierung für Classic McEliece innerhalb von OQS OpenSSL steht daher noch aus. Ähnliches gilt für den Signaturalgorithmus GeMMS, welcher noch nicht in *liboqs* integriert wurde.

### 3.1.3.2 Handshake mit PQC

Die OQS OpenSSL Implementierung ermöglicht innerhalb des TLS Handshakes die Verwendung der NIST Kandidaten sowie Alternativen für Schlüsselaustausch und Signatur.

KEM Verfahren sind als Post Quantum Only (PQ-Only) sowie als hybride Version verfügbar. Die Grundstruktur des TLS 1.3 Standards [213] wird beibehalten und für die PQC Verfahren werden neue *group*-Werte innerhalb der *supported_groups* Erweiterung definiert. Da OpenSSL keine übergreifende Anwendungsschnittstelle (API) für die Ergänzung der KEM Verfahren innerhalb der Komponente *libcrypto* bietet, ist eine Änderung der internen Logik, insbesondere innerhalb des *ssl* Ordners, notwendig. Die Umsetzung der hybriden Version orientiert sich an dem Entwurf von Stebila, Fluhrer und Gueron [238], siehe Abschnitt 3.1.2. Der Ablauf wird zudem in Abbildung 3.1 veranschaulicht.

Auch die Authentifizierung kann rein anhand von PQC Verfahren oder hybrid erfolgen. Für Signaturverfahren oder deren hybride Kombinationen werden neue Identifikatoren definiert. Die Anzahl der Algorithmen innerhalb der hybriden Kombinationen wurde zunächst ebenfalls auf zwei beschränkt. Bei Durchführung werden die beiden Signaturen sowie die beiden öffentlichen Schlüssel jeweils als Konkatenation übertragen, wobei die Signaturen über den gleichen SHA-2-Hashwert erfolgen.

Neben der reinen Authentifizierung bietet die modifizierte OpenSSL Version zudem Funktionen für die Generierung von X.509-Zertifikaten inklusive der Schlüssel und Signaturen.

Crockett, Paquin und Stebila [77] führen an, dass bei großen Signaturen und Zertifikaten innerhalb OpenSSL 1.1.1 zwei Aspekte zu beachten seien:

---

[3] https://httpd.apache.org, zuletzt aufgerufen 28.12.21
[4] https://curl.se, zuletzt aufgerufen 28.12.21
[5] https://www.nginx.com, zuletzt aufgerufen 28.12.21

- Die zunächst auf 102400 Bytes beschränkte Größe der *Certificate* Nachricht müsse während der Laufzeit auf $2^{24} - 1$ Bytes erhöht werden

- Die beschränkte Größe der *CertificateVerify* Nachricht müsse auf $2^{16} - 1$ Bytes erhöht werden, wobei die Signaturen von Picnic bei Sicherheitslevel drei oder höher noch größer seien und bei Nutzung eine Anpassung des Programmcodes erforderlich machen würden

Eine noch ausführlichere Beschreibung bezüglich der Implementierung von OQS OpenSSL 1.1.1 findet sich unter anderem in [196].

#### 3.1.3.3   OpenSSL 3 und Provider

Des Weiteren wurde im Jahr 2021 die neue Hauptversion OpenSSL 3.0.0[6] veröffentlicht. Sie ist nicht vollständig abwärtskompatibel, bietet jedoch einfache Möglichkeiten zur Integration neuer Algorithmen und weist im Vergleich eine deutlich verbesserte Kryptoagilität auf. Diese basiert unter anderem auf dem neu eingeführten Provider Konzept. Provider bündeln Algorithmenimplementierungen und können flexibel über eine Konfigurationsdatei oder innerhalb des Programmcodes aktiviert beziehungsweise deaktiviert werden. Aktuell bietet die Version fünf Standardwerte, welche jedoch durch zusätzliche, individuelle Provider ergänzt werden können. Unter anderem bietet OQS eine spezielle Providerimplementierung, welche die Nutzung quantensicherer Algorithmen über ein binäres Add-On erlaubt, ohne dass eine Veränderung der internen Logik notwendig ist.

#### 3.1.3.4   Alternative Implementierungen

Neben OQS gibt es weitere Implementierungen beziehungsweise Bibliotheken für die Anwendung von PQC Algorithmen. PQCrypto[7] bildet eine Art Sammlung für standalone PQC Implementierungen. Es handelt sich jedoch nicht um eine integrierbare Bibliothek und es werden keine Möglichkeiten für eine Performanzanalyse oder eine Integration in höherliegende Protokolle und Anwendungen bereitgestellt.

Neben OpenSSL ist auch ein Fork der TLS Implementierung BoringSSL verfügbar, welche speziell auf die Bedürfnisse von Google ausgerichtet ist.

### 3.1.4   Hohe Kryptoagilität

Das BSI gab die klare Handlungsempfehlung hinsichtlich einer möglichst sicheren und gleichzeitig zügigen Absicherung der bestehenden Infrastrukturen gegen Quantencomputer mittels PQC, insbesondere auch im Bezug auf TLS [267]. Da Updates und Weiterentwicklungen aufgrund der noch unausgereiften PQC Technologien sehr wahrscheinlich sind, sollten die Implementierungen so konzipiert werden, dass nachträgliche Änderungen schnell und reibungslos möglich sind. Hier spielt das relativ neue Forschungsgebiet der Kryptoagilität eine zentrale Rolle, welches auch bei TLS Anpassungen berücksichtigt werden sollte. Unter anderem stellen Ott, Peikert u. a. [193] Fragen zu Kryptoagilität und Vorgehensweise bei der Migration von PQC und betonen, dass verschiedenste Plattformen und Programmiersprachen berücksichtigt werden müssen, um eine umfassende Kompatibilität sicherzustellen. Insbesondere hybride PQC Verfahren bergen Herausforderungen, da sie meist mit erhöhtem Rechenaufwand, einer größeren zu übertragenden Datenmenge oder einer komplexeren Implementierung einhergehen.

---

[6] https://www.openssl.org/docs/man3.0/man7/migration_guide.html, zuletzt aufgerufen 03.02.22

[7] https://pqcrypto.eu.org/index.html, zuletzt aufgerufen 29.12.21

## 3.2 EVALUIERUNG QUANTENSICHERER TLS VERSIONEN

Die Algorithmen des NIST Verfahrens weisen ganz unterschiedliche Charakteristiken auf, welche die Migration in bestehende Systeme erschweren und einen Eins-zu-eins-Austausch meist unmöglich machen. Insbesondere größere öffentliche Schlüssel, größere Signaturen und Chiffrate oder eine langwierigere Berechnung können zu Problemen führen. Viele der Ansätze und zugrundeliegenden mathematischen Probleme sind zudem vergleichsweise jung und ihr Verhalten noch nicht in Gänze erforscht. Laut [240] sei daher eine baldige und umfassende Analyse und Evaluierung von PQC Algorithmen und deren Migration von großer Bedeutung. Insbesondere müsse ein besseres Verständnis für die Einflüsse der Charakteristiken innerhalb der unterschiedlichen Anwendungsszenarien gewonnen werden. TLS sowie die darunterliegenden Protokolle TCP und IP verfügen beispielsweise über diverse Mechanismen, um eine fehlerfreie und gleichzeitig möglichst performante Übertragung zu gewährleisten. Kampanakis u. a. [141] führen vier Aspekte auf, welche die Übertragung der TLS Daten beeinflussen und somit eine Einschätzung bezüglich des Verhaltens bei Einführung der PQC Algorithmen erschweren:

- **TCP Segmentierung & TLS Fragmentierung**: Zu große Datenpakete würden für die Weiterverarbeitung mit IP segmentiert. TLS ermögliche regulär Größen bis 16 Kilobytes. Andernfalls werde der sogenannte *Record Fragmentation* Mechanismus angewandt [82, 247]. Die Maximum Transmission Unit (MTU) innerhalb des Netzwerkpfads sei jedoch durchschnittlich auf etwa 1500 Bytes beschränkt, sodass TCP nochmals eine Segmentierung vornehme. In beiden Fällen käme es zu zusätzlichen Paketen. Zudem müsse der Empfänger die Daten wieder zusammenfügen.

- **Overhead**: Müssten größere Datenmengen beim Aufbau sicherer Verbindungen übermittelt werden, sei der mögliche negative Einfluss dann besonders hoch, wenn die eigentlichen Nutzdaten verhältnismäßig gering seien. Hier seien Verfahren wie HTTP/2 von Vorteil, welche eine Art Multiplexing bieten und Daten auf eine Verbindung zusammenfassen.

- **Caching**: Indem die Kommunikationspartner Daten aufbewahren würden, welche möglicherweise für nachfolgende Verbindungen erneut benötigt würden, könne die zusendende und -verarbeitende Datenmenge in Zukunft reduziert werden. Der Partner müsse jedoch entsprechend informiert sein. Dies führe zunächst zu einem erhöhten Arbeits- und Speicheraufkommen und sei nur sinnvoll, sollten die aufbewahrten Daten anschließend tatsächlich benötigt werden. Zudem berge das Vorgehen Risiken, sofern veraltete beziehungsweise ungültige Daten gechached werden würden.

- **Kompression**: Die Kompression von Daten reduziere die Größe der zu übertragenden Daten, sorge jedoch für zusätzlichen Rechenaufwand bei Sender und Empfänger. Im Zweifel könne dieser Mechanismus sogar einen DoS auslösen. Eine Abwägung von Kosten und Nutzen sei hier besonders wichtig.

### 3.2.1  Evaluierung unter TLS 1.3

#### 3.2.1.1  Schlüsselaustausch

Paquin, Stebila und Tamvada [196] beschreiben diverse Experimente, welche von Google, dem Internetdienstanbieter Cloudflare als auch Amazon nach Einführung von TLS 1.3 im Jahr 2018 durchgeführt worden seien. Die Forscher hätten NTRU-HRSS, SIKE und BIKE in Kombination mit X25519-ECDH als hybride KEM Variante integriert und innerhalb der eigenen Infrastrukturen evaluiert. Bei der Implementierung habe man sich jeweils am Entwurf von Stebila und Gueron [239] orientiert und diesen auf eine modifizierte Version von BoringSSL angewendet. Auch Kwiatkowski u. a. [153] wählten diesen Ansatz und evaluierten SIKE und NTRU-HRSS unter realen Netzwerkbedingungen. Die Forscher kommen zu dem Ergebnis, dass sich die Performanz unter Verwendung von PQC bei etwa 5 Prozent der Verbindungen wesentlich verbessere. Bei ebenfalls 5 Prozent habe die Performanz hingegen enorm verschlechtert. Die Autoren resümieren, dass eine Analyse der Ursachen für die Performanzunterschiede bei ausschließlich realen Verbindungen nur schwer möglich sei, denn das Spektrum der teils unbekannten Einflussfaktoren sei zu groß. Auch wenn die Forscher im direkten Vergleich aufgrund der Performanz und den Entwicklungsmöglichkeiten schlussendlich NTRU forcieren, würden alle getesteten hybriden Versionen Nachteile bezüglich der Effizienz bei klassischen Verfahren aufweisen.

#### 3.2.1.2  Authentifizierung

Kampanakis und Sikeridis [142] evaluieren die benötigte Zeit für einen TLS Handshake sowie die Fehlerrate bei allen NIST PQC Signatur Kandidaten der zweiten Runde innerhalb der X.509-Zertifikate. Dabei greifen sie auf OQS OpenSSL Version 1.1.1c zurück. Die Autoren legen dar, dass die Größe von Signatur und Zertifikatskette die benötigte Handshakezeit beeinflusse und kommen zu dem Schluss, dass Dilithium und Fast-Fourier Lattice-based Compact Signatures over NTRU (FALCON) dennoch am besten geeignet seien. Sie betonen zudem, dass selbst eine leicht verminderte Geschwindigkeit bei Signaturoperationen den Durchsatz von Servern signifikant beeinflussen könne.

Sikeridis, Kampanakis und Devetsikiotis [235] veröffentlichen eine Performanzanalyse des TLS 1.3 Verbindungungsaufbaus bei Verwendung der NIST Kandidaten zur Authentifizierung. Dabei wurden Verbindungen zu Servern an unterschiedlichen Standorten genutzt, sodass eine Durchführung unter realen Netzwerkbedingungen erfolgte. Sie stellten fest, dass der Mehraufwand akzeptabel sei. Eine Nutzung sei für nicht-zeitkritische Systeme durchaus zu empfehlen.

Raavi u. a. [207] untersuchten den Einfluss auf die Performanz bei Integration der NIST PQC Kandidaten für digitale Signatur (FALCON, Dilithium und Rainbow) innerhalb der PKI. Auch hybride Varianten wurden berücksichtigt. Die Autoren untersuchten die Schlüsselgenerierung bei TLS beziehungsweise HTTP Servern sowie die Signatur und Zertifikaterstellung der CA hinsichtlich der benötigten Rechen- und Speicherressourcen und dem zeitlichen Einfluss. Generell sei eine Integration problemlos möglich, der Einfluss auf die Performanz variiere jedoch stark, abhängig von der Wahl des Algorithmus. Dilithium weise die geringsten Kosten bezüglich der Generierung der Schlüssel-Paare, der CSRs und der Zertifikate auf. Mit FALCON sei hingegen eine effiziente Zertifikatsverifizierung möglich. Im Vergleich zwischen hybriden und PQ-Only Varianten seien keine signifikanten Unterschiede festzustellen.

### 3.2.1.3 *Schlüsselaustausch & Authentifizierung*

Paul u. a. [198] evaluierten die Performanz von TLS 1.3 bei Verwendung von Kyber sowie unterschiedlichen hybriden Signaturverfahren innerhalb einer Zertifikatskette. Sie berücksichtigten SPHINCS+ Dilithium, FALCON sowie XMSS jeweils in Kombination mit klassischem RSA. Für die Integration wurde eine Version der TLS Implementierung wolfSSL[8] entsprechend modifiziert. Die Experimente wurden unter realen Netzwerkbedingungen durchgeführt, wobei Server an unterschiedlichen Standorten mit Latenzen zwischen etwa 25 und 250 Millisekunden bei der Übertragung gewählt wurden. Die Autoren schlussfolgern, dass eine Nutzung der Algorithmen durchaus empfohlen werden könne und der Einfluss auf die benötigte Zeit für einen TLS Verbindungsaufbau nicht signifikant beeinflusst werde. Nur bei Verwendung hashbasierter Algorithmen bei Root-Zertifikaten seien größere Verzögerungen aufgetreten.

Barton u. a. [27] evaluieren Kyber, FrodoKEM, Saber, NewHope, NTRU, BIKE und SIKE sowie diverse Signaturverfahren im Hinblick auf die Eignung hinsichtlich ressourcenbeschränkter Geräte. Die Implementierung ba-sierte auf der OpenSSL Version von OQS und Client und Server waren über ein einfaches, kabelgebundenes Local Area Network (LAN) verbunden.

---

[8] https://www.wolfssl.com, zuletzt aufgerufen 25.01.22

Zhang u. a. [268] integrieren und evaluieren Kyber512 sowie Crystals Dilithium hinsichtlich der Handshakeperformanz. Die Implementierung der TLS Funktionalitäten erfolgte mithilfe des Softwarewerkzeugs *Global Security Kit* von IBM. Die Testmessungen wurde jedoch in einem isolierten Kontext, ohne die Bereitstellung einer typischer Netzwerkverbindung durchgeführt. Die Autoren betrachten die benötigte Zeit für einen TLS Verbindungsaufbau im Vergleich zu einer klassischen Variante mit DH und stellten fest, dass die PQC Variante keinen Mehraufwand erfordere und eine hybride Durchführung ebenfalls problemlos möglich sei.

Sikeridis, Kampanakis und Devetsikiotis [234] integrierten PQC in TLS 1.3 sowie SSH und evaluierten die Handshake Performanz unter realen Netzwerkbedingungen. Für die Experimente wurden drei Server mit unterschiedlichen Entfernungen beziehungsweise Latenzen (37, 67 und 163 Millisekunden) verwendet. Evaluiert wurden die Signaturverfahren Dilithium und SPHINCS+ in Kombination mit den KEMs Kyber-512 und NewHope-512-CCA für Sicherheitslevel 1 sowie Kyber-768 und NTRU-HRSS-701 für Sicherheitslevel 3. Als Kontrolle und für hybride Varianten wurden zudem die klassischen Verfahren RSA-2048 sowie ECDH mit P-256 herangezogen. Der Verbindungsaufbau sei durchweg fehlerfrei möglich gewesen und bei Verwendung der KEMs sei im Mittel eine geringe Verzögerung von 0.1 bis 5 Millisekunden im Vergleich zum klassischen Verfahren aufgetreten. Auch die Verwendung von hybriden anstelle der PQ-Only Varianten beeinflusse die Performanz nicht merklich. Während die Performanz bei Authentifizierung mittels Dilithium ebenfalls positiv zu bewerten sei, habe SPHINCS+ hingegen zur einer Latenzsteigerung von 150 bis 190 Prozent geführt. In weiteren Untersuchungen stellen die Autoren fest, dass die Verzögerung auf eine größere Anzahl zu sendender Pakete und die daraus resultierenden zusätzlichen Round-Trips (RTs) zurückzuführen sei. Die Anzahl der RTs werde maßgeblich durch die Größe des initialen Congestion Windows *initcwnd* von TCP beeinflusst, welches die Menge der Pakete bestimmt, die zu Beginn der Verbindung versendet werden können, ohne dass eine Bestätigung erforderlich ist. Die Größe werde in der Regel als ein Vielfaches der Maximum Segment Size (MSS) angegeben, wobei 1 MSS = 1460 Bytes, entsprechend der MTU von 1500 Bytes. In der Regel verwende Linux eine initiale Größe von 10 MSS. Durch eine Steigerung von *initcwnd* sei eine deutliche Verringerung der Handshake Dauer möglich.

Paquin, Stebila und Tamvada [196] präsentieren ein Framework für die Evaluierung von PQC KEM und Signatur Algorithmen unter realen Netzwerkbedingungen innerhalb von TLS 1.3. Sie emulieren eine Client-Server-Netzwerkverbindung auf einem Rechner mittels verschiedener Funktionalitäten des Linux Kernels, um die Variablen präzise und unabhängig voneinander zu kontrollieren und den Einfluss auf die Performanz der PQC Algorithmen zu bestimmen. Für die Integration von PQC in TLS und insbesondere die Implementierung der hybriden KEM wird auf die OpenSSL Vari-ante von OQS zurückgegriffen, beschrieben in Abschnitt 3.1.3. Die Autoren vergleichen die benötigte Zeit für einen Handshake bei hybridem Schlüsselaustausch mittels Kyber, SIKE und FrodoKEM in Kombination mit klassischen elliptischen Kurven sowie bei PQ-Only Signatur anhand von Dilithium, qTesla und Picnic jeweils mit den klassischen TLS 1.3 Algorithmen. Für die Nachbildung realer Netzwerkverbindungen wird die Paketverlustrate von 0 bis 20 Prozent sowie die zeitliche Verzögerung bei Übertragung in vier Stufen ($5.5ms, 31.1ms, 78.6ms, 195.6ms$) variiert. Ergänzend führen sie einen Abgleich zwischen den erzielten Ergebnissen und Experimentdaten von realen Servern über verschiedene Länder hinweg durch.

Die Autoren stellen fest, dass die *HandshakeCompletionTime* bei schnellen bis durchschnittlichen Netzwerkverbindungen maßgeblich durch die Performanz der Algorithmen Operationen beeinflusst wird. Bei Paketverlustraten ab etwa drei bis fünf Prozent hätten jedoch aufgrund der Fragmentierung größere Datenmengen (FrodoKEM) einen nennenswerten Einfluss. Ein ähnliches Verhalten ließe sich auch bei Signaturen beobachten.

### 3.2.2 Evaluierung unter TLS 1.2

Da TLS 1.3 erst in 2018 veröffentlicht wurde, beziehen sich viele Arbeiten auf die Vorgängerversion 1.2.

Bereits 2014 integrierten Chang u. a. [67] ein gitterbasiertes KEM sowie das Signaturverfahren Rainbow in PolarSSL. Auch Bürstinghaus-Steinbach u. a. [59] integrierten den KEM Algorithmus Kyber sowie das Signaturverfahren SPHINCS+ in mbed TLS. Im Vergleich zur klassischen Variante mit elliptischen Kurven habe sich unter anderem gezeigt, dass Kyber keinen größeren negativen Einfluss auf die Performanz des Systems habe, während bei SPHINCS+ Signaturgröße und Rechenzeit, gerade im Kontext der eingebetteten Systeme, durchaus problematisch seien. Hülsing, Rijneveld und Schwabe [129] integrierten SPHINCS+ auf einem sehr beschränkten System und kommen zu ähnlichen Ergebnissen.

Bos u. a. [51, 52] sowie Stebila und Mosca [240] Ring-LWE-basierte KEM-Algorithmen in TLS und stellten fest, dass der Verlust bezüglich der Performanz generell nicht sonderlich groß sei. Aufgrund der effizienten Berechnung und hinnehmbarer Schlüsselgrößen sei eine Verwendung gitterbasierter Kryptografie, insbesondere im Hinblick auf LWE, durchaus denkbar. Selbst die Performanz der hybriden Variante sei hinnehmbar.

Kampanakis u. a. [141] evaluierten die Authentifizierung mit hybriden Zertifikaten und hashbasierten PQC Algorithmen im Kontext von TLS. Die Autoren konnten zeigen, dass aufgrund der TLS *Record Segmentation* trotz der erweiterten Größe der Zertifikate respektive Zertifikatsketten eine erfolgreiche Übertragung möglich sei. Der messbare negative Einfluss auf die Anzahl der zu übertragenden Pakete und die benötigte Zeit bezüglich der Handshake Durchführung lägen in einem hinnehmbaren Rahmen. Problematisch seien die Verzögerungen und der erhöhte Rechenaufwand höchstens bei zeitkritischen oder stark ressourcenbeschränkten Anwendungen.

## 3.3    VERWANDTE THEMENGEBIETE

### 3.3.1    *Schlüsselaustausch mittels Quantentechnologie in TLS*

Neben der Integration von Algorithmen des NIST Standardisierungsprozesses bezüglich PQC, könnten auch QKD Verfahren innerhalb von TLS genutzt werden. QKD nutzt Quanteneffekte, um einen sicheren Schlüsselaustausch zwischen Kommunikationspartners zu ermöglichen. Aktuell befindet sich die Technologie noch in der Entwicklungsphase und funktioniert vor allem über kürzere Distanzen. Eine nähere Beschreibung findet sich in Abschnitt 2.2.1.

Erste Ansätze sind unter anderem in Pivk, Kollmitzer und Rass [201], Faraj [103] und Faraj [104] beschrieben. Die Autoren präsentieren Konzepte bezüglich einer außergewöhnlich sicheren Verschlüsselung und Authentifizierung.

Elboukhari, Azizi und Azizi [100] untersuchen die Nutzung von QKD innerhalb eines LAN mittels einer TLS 1.2 Integration. Hierbei wird das Handshake Protokoll erweitert und zudem wird eine neue Komponente, das „QKD Configuration Protocol", ergänzt. Mithilfe der Ergänzung ermöglichen die Autoren die Integration und Bedienung eines optischen Kanals sowie eines optischen Modems, welche der Verarbeitung der für die Technologie notwendigen Photonen dienen. Die Autoren kommen zu dem Schluss, dass dieser neue Ansatz den Ablauf des TLS Handshakes vereinfachen könne. Die Komplexität der ausgetauschten Nachrichten werde verringert und Zertifikate oder eine PKI seien nicht mehr notwendig. In Kombination mit der stetigen Veränderung der Geheimnisses mit jeder neuen Verbindung werde eine außergewöhnlich hohe Sicherheit zu vergleichsweise geringen Kosten gewährleistet. Des Weiteren wird der Vorschlag in einer Arbeit von Elboukhari, Azizi und Azizi [101] nochmals detaillierter, in einzelnen Schritten, beschrieben und durch ein konkretes Anwendungsbeispiel ergänzt.

Bei allen aufgeführten Arbeiten werden für die Integration der QKD Mechanismen Abläufe und Komponenten innerhalb des TLS Protokolls speziell angepasst und eine Kompatibilität zu anderen kryptografischen Verfahren wird nicht gewährleistet.

### 3.3.2    *Quantensichere Kryptografie in anderen Protokollen*

Kampanakis u. a. [143] sowie Crockett, Paquin und Stebila [77] beschreiben Möglichkeiten zur Integration von PQC Algorithmen innerhalb des Protokolls SSH, welches der Absicherung von Netzwerkverbindungen dient. Das prinzipielle Vorgehen bei einem Verbindungsaufbau sei in seinen Grundzügen gleich dem Handshake unter TLS. SSH habe jedoch noch geringere Ansprüche im Bezug auf die Dauer des Verbindungsaufbaus und die Nachrichtengröße, da es sich grundsätzlich um Eins-zu-eins-Nachrichten handle und ein Neuaufbau von Verbindungen nur unregelmäßig und vergleichs-weise selten erfolge. Bei Classic McEliece sei es aufgrund der beschränkten Paketgröße bei SSH von $2^{18}$ Bytes ebenfalls zu Problemen gekommen.

Kampanakis u. a. [141] evaluieren die Performanz hybrider Authentifizierung mittels hashbasierter Signaturen in X.509 Zertifikaten bei QUIC. Die Autoren resümieren, dass die praktische Verwendung, trotz größerer Datenmengen, durchaus möglich sei. QUIC sei hier besonders von Vorteil, da inkludierte Methoden wie Kompression und Caching mögliche Nachteile kompensierten.

Müller u. a. [184] befasst sich mit der Integration von PQC innerhalb von Domain Name System Security Extensions (DNSSEC). Bei DNSSEC handelt es sich um Erweiterungen des Dienstes DNS [87, 88], welcher insbesondere innerhalb des Internets für die Auflösung von Domain-Namen genutzt wird. DNSSEC umfasst diverse Erweiterungen, welche die Authentizität und Integrität der Daten sichern, indem die Quellen authentifiziert werden. Da DNS Server viele Anfragen zeitgleich beantworten müssen, die Nachrichtenpakete aufgrund der Spezifikation und der darunterliegenden Protokolle vergleichsweise klein sind, auch Endgeräte mit kritischen Echtzeitanforderungen bedient werden müssen und bei einem Großteil der Anfragen eine ganze Kette an zusammenhängenden Zertifikaten geprüft werden muss, sind für eine performante Nutzung von DNSSEC kurze Signaturen und eine effiziente Überprüfung dieser unabdingbar. Dies birgt bei Integration von neuartigen PQC Algorithmen besondere Herausforderungen. In Müller u. a. [184] wurden die NIST PQC Kandidaten zur Authentifizierung aus Runde drei in DNS beziehungsweise DNSSEC integriert und evaluiert. Drei Kandidaten scheinen zumindest teilweise geeignet, wenngleich eine Eins-zu-eins-Ersetzung nicht praktikabel erscheint.

Eine Reihe von Arbeiten beschäftigen sich mit PQC im Kontext der Technologie Blockchain. Dabei handelt es sich um eine Art abgesicherte Verkettung von Datenblöcken, welche fortlaufend weitergeführt wird und über ein gesamtes Netzwerk hinweg verteilt erfolgen kann, siehe auch Narayanan u. a. [186]. Die Datenblöcke beinhalten diverse Transaktionen, welche aktuell durch klassische kryptografische Signaturalgorithmen gesichert werden. Die Arbeit von Gao u. a. [113] sowie Zhang u. a. [270] präsentieren Ansätze zur digitalen Signatur mithilfe von Gittern. Li u. a. [160] sowie Yin u. a. [264] entwickelten und integrierten spezielle Signaturalgorithmen, welche auf der sogenannten „Bonsai Tree" Technologie [66] beruhen. Die Evaluierung zeigte jedoch, dass die Algorithmen entweder vergleichsweise ineffizient arbeiten oder die Länge der Signaturen enorm vergrößert ist, sodass diese als nicht praktikabel erscheinen. Allende u. a. [15] stellen in ihrer Arbeit ein zweischichtiges System vor, welches sie in libSSL integrieren. Als PQC Algorithmus wird Falcon-512 verwendet. Generell seien alle Kandidaten sehr ressourcenfordernd, Falcon sei im Vergleich jedoch kompakt und die angebotene Implementierung enthalte alle notwendigen Funktionalitäten. Im Experiment zeige sich, dass die Implementierung effizient, robust und skalierbar arbeite. Andererseits variiere die Performanz stark, abhängig von den unterliegenden Technologien und Plattformen.

Auch Virtual Private Network (VPN) und der Schlüsselmanagementdienst Internet Key Exchange (IKE) [144] wurden untersucht. Kampanakis u. a. [141] evaluieren die Performanz bei Verwendung hybrider X.509-Zertifikate mit hashbasierten PQC Verfahren und kommen zu dem Entschluss, dass Internet Key Exchange Version 2 (IKEv2) durch die Integration der PQC Verfahren nicht nennenswert negativ beeinflusst werde. Zudem wurden bereits vorläufige sowie offizielle Standards bezüglich der PQC Integration veröffentlicht [108, 109].

Bei ressourcenbeschränkten Geräten, insbesondere Bereich IoT ist die Performanz aufgrund von Echtzeitanforderungen besonders wichtig. Hinzu kommt der Bedarf an energieeffizienten Technologien. O. Saarinen [191], Roma, Tai und Hasan [216], sowie Banerjee und Hasan [23] evaluieren in ihren Arbeiten PQC Algorithmen hinsichtlich der Verwendung innerhalb eingebetteter Geräte. Dabei legen alle den Fokus auf den Energiebedarf, welcher speziell bei eingebetteten Geräten gering gehalten werden sollte.

## 3.4 ZUSAMMENFASSUNG & SCHLUSSFOLGERUNG

Die NIST, das BSI sowie weitere wichtige Institutionen im Kontext der IT-Sicherheit verweisen auf die Bedeutsamkeit und auch die Dringlichkeit im Bezug auf die Evaluierung quantensicherer Verfahren im Anwendungskontext [11], [267]. Dabei spielt insbesondere das Protokoll TLS eine wichtige Rolle, da es bei circa 60% aller Internetverbindungen Verwendung findet [196]. Untersucht werden müsse unter anderem der konkrete Einfluss hybrider Varianten sowie der jeweiligen Sicherheitsstufen.

In den vorangegangenen Abschnitten wird ersichtlich, dass sich in den letzten Jahren eine Reihe von Arbeiten mit der Implementierung, Migration und Evaluierung von PQC Algorithmen im Kontext von TLS 1.3 auseinandergesetzt haben. Auf alle Arbeiten hinsichtlich der hier fokussierten Verwendung von KEM Verfahren trifft jedoch mindestens einer der folgenden Kritikpunkte zu:

| | |
|---|---|
| Keine Berücksichtigung Netzwerkcharakteristik | [268] [27] |
| Netzwerkparameter nicht unabh. betrachtet | [153] [198] [234] |
| Exklusive Evaluierung PQ-Only oder hybrid | [196] [153] [198] [27] |
| Evaluierung weniger Kandidaten | [196] [153] [198] [234] [268] |
| Keine Variation Sicherheitslevel | [196] [153] [198] [234] [268] |

Paquin, Stebila und Tamvada [196] präsentieren ein Framework, um Netzwerkbedingungen zu emulieren und den Verbindungsaufbau von TLS 1.3 mittels PQC anhand von OQS OpenSSL durchzuführen. Sie bieten somit einen interessanten Ansatz für die Analyse von Einflüssen unterschiedlicher Netzwerkparameter und -konfigurationen auf die Algorithmen oder deren unterschiedliche Sicherheitslevel. Aufgrund der Emulation ist eine weitestgehend dezidierte Kontrolle und Steuerung der Komponenten und Abläufe möglich [256]. Die Autoren evaluieren jedoch ausschließlich drei hybride KEM

Verfahren beziehungsweise drei Signaturalgorithmen und variieren lediglich Paketverlust sowie Delay.

Die vorliegende Arbeit greift den Ansatz auf. Zum einen soll das Framework so erweitert werden, dass Algorithmen und Netzwerkszenarien automatisiert eingelesen und evaluiert werden können. Zum anderen soll anschließend die Evaluierung ergänzt werden, um bis dato vernachlässigte Aspekte wie zusätzliche Algorithmen, Sicherheitslevel und weitere Netzwerkparameter zu berücksichtigen. Des Weiteren sollen die hybriden Ansätze den PQ-Only Verfahren gegenüber gestellt werden.

# VORGEHEN BEZÜGLICH FRAMEWORK & EVALUIERUNG

Das folgende Kapitel wurde in zwei Hauptteile gegliedert. Im ersten Abschnitt wird das zugrundeliegende Framework von Paquin, Stebila und Tamvada [196] analysiert und die vorgenommenen Erweiterungen beschrieben. Anschließend folgt der Aufbau respektive die Vorgehensweise bei Durchführung der Experimente zur Evaluierung der PQC Algorithmen. Dies beinhaltet die Darstellung der Testumgebung inklusive der verwendeten Hardware, die Auswahl der PQC Verfahren, die Konfiguration der Netzwerkparameter, die Zusammenstellung der Netzwerkszenarien sowie weitere Aspekte.

## 4.1 ANALYSE UND ERWEITERUNG DES FRAMEWORKS

Das original Framework von Paquin, Stebila und Tamvada [196] ist als GitHub Projekt[1] frei verfügbar und wurde aufgrund der vorgenommenen Anpassungen und Evaluierungen auf den Gitlab Server der Hochschule Darmstadt geklont[2].

### 4.1.1 Generelle Struktur

Grob kann der Aufbau beziehungsweise der Ablauf der Experimente wie folgt beschrieben werden. Mittels Linux Kernel Features, siehe Abschnitt 2.4.3, werden realistische Client-Server-Verbindungen emuliert. Für Server und Client werden zwei Netzwerk *namespaces* hinzugefügt und diesen werden unterschiedliche IP-Adressen zugewiesen. Während die Initiierung und Durchführung der Handshakes auf Clientseite direkt mittels der Funktionen aus OQS OpenSSL innerhalb des Experimentskripts erfolgt, werden die Server vorab mittels einer Implementierung von NGNIX[3] gestartet. Ziel ist die Messung der benötigten Zeit für einen Handshake bei Verwendung unterschiedlicher Algorithmen und unter variierenden Netzwerkbedingungen. Dafür werden während eines Experiments wiederholt Handshakes initiiert und die Verbindungen abgebrochen, sobald der Handshake beendet wurde. Die benötigte Zeit pro Handshake wird mittels einer modifizierten Version der OpenSSL Funktion *s_time* gemessen, siehe folgende Abschnitte.

Die Installation der benötigten Komponenten kann automatisiert, mittels eines Skript erfolgen. Die Umsetzung sowie die Komponenten werden im Folgenden beschrieben.

---

1 https://github.com/xvzcf/pq-tls-benchmark, zuletzt aufgerufen 06.02.22
2 https://code.fbi.h-da.de/aw/prj/athenepqc/pqc-in-tls/, zuletzt aufgerufen 22.01.22
3 https://www.nginx.com, zuletzt aufgerufen 22.01.22

### 4.1.2   Aufbau der Experimente

Das KEX Experiment bildet den Hauptteil für die Evaluierung der NIST PQC KEM Kandidaten und alle zugehörigen Dateien befinden sich innerhalb des Ordners „kex".

#### 4.1.2.1   Authentifizierung

Neben dem Schlüsselaustausch mittels quantensicherer Algorithmen muss eine Authentifizierung erfolgen. Hierfür werden diverse Parameter ausgehandelt respektive übertragen. Authentifizierung und Parameterauswahl erfolgen während der Experimente immer gleich, um eine ebenbürtige Gegenüberstellung zu gewährleisten. Für die Server-Authentifizierung wird ein ECDSA Zertifikat mit dem NIST Elliptic Curve (EC) Parameter *P-256* in Kombination mit der Hashfunktion SHA-384 gewählt. Öffentlicher Schlüssel und Signatur weisen bei ECDSA NIST P-256 eine Länge von 512 Bits auf und die Zertifikatskette besteht nur aus einer Abstufung von *Root* zum Server mit dem gleichen Algorithmus. Das Zertifikat muss vor der Durchführung der TLS Handshakes, das heißt vor Start des Experiments generiert werden. Die Generierung kann anhand von OpenSSL wie folgt vorgenommen werden:

```
###########################
# Generate ECDSA P-256 cert
###########################
# generate curve parameters
${OPENSSL} ecparam -out prime256v1.pem -name prime256v1

# generate CA key and cert
${OPENSSL} req -x509 -new -newkey ec:prime256v1.pem -keyout CA.key -out
    CA.crt -nodes -subj "/CN=OQS test ecdsap256 CA" -days 365

# generate server CSR
${OPENSSL} req -new -newkey ec:prime256v1.pem -keyout server.key -out
    server.csr -nodes -subj "/CN=oqstest CA ecdsap256"

# generate server cert
${OPENSSL} x509 -req -in server.csr -out server.crt -CA CA.crt -CAkey CA
    .key -CAcreateserial -days 365
```

Für die symmetrische Verschlüsselung wird zudem AES-256 mit Galois/-Counter Modus festgelegt.

Die beschriebenen Konfigurationen werden häufig in der Praxis verwendet und bieten sich daher für die Emulation realistischer TCP Handshakes an.

#### 4.1.2.2   Ablauf

Der Ablauf des Experiment wird im folgenden anhand von den beteiligten Skripten und Dateien beschrieben. Bash- und Pythonskripte befinden sich im Ordner „scripts".

- **run.sh** dient der Initiierung der Versuchsdurchläufe und startet die einzelnen Teilskripte, welche an dem Prozess beteiligt sind. Nachdem die Umgebung vorbereitet respektive Konfigurationen vorgenommen wurden, startet das Skript pro angegebenem Netzwerkszenario nacheinander eine Versuchsreihe anhand von *experiment.py*.

    1. **setup.sh** ruft zunächst *setup_ns.sh* auf, um die Netzwerkemulation zu konfigurieren. Anschließend wird das für die Authentifizierung des Servers benötigte ECDSA Schlüsselpaar und ein korrespondierendes Zertifikat anhand von OpenSSL generiert. Zudem wird die NGINX Applikation für die Bereitstellung der Serverinstanzen gestartet. Aufgrund der Konfiguration von NGINX, siehe auch Abschnitt 4.1.3, können die Server auf die OQS Version von OpenSSL zurückgreifen und PQC Algorithmen verwenden.

        a) **setup_ns.sh** dient der Konfigurationen der Netzwerkumgebung respektive -emulation mittels *ip-netns* und liegt zentral, außerhalb des Ordners „kex", da es auch für Signatur-Experimente genutzt werden kann. Es werden zunächst zwei *namespaces*, für Server und für Client, erstellt und mittels eines *veths*-Paares gekoppelt. Beide Instanzen erhalten einen eigenen IP-Adressraum.

    2. **experiment.py** ist zentrales Element und übernimmt die Steuerung der Handshakes sowie das Management der Messungen. Pro Szenario-Datei wird eine Versuchsreihe durchgeführt. Dabei besteht eine Versuchsreihe aus der Abfolge unterschiedlicher Netzwerkkonfigurationen, welche innerhalb der Szenario-Datei angegeben sind. Pro Konfiguration werden für jeden in *algorithms.csv* angegebenen Algorithmus $n$ Handshakes durchgeführt, wobei

        $$n = MEASUREMENTS\_PER\_TIMER \times TIMERS.$$

        Die Durchführung und Messung der Handshakes erfolgt mittels *s_timer.c* und wird innerhalb eines Threadpools mit der Größe *POOL_SIZE* parallel ausgeführt. Nach Durchführung der Messung werden die gemessenen Zeiten abhängig von Testszenario und Charakteristik des PQC KEM Algorithmus innerhalb von CSV-Dateien in „*results*" persistiert.

        a) **testscenarios** enthält die CSV-Dateien mit Netzwerkszenarien. Die Szenario-Dateien bestehen aus Tabellenzeilen, welche Netzwerkkonfigurationen entsprechend der *netem*-Attribute enthalten. Die konkreten Werte innerhalb einer Zeile sind in korrekter Reihenfolge in Tabelle 4.1 aufgeführt.

        b) **algorithms.csv** spezifiziert alle zu evaluierenden PQC KEM Algorithmen. Die erste Spalte ist der Bezeichnung der Algorithmenklasse vorbehalten, welche später für die Ordnerstruktur der Ergebnisse verwendet wird. Die Bezeichnung enthält aktuelle Informationen bezüglich Sicherheitslevel, hybrider Vor-

gehensweise sowie Kandidat oder Alternative. Alle weiteren Spalten enthalten die OQS OpenSSL beziehungsweise liboqs Identifikatoren für die gewünschten Algorithmen inklusive ihrer Parameterkonstellation wie beispielsweise *ntru_hrss*701. Die Anzahl der Spalten und somit die Anzahl der zu evaluierenden PQC Verfahren ist beliebig.

c) **networkmgmt.py** beinhaltet die aus *experiment.py* ausgelagerten Funktionen zur Konfiguration der *namespaces*. Somit erfolgt über *networkmgmt.py* die Steuerung der Emulation.

d) **s_timer.c** ist eine modifizierte Version der OpenSSL Datei *s_time.c* und wurde unverändert aus [196] übernommen. Während OpenSSLs *s_time.c* innerhalb einer festgelegten Zeitperiode möglichst viele Verbindungen aufbaut, misst *s_timer.c* die benötigten Millisekunden für einen Handshake. Dabei wird lediglich der Handshake durchgeführt und die Verbindung unterbrochen, bevor es zu der Übertragung von Nutzdaten kommt. Die Durchführung wird, entsprechend der übergebenen Anzahl aus *experiment.py*, wiederholt und zudem teilweise parallel durchgeführt. Die Schleife zur Messung und Durchführung der Handshakes ist wie folgt implementiert:

```c
while(measurements < measurements_to_make) {
    clock_gettime(CLOCK_MONOTONIC_RAW, &start);
    ssl = do_tls_handshake(ssl_ctx);
    clock_gettime(CLOCK_MONOTONIC_RAW, &finish);
    if (!ssl) {
        continue;
    }
    handshake_times_ms[measurements] = ((finish.tv_sec -
        start.tv_sec) * MS_IN_S) + ((finish.tv_nsec -
        start.tv_nsec) / NS_IN_MS);
    measurements++;
}
```

Die Zeitmessung wird anhand der Zeitdifferenz zwischen Beginn und Abschluss bestimmt. Die Zeiten werden über die systemeigene Funktion

```c
clock_gettime(clockid_t clk_id, struct timespec *tp);
```

ermittelt, welche die aktuelle Zeit anhand der mit $clk_id$ spezifizierten Systemclock bezieht. Bei $CLOCK\_MONOTONIC\_RAW$ handelt es sich dabei um eine Linux-eigene monotone Zeitzählung seit einem bestimmten Startzeitpunkt in der Vergangenheit. In

```c
SSL* do_tls_handshake(SSL_CTX* ssl_ctx)
```

erfolgt schließlich der Handshake über den OpenSSL Aufruf

```c
int return = SSL_connect(ssl);
```

3. **teardown.sh** dient lediglich dazu, temporär generierte Dateien zu entfernen, die NGINX Server-Prozesse zu stoppen und die Netzwerk *namespaces* zu entfernen.

4. **analyze.py** ermittelt aus den gemessenen Zeiten statistische Lageparameter, um eine einfachere beziehungsweise gezieltere Gegenüberstellung und Auswertung zu ermöglichen. Sie sind in Tabelle 4.2 beschrieben. Bei der Aggregation von Werten gehen grundsätzlich Informationen verloren, sodass schnell falsche Annahmen getroffen werden können. Eine klare Definition der Werte und die Berücksichtigung der individuellen Charakteristik ist daher wichtig. Das arithmetische Mittel, auch als Durchschnitt bezeichnet, ist der wohl bekannteste Lageparameter und intuitiv verständlich. Nachteil ist die eingeschränkte Robustheit. Ausreißer werden mit einkalkuliert und können zu einer Verzerrung führen. Mit der Standardabweichung kann die Streuung der Werte abgeschätzt werden. Der Median ähnelt dem Mittelwert, wird jedoch durch Ausreißer an den Rändern nicht beeinflusst. Diese Eigenschaft ist im hier beschriebenen Experiment von Vorteil, da es in Messreihen aufgrund der Zufallskomponenten bei der Emulation mittels Netem immer wieder zu großen Zeitverlusten kommen kann. Mit 0.75- und 0.95-Quantilen kann eine Abschätzung bezüglich der oberen 25 beziehungsweise 5 Prozent getroffen werden.

Die ermittelten Werte werden in CSV-Dateien innerhalb des Ordners *analyzed* geschrieben. *analyzed* weist eine zu *results* äquivalente Ordnerstruktur auf.

|  | Beschreibung | Einheit |
|---|---|---|
| *title* | — | — |
| *srv_pkt_loss* | Paketverlust | Prozent |
| *srv_delay* | Delay | Millisekunden |
| *srv_jitter* | Jitter | Millisekunden |
| *srv_duplicate* | Duplikate | Prozent |
| *srv_corrupt* | Noise | Prozent |
| *srv_reorder* | Reordering | Prozent |
| *srv_rate* | Paketverlust | Prozent |
| *cli_delay* | Delay | Millisekunden |
| *cli_jitter* | Jitter | Millisekunden |
| *cli_duplicate* | Duplikate | Prozent |
| *cli_corrupt* | Noise | Prozent |
| *cli_reorder* | Reordering | Prozent |
| *cli_rate* | Übertragungsrate | Megabits |

Tabelle 4.1: Aufbau Netzwerkkonfiguration innerhalb einer CSV-Datei eines Testszenarios in exakter Reihenfolge. „*srv*" für Verbindungen, welche an Server gerichtet sind, „*cli*" entsprechend für Clients. Eine ausführlichere Beschreibung der Parameter findet sich in Abschnitt 2.4.3.2

| Lageparameter | Identifier | Beschreibung (kontextbezogen) | Formel |
|---|---|---|---|
| Arithmetisches Mittel | *average* | Summe aller gemessenen Zeiten geteilt durch die Messanzahl $n$ | $\bar{x} = \frac{1}{n} \sum_{n=1}^{N} x_i$ |
| Std.abweichung | stdDev | Reale Streuung der Messwerte um Mittelwert | $\sigma = \sqrt{\frac{1}{n} \sum_{n=1}^{N} (x_i - \bar{x})^2}$ |
| Median | median | Auch Zentralwert oder 0.5-Quantil, ist nach Sortierung aller Werte der Messreihe derjenige, welcher genau in der Mitte steht. Bei einer Menge $V$ mit $n$ Messergebnissen wird folglich aus der sortierten Menge $V_{sort}$ der Wert an der Stelle $(n-1)/2$ gewählt | $m = V_{sort}[(n-1)/2]$ bzw. $m = \frac{V_{sort}[(n-2)/2] + V_{sort}[n/2]}{2}$ |
| 0.75-Quantil | 0.75-qtl | Auch 75%-Quantil. Äquivalent zum Vorgehen bei Median (0.5-Quantil) wird der Wert aus der sortierten Menge der Ergebnisse gewählt, sodass in der Regel 75% der Werte kleiner oder gleich und 25% größer oder gleich diesem sind | |
| 0.95-Quantil | 0.95-qtl | Siehe 0.75-Quantil | |

Tabelle 4.2: Beschreibung statistischer Lageparameter, welche mittels *analyze.py* aus den Experimentdaten erhoben werden. Die Beschreibungen richten sich nach den Definitionen innerhalb der Python Programmbibliothek *NumPy* (https://numpy.org, zuletzt aufgerufen 22.01.22), welche für Berechnungen herangezogen wurde

### 4.1.2.3 *Ergebnisstruktur*

Die Ergebnisse in Form der gemessenen Zeiten pro durchgeführtem Handshake werden entsprechend der Messreihenfolge in CSV-Dateien gespeichert und grundsätzlich im Ordner *results* abgelegt. Die Ablage innerhalb von *results* gliedert sich wie folgt:

1. Test-/Netzwerkszenario
    a) Sicherheitslevel
    b) Kandidaten und Alternativen
    c) PQ-Only und hybride Verfahren
        i. KEX Algorithmen

### 4.1.3 *Installation*

Für die automatische Installation ist das Skript *install-prereqs-ubuntu.sh* im Experimentordner hinterlegt. Folgende Abhängigkeiten werden für die Installation benötigt:

```
apt install -y git \
           build-essential \
           autoconf \
           automake \
           libtool \
           ninja-build \
           libssl-dev \
           libpcre3-dev \
           wget
```

Die benötigten Anwendungen für das Experiment, welche während der Installation geladen sowie eingerichtet werden, sind in Tabelle 4.3 inklusive der verwendeten Versionen beschrieben.

| Anwendung | Version |
|---|---|
| CMake[4] | Version 3.18, Build 3 |
| OQS liboqs[5] | Git Branch *main*, letzter Commit 2*a f*8*ad*7 (04.01.22) |
| OQS OpenSSL[6] | Git Branch *OQS-OpenSSL_1_1_1-stable*, letzter Commit 8494303 (16.12.21) |
| NGNIX[7] | 1.20.1 |

Tabelle 4.3: Für Experimente installierte und verwendete Anwendungen

### 4.1.4 *OpenSSL 3.0.0*

Wie in Abschnitt 3.1 beschrieben, erschien 2021 die überarbeitete OpenSSL Version 3.0.0. Sie ermöglicht eine Ergänzung von Funktionalitäten mittels sogenannter *Provider*. Auch OQS bietet eine entsprechende Implementierung. Der *oqsprovider*[8] ist auf GitHub frei verfügbar und kann in das original OpenSSL 3.0.0 Projekt[9] eingebunden werden. In den durchgeführten Experimenten wurde OpenSSL Version 1.1.1 verwendet und 3.0.0 nicht berücksichtigt. Innerhalb des Branchs *openssl3* befindet sich jedoch eine Implementierug der Experimente, welche zumindest in Teilen die Nutzung von OpenSSL 3.0.0 mittels Einbindung des *oqsproviders* ermöglicht. Zur Installation wurde das Skript *install* − *prereqs* − *ubuntu.sh* entsprechend angepasst. Zunächst wird der *oqsprovider* geklont. Innerhalb des *oqsproviders* kann anschließend das Skript *f ullbuild.sh* aufgerufen werden, welches die Installation und Konfiguration von OpenSSL 3.0.0 und *liboqs* übernimmt. Mithilfe

---

8 https://github.com/open-quantum-safe/oqs-provider, zuletzt aufgerufen 31.01.22
9 https://github.com/openssl/openssl/tree/openssl-3.0, zuletzt aufgerufen 31.01.22

des folgenden Programmcodes kann zudem der Provider für die Aufrufe des Clients innerhalb von *s_timer.c* aktiviert werden:

```c
const char* OQS_PROVIDER_PATH = "/home/pqc-in-tls/pq-tls-benchmark-
    framework/emulation-exp/code/tmp/oqs-provider/_build/oqsprov";

OSSL_PROVIDER_set_default_search_path(NULL, OQS_PROVIDER_PATH);

OSSL_PROVIDER *oqsprovider;
OSSL_PROVIDER *deflt;

/* Load Multiple providers into the default (NULL) library context */
oqsprovider = OSSL_PROVIDER_load(NULL, "oqsprovider");
if (oqsprovider == NULL) {
    fprintf(stderr, "Failed to load provider\n");
    goto ossl_error;
}
deflt = OSSL_PROVIDER_load(NULL, "default");
if (deflt == NULL) {
    fprintf(stderr, "Failed to load Default provider\n");
    OSSL_PROVIDER_unload(oqsprovider);
    goto ossl_error;
}
```

Für die Nutzung müssen zusätzlich die NGINX Server angepasst werden. Aktuell bietet NGINX keine korrespondierenden Funktionalitäten für die Aktivierung von Providern. Eine alternative Möglichkeit ist die globale Aktivierung innerhalb der OpenSSL Konfigurationsdatei. Diese kann wie folgt vorgenommen werden:

```
[providers_sect]
default = default_sect
oqsprovider = oqsprovider_sect

[default_sect]
activate = 1

[oqsprovider_sect]
module = /home/pqc-in-tls/pq-tls-benchmark-framework/emulation-exp/code/
    tmp/oqs-provider/_build/oqsprov/oqsprovider.so
activate = 1
```

## 4.2 umsetzung & durchführung der evaluierung

Nach der Integration und Erweiterung des zuvor beschriebenen Frameworks konnten mit dessen Hilfe diverse Experimente respektive Evaluierungen der NIST PQC Kandidaten vorgenommen werden. Die konkrete Durchführung inklusive der verwendeten Hardware und Software sowie den Netzwerkkonfigurationen und dem Auswertungsprozess wird im Folgenden erläutert.

4.2.1 *Gewählte Testumgebung*

Für die Tests wurde ein Dell[10] Precision 3630 Tower (0871) verwendet. Er basierte auf der x86_64 Architektur und war ausgestattet mit dem Prozessormodell Intel[11] *i7-8700* Prozessor mit einer Grundtaktfrequenz von 3.20 Gigahertz und einer maximalen Turbo-Taktfrequenz von 4.60 Gigahertz. Der Rechner verfügte über sechs Kerne, zwölf vCPUs und ermöglichte die Ausführung aller für die Nutzung der Algorithmen in liboqs benötigten Befehlssätze wie *bmi*1 oder *avx*2. Des Weiteren besaß er einen 32 Gigabyte Arbeitsspeicher und war mit dem linuxbasierten Betriebssystem Ubuntu 20.04.3 LTS ausgestattet.

4.2.2 *Gewählte Frameworkkonfiguration*

Für die Anzahl der NGINX Server wurde auf 4 gesetzt und für die maximale Anzahl paralleler Clients galt $POOLSIZE = 7$, sodass, äquivalent zum Vorgehen in [196], jeder Instanz möglichst eine eigene vCPU zugewiesen werden konnte. Clients und Server sollten möglichst unabhängig voneinander existieren und Störungen durch Scheduling sollten reduziert werden. Die Anzahl der Timer wurde zudem auf 10 gesetzt und die Anzahl der Messungen auf 20, sodass insgesamt pro Netzwerkkonfiguration und Algorithmus $10 \times 20 = 200$ Messungen durchgeführt wurden.

4.2.3 *Gewählte Netzwerkparameter & Testszenarien*

Für die Emulation eines herkömmlichen Netzwerks bei TLS Anwendung mittels NetEm wurden unterschiedliche Parameterkonstellationen gewählt. Die Auswahl wird im Folgenden anhand des jeweiligen NetEm Optionen beschrieben und im Bezug zu typischen Netzwerkverbindungen gestellt. Die verfügbaren Parameter sind in Tabelle 4.1 aufgeführt.

Generell muss die Übertragung von der Vermittlungsschicht des Senders bis zur Vermittlungsschicht des Empfängers mittels der NetEm Parametern dargestellt werden, während die Kommunikation von der OpenSSL Implementierung des Senders bis einschließlich der Vermittlungsschicht des Senders sowie von der Vermittlungsschicht des Empfängers bis zur OpenSSL Implementierung des Empfängers auf den Rechnern faktisch durchgeführt wird. Dabei handelt es sich um ein TCP/IP-Netz, wie es oftmals bei TLS Verbindungen Verwendung findet. Eine generelle Beschreibung des grundlegenden Aufbaus eines Netzwerks innerhalb des Internets und den Einflussfaktoren findet sich in Abschnitt 2.4.1.

Die hier beschriebenen Konfigurationen sind innerhalb der CSV-Dateien für Testszenarios definiert, wie in Abschnitt 4.1.2 beschrieben.

Generell erfolgte die Auswahl entsprechend folgender Zielsetzungen:

---

10 https://www.dell.com/, zuletzt aufgerufen 17.01.22
11 https://www.intel.com/, zuletzt aufgerufen 22.01.22

1. Rückschlüsse auf die Charakteristik und das Verhalten einzelner PQC Algorithmen, hybrider Verfahren und Sicherheitslevel anhand des Verhaltens bei Veränderung eines gezielt ausgewählten Netzwerkparameters

2. Eignung der Algorithmen bei Verwendung in verschiedenen, typischen Szenarien, welche eine tatsächliche Anwendung innerhalb des Internets widerspiegeln

Generell gibt es eine Vielzahl von Einflussfaktoren, welche bei der Einschätzung typischer Netzwerkparameter berücksichtigt werden müssen. Laut Camp, Boleng und Davies [60] sei die typische Konstellation der Netzwerkparameter wie Latenz, Jitter oder Anzahl erfolgreich versendeter Pakete maßgeblich von folgenden Aspekten abhängig:

- Endgeräte

- Übertragungsmedien

- Übertragungsdistanz

- Beteiligte Kommunikationsprotokolle

- Anzahl und Art der Zwischenknoten beziehungsweise Vermittlungsinstanzen

Insbesondere bei Mobilfunk und WLAN kann die Charakteristik der Verbindung aufgrund der Mobilität der Endgeräte stark variieren [60]. Zudem werden die Signale durch die räumliche Umgebung verändert. Mehrwegeausbreitung und Fading können das Signal dämpfen, verzögern und verfälschen [135].

Bei Variation eines gezielten Parameters wurden die Werte für Client und Server in der Regel gleichmäßig verändert. Alle weiteren simulierbaren Störungen wurden möglichst gering gehalten. Das heißt, prozentuale Anteile wie beispielsweise Paketverlustrate wurden auf Null gesetzt, für die Übertragungsrate wurde ein vergleichsweise hoher Wert von 500 Mbps angenommen und die Latenz orientierte sich an dem von Paquin, Stebila und Tamvada [196] verwendeten Wert von 2.684 Millisekunden für nah beieinander liegende Rechner bei sehr guter Verbindung.

#### 4.2.3.1   Latenz (scenario_delay.csv)

Bei der Einschätzung der herkömmlichen Latenz von Datenpaketen bei typischen Internetverbindungen herrscht eine große Varianz, welche auf diverse Faktoren zurückgeführt werden kann. Sie werden in Abschnitt 2.4.1 beschrieben. Je nach geografischer Entfernung und verwendeter Technologie werden beispielsweise die Werte 2, 10, 20, 30, 50, 100, 140 oder 200 Millisekunden für eine realistische Simulation verwendet [47, 76, 140, 196, 256, 258, 265, 271]. Dabei gelten Werte ab 100 Millisekunden meist als vergleichsweise hoch [47, 174]. In einem Beispiel von Biswal und Gnawali [47] wird für die Kommunikation innerhalb eines gewöhnlichen WLANs an einer Universität sowie innerhalb des T-Mobile LTE-Netzes in den USA jeweils ein durchschnittlicher Wert von etwa 30 Millisekunden gemessen. In der gleichen Arbeit kamen 3G Netze in den USA sowie Indien auf 94 beziehungsweise 153 Millisekunden. Im Bereich IoT und autonomes Fahren ist eine geringe Latenz von großer Bedeutung, um den Echtzeit-Anforderungen gerecht zu werden. Mei u. a. [175] geben für den Datenverkehr unter V2V eine durchschnittliche Latenz von 46.9 Millisekunden für LTE und 15.1 Millisekunden für 5G an. Die maximal akzeptable Latenz liege in diesem Kontext bei 100 Millisekunden.

Aufgrund dieser Beobachtungen wurde für die Versuchsreihe folgende Menge $D$ untersucht:

$$D = [0, 1, 2.5, 5, 7.5, 10, 15, 20, 25, 30, 40, 50, 60, 80, 10, 120]$$

### 4.2.3.2 Jitter (scenario_jitter_delay20ms.csv)

Jitter ist die Varianz bei zeitlicher Verzögerung der Datenpakete und geht daher mit der Latenz einher. Sie ist stark vom gewählten Übertragungsmedium sowie den auftretenden Störungen innerhalb des Netzes abhängig. Ashraf u. a. [19] führen als Beispiel an, dass bei weitestgehend fehlerfreien Funkverbindungen eine Streuung von etwa 500 Nanosekunden kaum zu vermeiden sei. Bei Auftreten diverser variabler Faktoren wie sich bewegenden Objekten könne es zudem zu einer unregelmäßigen Ab- beziehungsweise Umleiten von Funksignalen kommen, welche in punktuellen Verzögerungen von mehreren Millisekunden resultiere. Bei Verarbeitung und Weiterleitung der Daten an Zwischen- und Endknoten auf verschiedenen Netzwerkebenen kann es ebenfalls zu kurzzeitigen Verzögerungen kommen, insbesondere bei Überlastung. Isolierte Kabeltechnologien sind weniger betroffen. Für die Überprüfung wurde daher eine übliche Latenz von 20 Millisekunden gewählt und der Jitter schrittweise vergrößert. Folgende Menge $J$ wurde berücksichtigt:

$$J = [0, 0.1, 0.25, 0.5, 0.75, 1, 1.5, 2, 2.5, 3, 5, 7, 9, 12, 15, 20]$$

Somit ergibt sich für die tatsächliche Latenz der Datenpakete Werte im Intervall $[0, 40]$. Da sich eine auftretende Störungen meist über einen bestimmten Zeitraum erstreckt und mehrere Pakete beeinflussen kann, ist der Jitterwert in der Regel abhängig vom vorausgehenden Paket, sodass eine Korrelation von 25 Prozent zum vorherigen Datenpaket ergänzt wurde.

### 4.2.3.3 Übertragungsrate (scenario_rate_both.csv)

Um den Wertebereich realistischer Übertragungsraten bei HTTPS Nutzung unter verschiedenen Voraussetzungen umfassend abzustecken, wurde ins-gesamt ein Bereich zwischen 0.1 und 2000 Mbps gewählt. Die untersuchte Menge *R* beinhaltet daher folgende Werte:

$$R = [0.1, 0.25, 0.5, 1, 1.5, 2, 3, 4, 5, 7.5, 10, 15, 20, 30, 45, 60, 75, 100, 150,$$
$$250, 500, 1000, 2000]$$

Die Auswahl stützt sich auf verschiedene Ausarbeitungen. So wird beschrieben, dass für eine Evaluierung realistischer Szenarien bei HTTP die Werte $5, 10, 50, 100$ [140, 161, 174, 258] beziehungsweise zusätzlich 1 und 2 [47, 174] berücksichtigt werden sollten. Plaßmann [203] gibt als maximale Netto-Übertragungsraten für WLAN, je nach Standard, $5, 25$ und $100$ Mbps an. Bei verschiedenen Versuchen an Universitäten wurden für das Campus-WLAN Durchschnittswerte von $51, 30$ beziehungsweise $40$ Mbps gemessen [9, 47, 265]. Die oben beschriebenen Werte spiegeln sich auch bei Messungen im Bereich Mobilfunk wieder, wobei es gerade bei Mobilfunk der 2. und 3. Generation zu niedrigeren Raten bis $0.17$ kommen kann [47]. Mamman u. a. [166] messen LTE Raten von 25 bis 0.25 Mbps. IoT und andere ressourcenbeschränkte Geräte kommen teils auf noch geringere Werte. Bei optimaler Glasfaseranbindung im Heimnetz oder Verbindungen von Rechenzentren sind hingegen extrem hohe Raten von über 2000 Mbps möglich.

Im Bezug auf das Experiment sollte beachtet werden, dass die tatsächlich verfügbare Rate ab der simulierten Vermittlungsschicht etwas geringer ist als die des Gesamtnetzes, da Bits für die Header der unterliegenden Protokollschichten benötigt werden. Ethernetübertragung bietet beispielsweise nach Standard IEEE 802.3 [130] eine Framegröße zwischen 64 und 1518 Bytes, wobei für Header und Tail in der Regel 18 Byte vorgesehen sind. Bei einer Übertragungsrate von 30 Mbps könnten somit mindestens 2470 Pakete pro Sekunde übertragen werden. Nach Abzug des Headers bleiben bei annähernd optimaler Verteilung $30Mb - 2470 \times 18B \approx 3,7MB$ pro Sekunde für IP Pakete.

Da mitunter die durchschnittlichen Übertragungsraten für Up- und Download sowie bei Client und Server stark variieren und Client und Server während des TLS Handshakes unterschiedliche Daten und insbesondere unterschiedliche Datenmengen versenden, wurden die Datenraten in den Szenarien *scenario_rate_cli.csv* und *scenario_rate_srv.csv* jeweils unabhängig voneinander untersucht.

#### 4.2.3.4 *Paketverlust (scenario_packetloss.csv)*

In der Regel werden für den Paketverlust bei HTTPS Anwendungen Werte unterhalb von 5 Prozent angenommen [196]. In Studien sei für Google Server beispielsweise eine Paketverlustrate zwischen 0 und 2 Prozent festgestellt worden [258]. Ein Paketverlust von über 10 Prozent erscheine hingegen als sehr hoch und ungewöhnlich [76, 256]. In vielen Experimenten werden daher Werte von $0\,0.1, 0.5, 1, 2, 3, 5$ oder $10$ Prozent angenommen [47, 65, 256, 258, 265]. In einigen Versuchsreihen für WLAN und zellulare Netze kam es hingegen zu Werten von 12 [22] respektive 15 Prozent [116]. Für die durch-geführten Testdurchläufe mit variierendem Paketverlust wurde die Menge *PL* mit

$$PL = [0, 0.25, 0.5, 1, 1.5, 2, 3, 4, 5, 6, 7, 8, 9, 10, 11, 12, 13, 14, 15, 16, 17, 18, 19, 20]$$

verwendet. Erste Testdurchläufe mit

$$PL = [0, 0.25, 0.5, 1, 1.5, 2, 3, 4, 6, 8, 10, 12, 14, 16, 18]$$

ließen keine ausreichenden Rückschlüsse bezüglich der Entwicklung der Handshakezeiten zu, da die Sprünge innerhalb des Graphen, gerade im Hinblick auf die in Abschnitt 5.3 beschriebenen Ausreißer, zu groß waren.

### 4.2.3.5 Reordering (scenario_reorder.csv)

Da Pakete bei IP grundsätzlich unabhängig voneinander gesendet werden und je nach Netzwerk Zustand unterschiedliche Routen nehmen können, ist eine falsche Reihenfolge beim Empfänger durchaus denkbar. Yu, Xu und Yang [265] geben an, dass bei typischen WLAN und Mobilfunk Verbindungen zwischen 0 und 30 Prozent der Pakete in falscher Reihenfolge beim Empfänger eingehen. Um eine Vergleichbarkeit zu den Messungen bei Paketverlust zu gewährleisten, wurden für die Menge RO der zu testenden Prozentanteile die gleichen Werte angenommen, das heißt $RO = PL$

### 4.2.3.6 Duplikate (scenario_duplicate.csv) & korrupte Pakete (scenario_corrupt.csv

Zu den Anteilen von Duplikaten und fehlerhaften Paketen bei gewöhnlichen HTTPS beziehungsweise TLS Verbindungen konnten zunächst keine Arbeiten ausfindig gemacht werden. Es wird angenommen, dass die Zahlen dem Paketverlust ähneln. Insbesondere verwerfen TCP und IP lediglich fehlerhafte Pakete ohne weitere Maßnahmen anzustoßen, sodass sich die Situation für den Sender äquivalent zu verlorenen Paketen gestaltet. Der Anteil korrupter Pakete kann den Anteil verlorener Pakete somit nicht übersteigen. Für einen direkten Vergleich zum Paketverlust, wurden, ebenso wie bei Reordering, die gleichen Prozentwerte verwendet, sodass für die Mengen *DU* und *C* bei Duplikaten und korrupten Paketen gilt $DU = C = PL = RO$

### 4.2.4 Gewählte Algorithmen

Grundsätzlich wurden für die Experimente alle verfügbaren Algorithmen berücksichtigt. Neben den Kandidaten, welche von der NIST als besonders geeignet und ausgereift angesehen werden, ist auch eine Standardisierung von Alternativen noch nicht ausgeschlossen. Ein aktueller Vergleich zwischen allen Algorithmen bietet sich an, um Vor- und Nachteile ausfindig zu machen und die Einflüsse der darunterliegenden mathematischen Ansätze zu verstehen. Zudem können Einflüsse bei eventuellen Änderungen innerhalb der Implementierungen durch später durchgeführte Experimente besser nachvollzogen werden. Das KEM Verfahren Classic McEliece ist aktuell noch nicht innerhalb der modifizierten OpenSSL Version von OQS verfügbar, da die Größe des öffentlichen Schlüssels die zulässige Maximalgröße der verwendeten Erweiterungen innerhalb des *key_share* deutlich übersteigt. Siehe Abschnitt 3.1.2.1. Für die Integration wäre eine umfassendere Anpassung der internen Logik bei OpenSSL notwendig, sodass aus Verfahren nicht berücksichtigt wurde. Für alle anderen KEM Algorithmen wurden für alle verfügbaren Parameterkonstellationen respektive Sicherheitslevel und hybride Varianten Experimente durchgeführt.

Für einige Algorithmen existieren spezielle Implementierungen, welche bei entsprechend verfügbarer Hardwarebeschleunigung von Vorteil sind. Kyber-90s nutzt beispielsweise AES-256 mit CTR Modus und SHA-2 anstelle von SHAKE bei Kyber. Ähnlich bei FrodoKEM. In dieser Arbeit wurden jedoch zunächst nur die Hauptversionen Kyber und FrodoKEM mit SHAKE berücksichtigt. Des Weiteren existieren für für Sike und NTRU Prime Parametersets (*sikep*434, *ntrulpr*653, *sntrup*653), deren Einstufung bezüglich des Sicherheitslevels für die NIST noch unklar ist [11] oder es existieren mehrere Konfigurationen für die gleiche Stufe. Diese wurden ebenfalls nicht berücksichtigt.

### 4.2.5   *Einordnung der Handshakedauer*

Um die gemessenen Zeiten für TLS Handshakes einordnen und bewerten zu können, muss der entsprechende Kontext berücksichtigt werden. TLS wird in der Regel innerhalb des Internets verwendet, um Inhalte von Webseiten und -anwendungen zu laden. Zudem existieren vereinzelt weitere Anwendungsgebiete, beispielsweise im Bereich IoT.

Ein performanter TLS Verbindungsaufbau ist von Bedeutung, um die Vermittlung der Nutzdaten nicht zu verzögern. Gerade im Bereich der Webanwendungen gilt die verzögerte Datenvermittlung als eine der hauptsächlichen Ursachen für die Unzufriedenheit von Endanwendern [111, 112, 131, 208, 257]. Als maximale Grenze werden 2 bis 3 Sekunden angegeben [112, 188], wobei zahlreiche Protokolle und Prozesse beteiligt sind und gewisse Zeit in Anspruch nehmen. Der TLS Handshake selbst benötigt nur einen Bruchteil der Gesamtzeit. Der erfolgreiche Verbindungsaufbau ist jedoch Grundvoraussetzung für viele weitere Schritte. Zudem fanden Sy u. a. [243] heraus, dass über 95 Prozent der Webseitenaufrufe mehr als eine TLS Verbindung aufbauen würden, da beispielsweise andere Webinhalte integriert würden. Bei etwa 25 Prozent der Webseitenaufrufe seien mehr als 25 TLS Handshakes durchgeführt worden. Somit wirkt sich die durch PQC verursachte Verzögerung bei einem Handshake mehrfach auf den Webseitenaufruf aus. Dies bestätigen Naylor u. a. [187] in ihren Analysen verschiedener Webseitenaufrufe. Sie kommen zu dem Ergebnis, dass die Handshakedauer einen merklichen Einfluss auf die Gesamtzeit des Webseitenaufrufs habe. In annähernd der Hälfte aller Fälle verzögere die Verwendung von TLS den Aufruf mit HTTP um mehr als 500 Millisekunden und wirke sich damit mit einem Faktor von 1.5 auf die ursprüngliche Gesamtzeit aus. Verzögerungen von mehr als 500 Millisekunden durch Verwendung von TLS würden die UX deutlich einschränken. Während die symmetrische Verschlüsselung vergleichsweise performant durchführbar sei, würde insbesondere der Handshake die benötigte Zeit vergößern. Bei durchschnittlich 2 bis 5 durchgeführten Handshakes [243] ergibt sich in etwa eine erstrebenswerte Zeitspanne von 0 bis 100 Millisekunden. Bei einer Handshakedauer von mehreren hundert Millisekunden kann es hingegen zu Einschränkungen der UX kommen. Dieser Wert schwankt stark, je nach Zusammensetzung der Webseite und Netzwerkcharakteristik. Eine Überschreitung sollte aber vermieden werden.

In Bereichen mit Echtzeitanforderungen spielt die performante Datenvermittlung eine noch tragendere Rolle. Um dennoch Authentizität und Vertraulichkeit zu gewährleisten, ist ein Einsatz von TLS in einigen Fällen sinnvoll [205, 248]. Die Ansprüche bezüglich der Dauer für Verbindungsaufbau und Übertragung der Nutzdaten sind sehr unterschiedlich, sodass hier keine dezidierte Angabe erfolgen kann. Generell gilt, je performanter der Handshake, desto besser.

# 5 ERGEBNISSE & DISKUSSION

Im folgenden Kapitel werden die beobachteten Ergebnisse beschrieben und diskutiert. Die gemessenen Zeiten für alle Algorithmen- und Netzwerkkonstellationen sowie die berechneten Lageparameter befinden sich innerhalb des Projektordners *PQC in TLS*[1]. Dort befinden sich ebenfalls alle angefertigten grafischen Auswertungen. Die Darstellungen der im Folgenden diskutierten Verläufe sind im Anhang beigefügt und besonders bemerkenswerte Grafiken sind in den Textverlauf eingebunden. Die Charakteristik der Algorithmen und die Einschätzungen der NIST, auf welche sich im folgenden Kapitel mehrfach bezogen wird, sind in Abschnitt 2.2 aufgeführt. Insbesondere im Statusbericht zur zweiten Runde des PQC Auswahlverfahrens [11] schildert die NIST ihre Einschätzungen und Erwartungen bezüglich der Kandidaten und Alternativen.

Die angeführten Werte für typische Netzwerkcharakteristiken und Szenarien im Bezug auf mit TLS abgesicherte Verbindungen beziehen sich grundsätzlich auf die Analyse in Abschnitt 4.2.3.

Aufgrund der sehr großen Differenzen zwischen den gemessenen Zeiten und der durchweg hohen Varianz, gerade bei minderwertiger Netzwerkqualität, erschien die Auswertung des arithmetischen Mittels meist nicht als sinnvoll. Der Fokus der Auswertungen liegt in der Regel auf dem Median. Wird kein konkreter Lageparameter benannt, so ist grundsätzlich von einer Verwendung dessen auszugehen. Das arithmetische Mittel, das 0.75-Quantil sowie das 0.95-Quantil werden teils zum Vergleich hinzugezogen.

Für die Einordnung der gemessenen Zeiten wurde in Abschnitt 4.2.5 ausgearbeitet, dass eine Handshakedauer unter 80 bis 100 Millisekunden erstrebenswert ist. Verzögerungen von mehreren hundert Millisekunden sollten hingegen vermieden werden.

Zunächst kann festgestellt werden, dass die Algorithmenperformanz bei gleichbleibend guter Netzwerkqualität ebenfalls konstant bleibt, wie in den Abbildungen A.2, A.1a und 5.2 ersichtlich wird. Verschlechtert sich die Netzwerkqualität, beispielsweise aufgrund einer steigenden Wahrscheinlichkeit bezüglich fehlerhafter oder verlorener Pakete wie in Abbildung A.10, verzögert sich die Durchführung der Handshakes und auch die Verhältnisse zwischen den Algorithmen verändern sich. Die Unterschiede sind zunächst auf die Größen von öffentlichen Schlüsseln und Chiffraten sowie die Effizienz der Algorithmenoperationen zurückzuführen. Wie bereits Paquin, Stebila und Tamvada [196] mittels der Netzwerk Anwendung *tcpdump*[2] feststellen konnten, ist bei größeren zu übertragenden Datenmengen eine Segmentierung in mehrere Pakete erforderlich. Bei vermindertem Durchsatz, beispielsweise aufgrund von geringerer Übertragungsrate, Latenz oder verlorenen und fehlerhaften Paketen, nimmt die benötigte Zeit zu.

---

1 https://code.fbi.h-da.de/aw/prj/athenepqc/pqc-in-tls, zuletzt aufgerufen 02.02.21
2 https://wiki.ubuntuusers.de/tcpdump/, zuletzt aufgerufen 04.02.22

## 5.1 netzwerkcharakteristiken

### 5.1.1 *Gleichbleibende Netzwerkparameter*

Während der Messreihen bei gleichbleibendem Verbindungszustand wurden keinerlei Veränderungen innerhalb der NetEm Konfigurationen vorgenommen und keinerlei Störungen emuliert. Die Übertragungsrate lag bei 500 Mbps und es wurde eine konstante Verzögerung von 2.648 Millisekunden pro Paket hinzugefügt. Wenn im Folgenden von einer sehr guten Netzwerk- oder Verbindungsqualität gesprochen wird, so nimmt dies Bezug auf diese geschilderten Konditionen.

Bei Durchführung der Testdurchläufe zeigte sich, dass auch die gemessenen Zeiten für eine Handshakedurchführung relativ konstant verlaufen. Abbildung 5.1 zeigt den Median für alle getesteten KEM Algorithmen bei Sicherheitsstufe 1. Die Werte variieren nur leicht im Bereich von ein bis zwei Millisekunden und im Gesamten lässt sich weder eine Steigung, noch ein Gefälle erkennen. Da die Varianz innerhalb einer Messreihe nur gering ist, verhalten sich Median und arithmetisches Mittel ähnlich, siehe Abbildung A.1. Auch die Abweichungen der Verläufe bei Betrachtung des 0.95-Quantils sind im Vergleich zu anderen Testszenarien gering. Abbildung A.1c zeigt, dass es sich in der Regel nur um einige Millisekunden handelt.

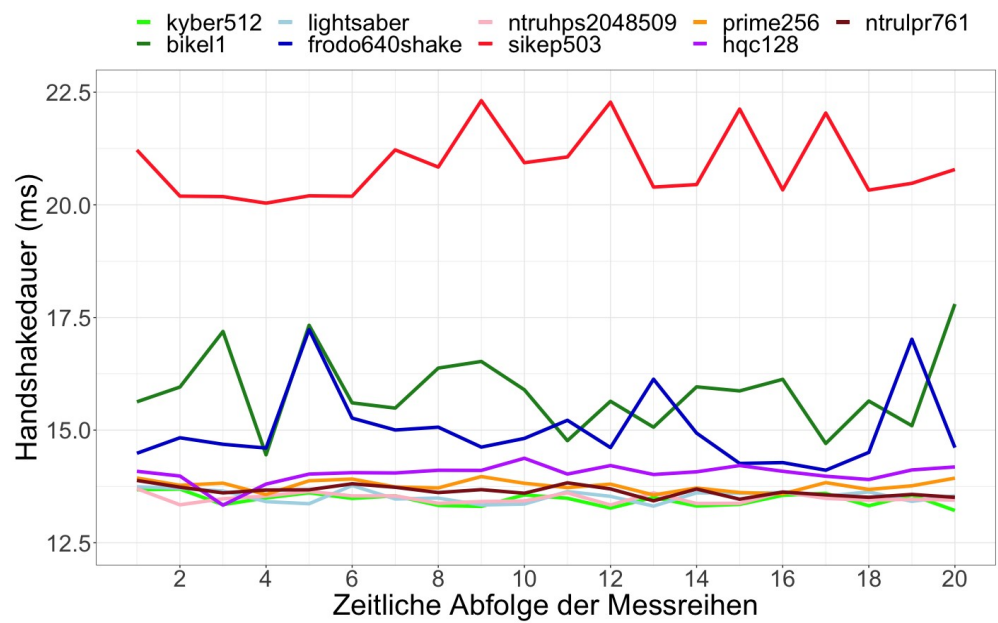

Abbildung 5.1: Median der Handshakedauer in Millisekunden bei 20 Messreihen mit unveränderten Netzwerkparametern für alle evaluierten Kandidaten und Alternativen unter Sicherheitsstufe 1

Beim Vergleich der KEMs fällt auf, dass die Kandidaten sowie NTRU Prime allesamt eine sehr gute Performanz aufweisen und größtenteils das klassische ECDH Verfahren zeitlich unterbieten können. Die Alternativen BIKE und FrodoKEM sowie insbesondere SIKE benötigen hingegen teils einige Millisekunden mehr. Dies ist auf die unterschiedliche Effizienz der Algorithmenoperationen zurückzuführen. Insbesondere SIKE benötigt zwar geringe Datenmengen, die Berechnungen für Schlüssel und Chiffrate sind jedoch aufwändig. Die Annahme bestätigt sich auch bei theoretischer Betrachtung. Sei $t_{Ges}$ die insgesamt benötigte Zeit, so ergibt sich diese anhand der Summe aus der Zeit $t_{S\_Ges}$ für das Senden aller Pakete und der Zeit $t_{Op}$ bezüglich der Datenverarbeitung inklusive der KEM Operationen:

$$t_{Ges} = t_{S\_Ges} + t_{Op} \qquad \text{mit } t_{S\_Ges}, t_{Op} \in \mathbb{R}_0^+$$

Ferner kann die Gesamtsendezeit $t_{S\_Ges}$ bestimmt werden anhand der Anzahl $n$ der zu sendenden Pakete geteilt durch die Senderate $r$ in Paketen pro Sekunde beziehungsweise Millisekunde:

$$t_{S\_Ges} = \frac{n}{r} \qquad \text{mit } n \in \mathbb{N},\ r \in \mathbb{R}_0^+$$

Für die Gesamtzeit gilt nun:

$$\begin{aligned} t_{Ges} &= t_{S\_Ges} + t_{Op} \\ &= \frac{n}{r} + t_{Op} \qquad \text{mit } n \in \mathbb{N},\ r, t_{Op} \in \mathbb{R}_0^+ \end{aligned}$$

Bei konstantem $r$, $n$ und hinreichend großem $r$ ist somit der Summand $t_{Op}$ entscheidend im Bezug auf $t_{Ges}$.

Abbildung 5.1 zeigt, dass $t_{Ges}$ in der Praxis schwankt und von weiteren äußere Einflüssen abhängt. Dies sollte bei der Analyse grundsätzlich beachtet werden. Die Unregelmäßigkeiten können durch verschiedene Ursachen hervorgerufen werden. Möglich ist beispielsweise eine Beeinflussung aufgrund von weiteren Prozessen innerhalb der Testumgebung oder Kontroll- und Steuerungsmechanismen der unterliegenden Netzwerkschichten, siehe Abschnitt 5.3. Es fällt auf, dass die Varianz bei den langsameren Algorithmen FrodoKEM, Bike und SIKE stärker ausgeprägt ist. Dies könnte darauf zurückzuführen sein, dass bei längerer Durchführung auch die Wahrscheinlichkeit steigt, dass zwischenzeitlich eine Verzögerung durch parallele Berechnungen oder ähnliches auftritt.

Beim Vergleich der Sicherheitslevel wurde die grafische Darstellung zur besseren Übersicht zwischen Kandidaten und Alternativen aufgeteilt. Abbildung 5.2a zeigt den Medianverlauf der Kandidaten für alle Sicherheitslevel. Der Graph wirkt sehr unübersichtlich, da die Verläufe sehr dicht beieinander liegen. Die Differenz zwischen den jeweiligen Zeiten der Sicherheitslevel für einen Kandidaten beträgt meist unter 0.5 Millisekunden. Daraus folgt, dass bei sehr guter Verbindungsqualität die Wahl des Sicherheitslevels eines Kandidaten keinen signifikanten Einfluss auf dessen Performanz hat. Da die Performanz in diesem Kontext insbesondere von der Effizienz der KEM

Operationen abhängt, folgt weiter, dass sich die Gesamteffizienz der Operationen eines Kandidaten mit steigendem Sicherheitslevel nicht signifikant verschlechtert. Bei den Alternativen gestaltet sich die Situation anders, siehe Abbildung 5.2b. Mit Ausnahme der NTRU Prime Versionen weisen die Verfahren, je nach Sicherheitslevel, unterschiedliche Zeiten auf. Insbesondere bei FrodoKEM und SIKE wird dies deutlich. Die durchschnittliche Differenz zwischen Level 1 und 5 beträgt circa 20 Millisekunden. Ergo sollte beachtet werden, dass sich ein Wechsel des Sicherheitslevels je nach Algorithmus völlig unterschiedlich auf die Gesamtperformanz des Handshakes auswirken kann. Ferner sollte die Wahl eines PQC KEM innerhalb von TLS 1.3 im Bezug auf die Performanz auch vom angestrebten Sicherheitslevel abhängig gemacht werden.

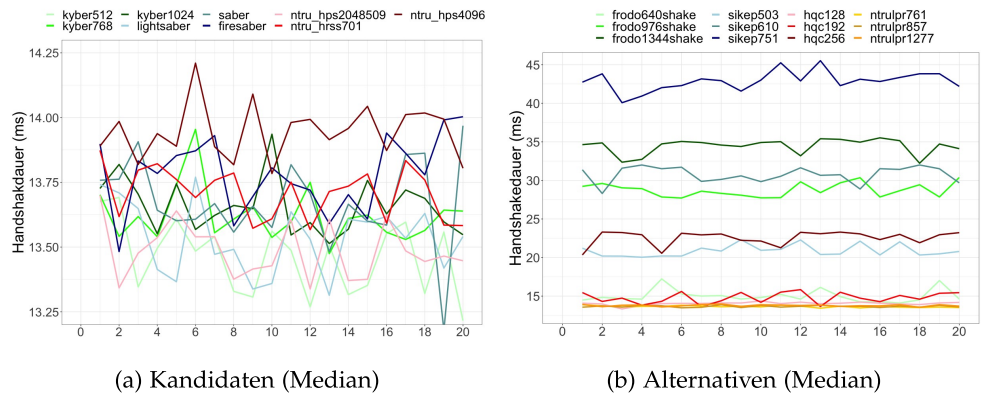

(a) Kandidaten (Median)      (b) Alternativen (Median)

Abbildung 5.2: Handshakedauer in Millisekunden in Abhängigkeit zur Nummerierung der durchgeführten Messung bei unveränderten Netzwerkparametern und Betrachtung des Median für alle Sicherheitslevel bei Kandidaten und Alternativen

Für die Betrachtung hybrider Versionen wurden Kandidaten und Alternativen ebenfalls getrennt dargestellt. Abbildung 5.3a zeigt Messungen für hybride und PQ-Only Versionen der Kandidaten sowie für das klassische ECDH unter Sicherheitslevel 1. Ähnlich wie bei den Messungen zu unterschiedlichen Sicherheitsleveln kann bezüglich der Kandidaten nur bei näherer Betrachtung ein Unterschied ausgemacht werden. Wie bereits zu Beginn angemerkt, sind die PQ-Only Varianten in der Regel performanter als das ebenbürtige ECDH Verfahren mit *prime*256*v*1. Die hybriden Varianten als Kombination aus PQC und ECDH *prime*256*v*1 sind hingegen um wenige hundert Mikrosekunden verzögert.

Bei Alternativen, welche bereits als PQ-Only dem klassischen *prime*256*v*1 unterlegen waren, ist im Bezug auf den zeitlichen Unterschied zwischen PQ-Only und hybrid kein klarer Trend erkennbar. Die korrespondierenden Paare wie *SIKEp*503 und *p*256_*SIKEp*503 scheinen im Schnitt gleichauf zu liegen. Lediglich die Versionen von NTRU Prime, welche als PQ-Only Variante besser performen als *prime*256*v*1, werden bei hybrider Ausführung grundsätzlich verlangsamt. Es scheint, dass bei optimalen Netzwerkbedingungen keine zeitlichen Einschränkung bei Verwendung der hybriden Variante zu

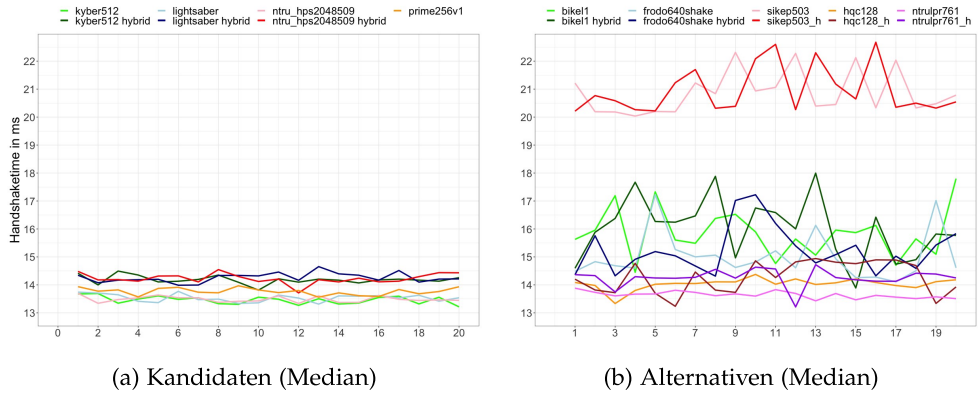

(a) Kandidaten (Median)  (b) Alternativen (Median)

Abbildung 5.3: Handshakedauer in Millisekunden auf der X-Achse in Abhängigkeit zur Nummerierung der durchgeführten Messung bei unveränderten Netzwerkparametern und Betrachtung des Median für hybride und PQ-Only Varianten bei Kandidaten und Alternativen auf Sicherheitslevel 1. Hybride Varianten werden innerhalb der Legende durch „hybrid" beziehungsweise ein „h" gekennzeichnet

erwarten sind, sofern der gewählte Algorithmus als PQ-Only langsamer performt als das kombinierte klassische Verfahren (ECDH). Als Handshakedauer würde in etwa das Maximum aus den Handshakezeiten der kombinierten Algorithmen resultieren. Für Sicherheitslevel 3 und 5 zeigt sich bei den Kandidaten tatsächlich ein ähnliches Bild, siehe Abbildung A.2c und A.2e. Bei Betrachtung von SIKE und FrodoKEM kann die Annahme jedoch nicht bestätigt werden. Die Algorithmen sind bei hybrider Ausführung in höheren Sicherheitsleveln teils um bis zu 10 Millisekunden verlangsamt, siehe Abbildung A.2d und A.2f. Es kann also einzig festgehalten werden, dass die Kandidaten sowie NTRU Prime bei hybrider Ausführung zwar eine reduzierte Performanz aufweisen, welche jedoch im Verhältnis zur gesamten Handshakedauer vernachlässigt werden kann. Dem Vergleich zum klassischen ECDH Verfahren können auch hybride Varianten der Kandidaten sowie von NTRU Prime durchaus standhalten, sofern eine hohe Verbindungsqualität zu erwarten ist.

### 5.1.2 Latenz

Bei Hinzufügen einer künstlichen Latenz pro versendetem Paket sollte auch eine Verzögerung des Handshakes zu erwarten sein. Wir betrachten, wie bereits im vorherigen Abschnitt, die Entwicklung der Handshakedauer bei sich verändernder Latenz zunächst theoretisch und unter der vereinfachten Annahme, dass keine weiteren äußeren Einflüsse oder Steuerungsmaßnahmen unterliegender Netzwerkschichten auf die Messungen einwirken. Wie bereits beschrieben, bleibt die Übertragungszeit $t_{S\_Ges}$ bei gleichbleibender Datenrate sowie gleichbleibender Anzahl $n$ zusendender Pakete ebenfalls konstant. Bei Ergänzung der benötigten Zeit $t_{Op}$ für die Datenverarbeitung bei Client und Server ergibt sich:

$$t_0 = t_{S\_Ges} + t_{Op} \qquad \text{mit } t_{S\_Ges}, t_{Op} \in \mathbb{R}_0^+$$

Wird nun eine weitere, variable Verzögerung $t_{Delay\_P}$ beim Versenden eines jeden Pakets hinzugefügt, ergibt sich bei $x$ Paketen eine zusätzliche Gesamtverzögerung $t_{Delay}(t_{Delay\_P}, x)$, sodass:

$$t_{Ges}(x) = t_{Delay}(t_{Delay\_P}, x) + t_0 \qquad \text{mit } x, t_0, t_{Delay\_P} \in \mathbb{R}_0^+$$

Da für das durchgeführte Experiment $x, t_{Delay\_P}, t_{Delay}(x) \in \mathbb{R}^+$, ist die Funktion streng monoton steigend.

Diese Annahme spiegelt sich zunächst in den durchgeführten Experimenten wieder, wie anhand von Abbildung 5.4 deutlich wird. Der Graph zeigt die benötigte Zeit in Millisekunden in Abhängigkeit zur konfigurierten Verzögerung bei Client und Server. Alle Kandidaten sowie das klassische ECDH Verfahren liegen im Vergleich nah beieinander und weisen eine streng monotone Steigung auf.

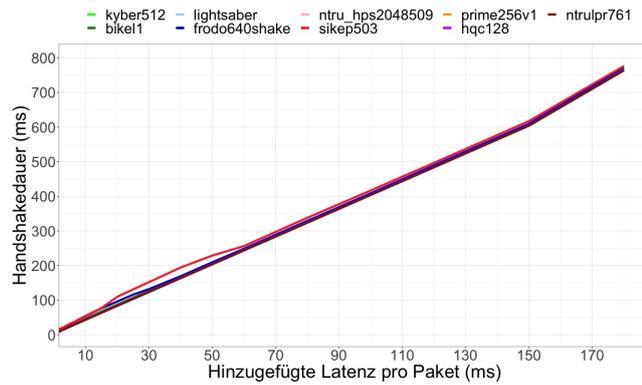

Abbildung 5.4: Median der Handshakedauer bei ansteigender Latenz, welche pro Paket hinzugefügt wurde. Werte in Millisekunden und Abbildung aller evaluierten Kandidaten und Alternativen für Sicherheitslevel 1

Nun könnte erwartet werden, dass Algorithmen mit größeren zusendenden Datenmengen im Vergleich eine verlangsamte Handshakedauer aufweisen, da NetEm jedes Paket um den konfigurierten Delay $t_{Delay\_P}$ verzögert. Bei $x$ Paketen würde sich somit in etwa ein Gesamtdelay $t_{Delay}$ aus dem Produkt ergeben:

$$t_{Delay}(n) = t_{Delay\_P} \times n \qquad \text{mit } t_{Delay\_P} \in \mathbb{R}^+, n \in \mathbb{N}$$

Aus dieser Überlegung würde auch resultieren, dass die Steigung des Gesamtdelays für Algorithmen mit großen Datenmengen vergrößert sein müsste. Wird jedoch Abbildung 5.4 betrachtet, so ist kein Unterschied erkennbar. Die Handshakedauer bei unterschiedlichem PQC KEM unterscheidet sich bei Sicherheitslevel 1, ähnlich dem konstanten Verlauf aus Abschnitt 5.1.1, hauptsächlich aufgrund der Effizienz der KEM Operationen. Dies macht die vergrößerte Aufnahme für 42.5 bis 47.5 Millisekunden in Abbildung 5.5 deutlich. Die Kandidaten sowie NTRU Prime, HQC und das klassische ECDH unterscheiden sich in solch geringem Maß, dass ihre Graphen sich gegenseitig verdecken. BIKE, FrodoKEM und SIKE liegen etwas darüber, die Entwicklung der Graphen in Abhängigkeit zur ergänzten Latenz verläuft jedoch trotz unterschiedlicher Schlüssel- und Chiffratgrößen annähernd äquivalent. Erst bei Sicherheitslevel 3 und 5, Abbildung A.3, ist eine vergrößerte Steigung der Handshakedauer bei FrodoKEM beziehungsweise bei FrodoKEM und HQC ersichtlich. Die Graphen von HQC und FrodoKEM setzten sich deutlich ab und der Abstand nimmt mit steigender Latenz zu.

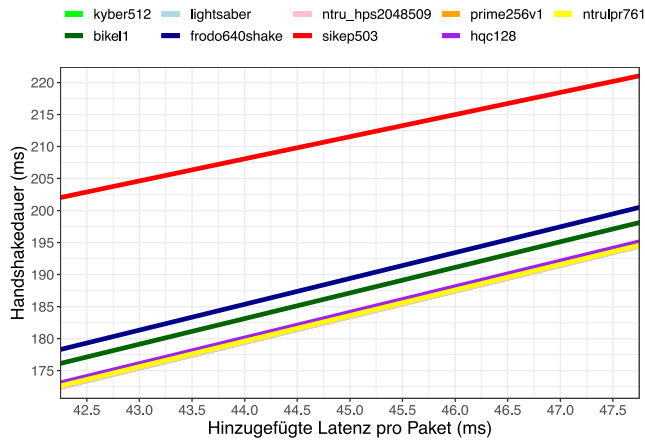

Abbildung 5.5: Median der Handshakedauer bei ansteigender Latenz im Bereich von 42.5 bis 47.5 ms. Abbildung aller evaluierten Kandidaten und Alternativen für Sicherheitslevel 1

Eine Erklärung ergibt sich bei Betrachtung der beteiligten Netzwerkkomponenten. Für eine effiziente, möglichst fehlerfreie Datenübermittlung werden verschiedenen Kontrollmechanismen auf den zwischenliegenden Netzwerkprotokollen genutzt. Eine Beschreibung der beteiligten Komponenten findet sich in Abschnitt 2.4.1. Das Protokoll IP ist zustandslos und nicht verbindungsorientiert. TCP hingegen regelt die Segmentierung, die Flusskontrolle sowie die Fehlerbehandlung und sorgt bei Verzögerung oder Verlust von Segmenten für entsprechende Maßnahmen. Mechanismen zur Flusskontrolle beziehungsweise -steuerung schätzen unter anderem mittels der Bestätigungsnachrichten (*ACK*) des Empfängers die maximalen Kapazitäten der Verbindung ab. Anschließend kann eine Bündelung einzelner Nachrichtenpakete oder eine Zeitanpassung bis zu einem erneuten Senden eines Segments bei Ausbleiben der Bestätigung erfolgen.

Insbesondere TCP hat somit einen wesentlichen Einfluss auf die Auswirkungen der ergänzten Latenz. Auf der einen Seite entstehen zusätzliche Verzögerungen, da eingehende Pakete vom Empfänger mittels eines weiteren *ACK* Pakets bestätigt werden müssen. Auf der anderen Seite werden Verzögerungen einzelner Pakete eingespart, da mehrere Pakete annähernd gleichzeitig gesendet werden können, bevor eine Bestätigung erforderlich ist. Die Anzahl gleichzeitig vermittelbarer Pakete wird durch die Größe des soge-nannten *Congestion Windows* (*cwnd*) festgelegt. Die Mehrheit der TCP Mechanismen sind adaptiv und werden entsprechend der verfügbaren Netzwerkqualität konfiguriert. Bei der im Experiment gewählten Übertragungsrate von 500 Mbps und einem Paketverlust von 0 Prozent können viele Pakete annähernd zum gleichen Zeitpunkt gesendet werden, sodass Algorithmen mit größeren Schlüsseln und Ciphertexten nicht zwangsläufig zusätzliche TCP RTTs benötigen. Der Unterschied wird folglich eliminiert. Ergo sollte sich erst dann eine variierende Steigung der Handshakezeiten zwischen den PQC Algorithmen ergeben, wenn die Anzahl zusätzlich zusendender Pakete die Anzahl gleichzeitig vermittelbarer Pakete übersteigt. Dies wird anhand von Sicherheitslevel 3 und 5 bestätigt, denn hier sind Schlüssel- und Ciphertextgrößen nochmals vergrößert und somit die Steigung verändert.

Die Analyse der hybriden Varianten zeigt ein ähnliches Bild, Vergleich Abbildung A.4. Wie in Abschnitt 5.1.1 geschildert, sind die Unterschiede der Ergebnisse hybrider Varianten bei Sicherheitsstufe 1 gering und die Differenz der zu übertragenden Datenmengen ist nicht signifikant erhöht, sodass alle Algorithmen in den Abbildung A.4a und A.4b die gleiche zeitliche Entwicklung besitzen. Erst mit steigendem Sicherheitslevel vergrößert sich die Differenz der zu übertragenden Daten bei FrodoKEM und HQC im Vergleich zu anderen KEMs in einem solchen Maß, dass innerhalb der Graphen ein abweichender Verlauf erkennbar wird.

Paquin, Stebila und Tamvada [196] betrachten unterschiedliche Latenzen im Bezug auf eine steigende Paketverlustrate und stellen fest, dass sich die Auswirkungen des Paketverlusts bei vergrößerter Latenz verstärken würden, da bei großen Datenmengen eine höhere Anzahl Paketverluste zu erwarten sei, welche zusätzliche Übertragungen erforderlich machten. Diese Annahme kann im Experiment für alle Algorithmen bestätigt werden, siehe auch Abschnitt 5.1.5.

Aus den Beobachtungen bei veränderten Latenz mittels NetEm folgt, dass ein datenarmer KEM Algorithmus bei großen zeitlichen Verzögerungen innerhalb des Netzwerks nicht zwangsläufig eine performantere Durchführung des Handshakes ermöglicht. Erst bei deutlich größeren Datenmengen oder weiteren Einschränkungen, beispielsweise durch eine geringe Übertragungsrate oder erhöhten Paketverlust, wie von Paquin, Stebila und Tamvada [196] beschrieben, sind geringe Schlüssel- und Ciphertextgrößen für eine signifikant verkürzte Handshakedauer vorzuziehen.

Im Hinblick auf die einzelnen Kandidaten und Alternativen fällt auf, dass bei hinreichend geringem Datenaufkommen, beispielsweise bei niedrigem Sicherheitslevel, alle Algorithmen und insbesondere auch die hybriden Varianten dem Vergleich mit der klassischen ECDH Variante standhalten können. Wie in Abschnitt 4.2.3 beschrieben, sind Verzögerungen von 5 bis etwa 50 Millisekunden aufgrund der Netzwerkcharakteristik bei typischen Internetverbindungen durchaus denkbar. Bei Sicherheitslevel 1 in diesem Bereich ergibt sich bezüglich der durchgeführten Experimente eine Gesamtverzögerung von etwa 25 bis 200 Millisekunden. Die Performanz des Handshakes bei Verwendung von HQC oder FrodoKEM ist jedoch, abhängig vom ange-strebten Sicherheitslevel und der vorherrschenden Netzwerkcharakteristik, eingeschränkt und eine generelle Verwendung dieser Algorithmen, unabhängig vom Anwendungskontext, kann nicht empfohlen werden. Bei FrodoKEM ergab sich unter Sicherheitslevel 3 beispielsweise bei 50 Millisekunden ergänzter Latenz eine Gesamtverzögerung im Median von etwa 400 Millisekunden. Im Vergleich zur klassischen Variante oder den NIST KEM Alternativen ist dies eine Verdopplung der benötigten Zeit. Eine solche Verzögerung ist bei Aufruf einer Internetseite beispielsweise durchaus wahrnehmbar und sollte vermieden werden.

### 5.1.3 Jitter

Durch die Angabe eines weiteren Werts bei Latenzkonfiguration innerhalb von NetEm kann ein zeitlicher Bereich für die Verzögerung definiert werden. Jedem Paket wird mittels Normalverteilung eine zufällige Verzögerung aus diesem Bereich zugewiesen. Somit kann eine typische Verbindung mit leicht variierenden Latenzen emuliert werden. Im durchgeführten Experiment wurde eine Latenz von 20 Millisekunden angenommen. Diese repräsentiert, entsprechend Abschnitt 4.2.3, einen bei HTTPS Verbindungen herkömmlichen Wert. Beispielsweise, wenn Server und Client weitestgehend über Kabel kommunizieren und einige 100 Kilometer voneinander entfernt sind [196]. Abbildung 5.6 zeigt den Median bezüglich der benötigten Zeit für einen Handshake bei steigender Abweichung der Latenz in Millisekunden, siehe auch Abschnitt 2.4.3.2. Ähnlich zu den Beobachtungen im vorherigen Abschnitt ist eine gleichmäßige Steigung erkennbar. Die Leicht variierenden Zeiten relativieren sich gegenseitig und beeinflussen die TCP Mechanismen nicht. Bezüglich der Algorithmen und ihren unterschiedlichen Konstellationen können die gleichen Beobachtungen gemacht werden, wie in Abschnitt 5.1.2, sodass diese an dieser Stelle nicht noch einmal ausgeführt werden. Die Grafiken sind unter A.5 und A.6 abgebildet.

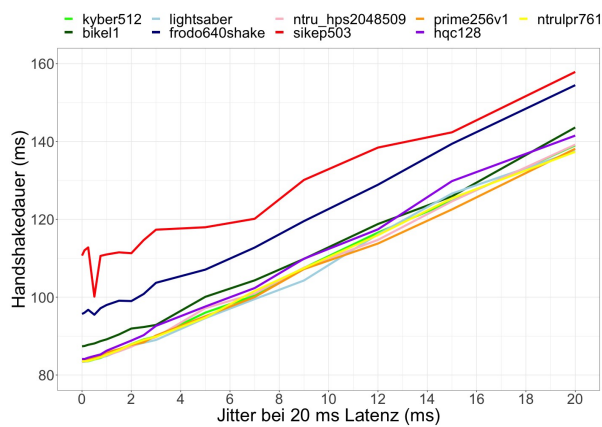

Abbildung 5.6: Median der Handshakedauer bei 20 ms Latenz und ansteigendem Jitter. Werte in Millisekunden und Abbildung aller Kandidaten und Alternativen für Sicherheitslevel 1

5.1.4 *Übertragungsrate*

Ohne störende äußere Einflüsse ergibt sich die für die Übertragung benötigte Zeit $t_S$ in Millisekunden aus der zusendenden Datenmenge $m$ in Bits pro verwendeter Datenrate $x$ in Kbps, sodass folgende gebrochen-rationale Funktion im Bezug auf die Messverläufe zu erwarten wäre:

$$t_S(x) = \frac{m}{x} \qquad \text{mit } x \in [0.1, 2000],\ m \in \mathbb{R}_0^+$$

Bei Betrachtung der Handshakedauer muss neben der Nachrichtenübertragung die benötigte Zeit $t_{Op}$ für die gesamte Datenverarbeitung bei Client und Server einbezogen werden, sodass

$$\begin{aligned} t_{Ges}(x) &= t_S(x) + t_{Op} \\ &= \frac{m}{x} + t_{Op} \qquad \text{mit } x \in [0.1, 2000],\ m \in \mathbb{R}_0^+ \end{aligned}$$

Wirft man einen Blick auf die Gesamtmenge aller getesteten Werte von 2000 bis 0.1 Mbps für den vom Client ausgehenden Nachrichtenverkehr, während die Rate mit 500 Mbps beim Server konstant bleibt, so bestätigt sich diese Annahmen. Es fällt auf, dass bei höheren x-Werten lange Zeit fast keine Änderung zu verzeichnen ist. Bei hinreichend geringer Übertragungsrate verzögert sich die Handshakedauer hingegen enorm, wie beispielsweise in Abbildung 5.7a für Sicherheitslevel 1 dargestellt.

Generell kann das unterschiedliche Verhalten anhand der entsprechend zu übertragenden Datenmengen begründet werden. Abbildung 2.2 veranschaulicht den Vergleich zwischen Schlüssel- und Chiffratgrößen und macht deutlich, dass viele der Algorithmen relativ nah beieinander liegen, während HQC leicht entfernt und FrodoKEM weit abgeschlagen ist. Um das Verhalten für Sicherheitslevel 1 genauer zu analysieren, müssen aufgrund der enormen Wertunterschiede gezielte Abschnitte des Graphen betrachtet werden:

- Abbildung 5.7c spiegelt die Entwicklung für sehr geringen Übertragungsraten zwischen 100 und 2000 Kbps wieder. Bei 100 Kbps sind große Differenzen zwischen den Algorithmen, abhängig von den zu übertragenden Datenmengen, erkennbar. Das klassische Verfahren mittels ECDH schneidet mit Abstand am besten ab, benötigt aber dennoch im Median über eine halbe Sekunde. SIKE profitiert von seiner geringen Schlüssel- und Chiffratgröße, während die darauf folgenden Kandidaten Kyber, Saber und NTRU dicht beieinander liegen. FrodoKEM sticht extrem heraus und benötigt im Median über 6 Sekunden. Mit steigender Rate wird der gravierende Abstand von FrodoKEM zwar geringer, kann sich jedoch nicht ausreichend annähern. Die anderen Verfahren gleichen sich ab etwa 0.5 Mbps jedoch stark an und bei 1.5 Mbps fällt der Unterschied zwischen klassischer Variante und den NIST Kandidaten auf unter 30 Millisekunden, was hinnehmbar erscheint [196]. Problematisch ist vorrangig der Bereich unter einem Mbps auf der Vermitt-lungsschicht beziehungsweise IP-Ebene, wenngleich dieser insbesondere im Bereich Mobilfunk sowie IoT durchaus realistisch ist, siehe auch Abschnitt 4.2.3. Möglicherweise könnte ein der Übertragungsrate angepasster Wechsel zwischen den bewährten NIST Kandidaten und SIKE eine Lösung dieses Problems im Bereich zwischen etwa 0 und 1 Mbps bieten. Zumal sich der Effekt bei steigendem Sicherheitslevel teils verstärkt, wie aus A.9 hervorgeht.

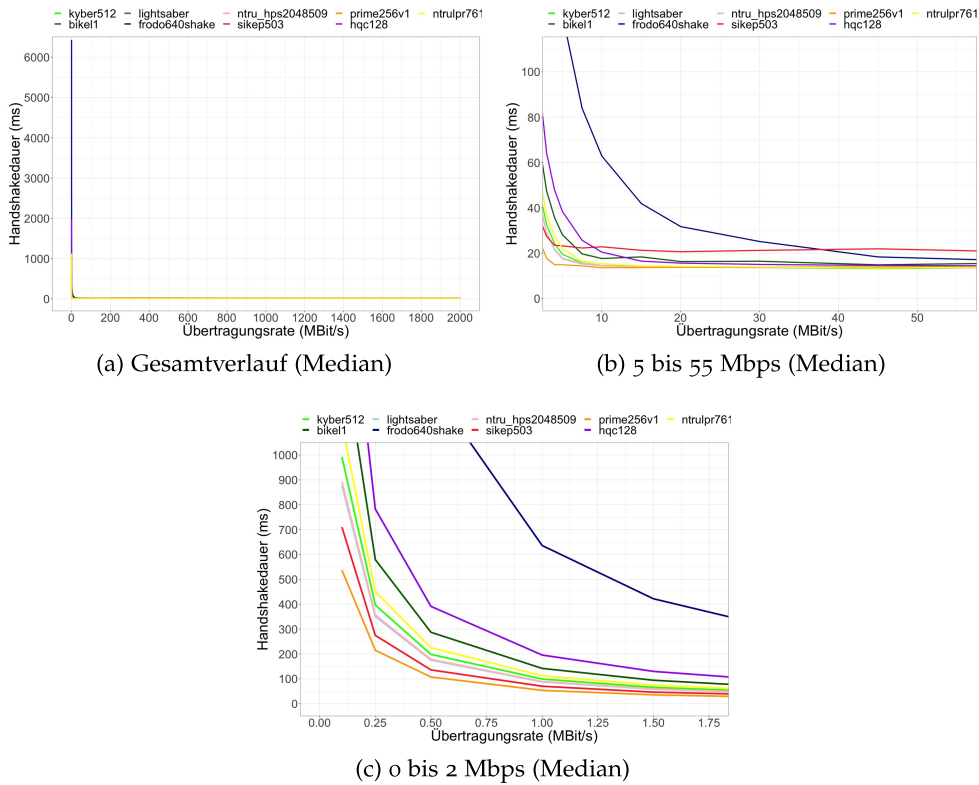

Abbildung 5.7: Handshakedauer in Millisekunden in Abhängigkeit zur clientseitig veränderten Übertragungsrate in Mbps bei Betrachtung des Median für alle evaluierten Kandidaten und Alternativen auf Sicherheitslevel 1.

- In Abbildung 5.7b ist der Verlauf zwischen 5 und 55 Mbps erkennbar. Im Bereich 5 und 10 Mbps nimmt das Gefälle aller Algorithmen, mit Ausnahme von FrodoKEM, stark ab. Es sind nur noch leichte Unterschiede von zunächst etwa 20 sowie später 10 Millisekunden bei der mittleren Handshakezeit zu beobachten und mit Ausnahme von FrodoKEM lässt sich das aus Abbildung 5.1 bekannte Verhalten bei hoher Netzwerkqualität erkennen. Der gezeigte Bereich umfasst die üblichen Werte bei HTTP- respektive TLS Verbindungen, siehe Abschnitt 4.2.3, und gibt daher am ehesten Aufschluss bezüglich der Eignung der Algorithmen bei herkömmlichen Übertragungsraten. Einhergehend mit den Empfehlungen der NIST können die PQC Kandidaten Kyber, Saber und NTRU mit der klassische Variante mithalten und weisen eine gute

Performanz auf. Des Weiteren befindet sich auch die Alternative NTRU Prime auf diesen Level und die Alternativen HQC und BIKE nähern sich den guten Werten mit steigender Übertragungsrate an.

Bei Betrachtung der sich verändernden Übertragungsrate für vom Server ausgehende Verbindungen, siehe Abbildungen A.7c, A.7d A.8b, fällt auf, dass der Graph im Gesamtbild eine ähnliche Form annimmt. Gerade die Werte im sehr kleinen Bereich zwischen etwa 0 und 2 Mbps sind jedoch bei allen Algorithmen erhöht. Die akzeptablen Raten verschieben sich entsprechend und HQC sowie FrodoKEM weisen bereits bei etwa zwei Mbps eine zu große Verzögerung auf. Der Median der gemessenen Zeiten bei 0.5 Mbps für BIKE, Kyber, SIKE oder die klassischen Vergleichsvariante ECDH steigt jeweils um etwa 30 bis 50 Prozent an. Insbesondere die Testdurchläufe mit HQC sind deutlich verlangsamt. Bei Vergleich von Client- und Serverrate zeigt sich eine Steigerung des gemessenen Median von fast 90 Prozent. Zudem gleicht sich die Handshakedauer von HQC erst zwischen 15 und 20 Mbps den anderen Algorithmen an. Bei FrodoKEM hingegen beträgt die Steigerung bei 0.5 Mbps lediglich rund 8 Prozent. Der Abstand zu den anderen Algorithmen ist deutlich geschmälert. Diese Entwicklungen lassen sich auf die unterschiedlichen Nachrichteninhalte bei KEMs zurückführen, siehe auch Abschnitt 2.1.2.2:

(A) Der Client sendet zunächst innerhalb des *key_share* den vorläufigen öffentlichen Schlüssel $pk_C$. Bei Reduktion der Rate bei vom Client ausgehenden Nachrichten ist somit die Größe des öffentlichen Schlüssels entscheidend.

(B) Der Server hingegen nutzt $pk_C$ zur Erzeugung eines Ciphertextes *c* sowie des Schlüssels *k* mittels *Encaps()*. Anschließend überträgt er *c* an den Client, sodass dieser anhand von *c* und seinem privaten Schlüssel $sk_C$ *k* bestimmen kann. Bei Reduktion der Serverrate ist somit die Größe der Ciphertexte von Interesse.

Abbildung 2.2 veranschaulicht, dass bei den meisten Algorithmen ein relativ ausgeglichenes Verhältnis zwischen Schlüssel und Chiffrat gegeben ist. Die Chiffratgröße ist nur leicht erhöht. Bei HQC überwiegt jedoch die Chiffratgröße, wodurch die deutliche Verzögerung bei Reduktion der Übertragungsrate beim Server verursacht wird.

Da die Chiffratgrößen bei PQC Kandidaten und Alternativen im Gegensatz zur klassischen Variante mit elliptischen Kurven leicht erhöht sind, folgt daraus, dass die vom Server ausgehende Übertragungsrate beziehungsweise der „Download Link" des Client eine leicht übergeordnete Rolle im Bezug auf die Performanz spielt. Wenngleich der Client „Upload Link" nicht vernachlässigt werden sollte. FrodoKEM ist weiterhin nicht für eine Verwendung bei sehr geringen Übertragungsraten von 100 bis 2000 Kbps geeignet und setzt sich schon bei etwa 25 bis 30 Mbps von den anderen ab. Auch HQC sollte bei reduzierter Rate unter 15 bis 20 Mbps bei serverseitigen Nachrichten nicht genutzt werden. Gerade im Hinblick darauf, dass 20 bis 30

Mbps noch erwartbare Werte innerhalb der herkömmlichen Netze sind [265], ist diese Bilanz für HQC und FrodoKEM ernüchternd. Zumal sich die Effekte bei äquivalenter Veränderung von Server- und Clientrate, abgebildet in A.7f und A.8b, nochmals verstärken, da sich eine Art Kombination der oben beschriebenen Verhaltensmuster ergibt. FrodoKEM unterscheidet sich bereits bei Werten über 35 bis 40 Mbps und auch HQC gleicht sich erst zwischen 25 und 30 an. Obwohl für alle Algorithmen bei 500 Kbps sehr langsame Handshakes gemessen wurden, ist doch bemerkenswert, dass die Differenz zwischen den Kandidaten und dem klassischen Verfahren trotz enorm vergrößerter Datenmenge nur 150 bis 250 Millisekunden beträgt. Ab einer Rate unter 3 bis 4 Mbps scheint zudem SIKE eine sehr geeignete Alternative. Bei beidseitiger Reduktion war jedoch bei 100 Kbps für keinen der verfügbaren Algorithmen eine erfolgreiche Durchführung möglich, da die konfigurierten Zeitschranken von TLS sowie TCP überschritten wurden.

Bei Steigerung des Sicherheitslevel ändert sich das Verhältnis zwischen Schlüssel- und Chiffratgröße bei allen Algorithmen im Wesentlichen nicht. Für das Verhältnis von Client- und Serverrate können daher die gleichen Aussagen getroffen werden. Bei beispielhafter Analyse der beidseitig veränderten Übertragungsraten bei Sicherheitslevel 3, Abbildung 5.8, sind, wie zu erwarten, alle Handshakezeiten verlangsamt. Bei Übertragungsraten unter 50 bis 60 Mbps setzt sich FrodoKEM ab und unter 35 bis 45 entfernt sich auch HQC deutlich von den Kandidaten. Es sollte beachtet werden, dass auch BIKE im Bereich unter etwa 10 Mbps dem Vergleich mit den performanteren Algorithmen nicht mehr standhalten kann. Die Kandidaten hingegen ähneln sich so sehr, dass *kyber*768 im Graphen durch *saber* verdeckt wird.

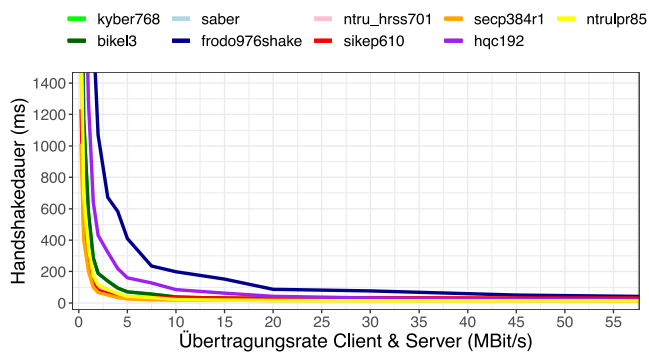

Abbildung 5.8: Median der Handshakedauer bei beidseitig veränderter Übertragungsrate und unter Berücksichtigung aller evaluierten Kandidaten und Alternativen für Sicherheitslevel 3

Die hybriden Verfahren der Kandidaten und Alternativen, Abbildung A.9a, A.9b, unterscheiden sich im Wesentlichen kaum von ihrer jeweils korrespondierenden PQ-Only Variante. Erst bei näherer Betrachtung im Bereich zwischen 0 und 2 Mbps, Abbildung 5.9, werden Unterschiede deutlich. Die Differenz vergrößert sich mit sinkender Übertragungsrate. Im Verhältnis zur benötigten Gesamtzeit bezüglich des Handshakes ist diese dennoch vernachlässigbar.

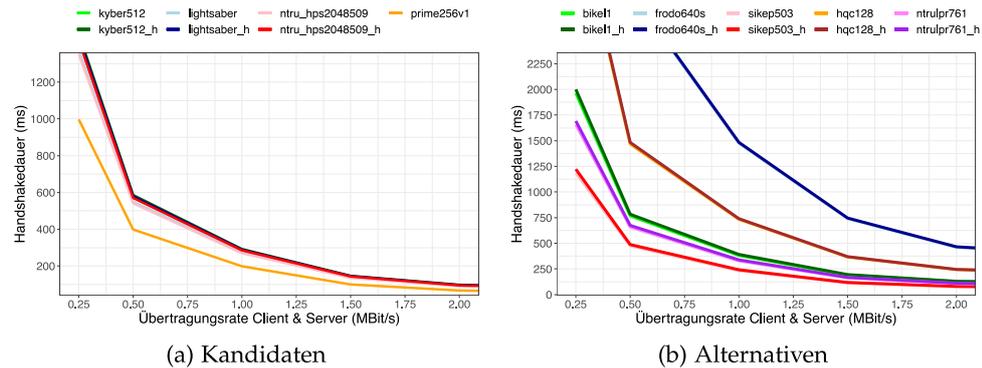

(a) Kandidaten  (b) Alternativen

Abbildung 5.9: Median der Handshakedauer bei beidseitig veränderter Übertragungsrate von 0 - 2 Mbps unter Berücksichtigung aller evaluierten Kandidaten und Alternativen als hybride und PQ-Only Version für Sicherheitslevel 1

Bei Sicherheitslevel 3 und 5 vergrößert sich bei einigen Algorithmen die Differenz im Vergleich zur nächst geringeren Sicherheitsstufe. Ein allgemeingültiges Verhaltensmuster ist aber nicht ableitbar. Abbildung 5.10a zeigt beispielsweise die Handshakezeiten für alle drei Kyber Varianten bei hybrider sowie nicht-hybrider Vorgehensweise im Bereich von 0 bis 2 Mbps. Dabei weißt Sicherheitslevel 3 die größte Differenz auf. Dies ist auch bei arithmetischem Mittel und 0.95-Quantil zu erkennen, siehe 5.10b, 5.10c.

Generell sind die Unterschiede zwischen Median, 0.75- und 0.95-Quantil im Vergleich zu Paketverlust oder korrupten Paketen gering. Gleiches gilt für die Standardabweichung und den Vergleich zwischen arithmetischem Mittel und Median. Es scheint, als würden keine beziehungsweise nur wenige äußere Einflüsse auf die Messungen einwirken. Die Veränderung der Zeiten wird weitestgehend durch die Charakteristik der Algorithmen bestimmt.

### 5.1.5 Paketverlust

Wie bereits in Abschnitt 5.1.2 angeführt, ist für die Kontrolle und Steuerung des Datenverkehrs insbesondere das Protokoll TCP zuständig. Gehen Pakete verloren, so werden diese durch TCP neu gesendet, sobald eine bestimmte Zeit ohne Empfangsbestätigung beim Sender verstrichen ist.

Der Paketverlust wurde für bestimmte Algorithmen auch in der Arbeit von Paquin, Stebila und Tamvada [196] untersucht. Wie die Autoren bereits feststellen konnten, unterscheidet sich der Median der gemessenen Zeiten aller Algorithmen bei sehr geringem Paketverlust unter etwa 3 bis 4 Prozent sowie geringer Latenz einzig anhand der Geschwindigkeit des jeweiligen Algorithmus. Lediglich die Handshakedauer von FrodoKEM verzögert sich aufgrund der großen zusendenden Datenmengen bereits zu einem frühen Zeitpunkt. Dies konnte in den durchgeführten Experimenten für alle Kandidaten und Alternativen bestätigt werden, siehe Abbildung 5.11a. Es zeigt sich, dass der Graph von FrodoKEM bereits ab etwa 1.5 Prozent ansteigt

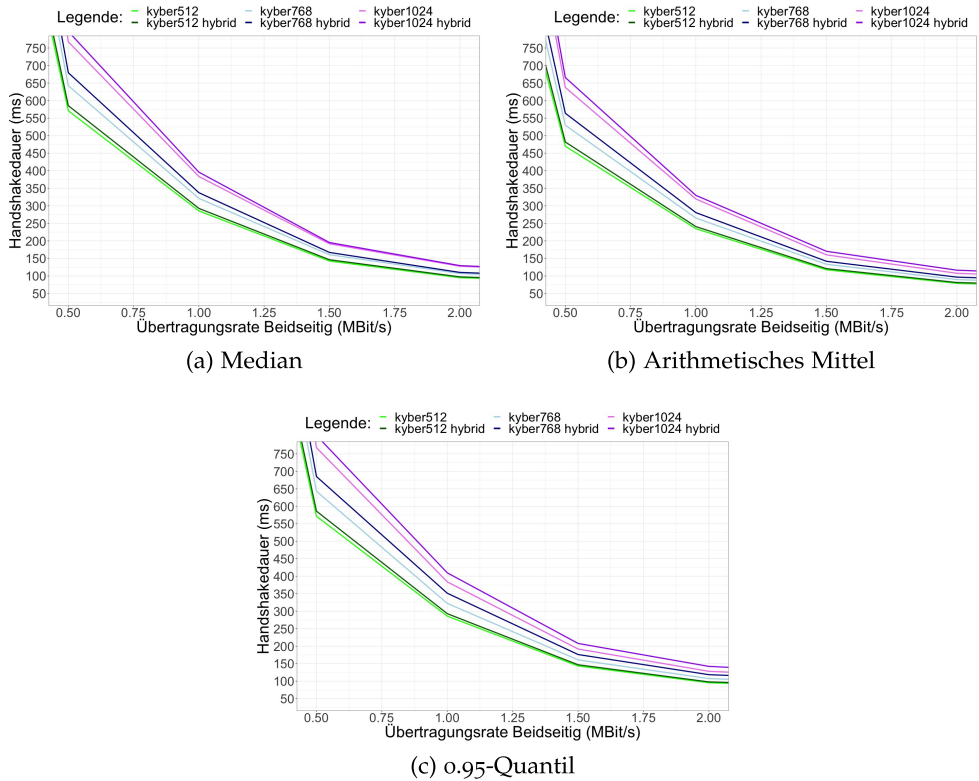

**Abbildung 5.10:** Handshakedauer in Millisekunden auf der X-Achse in Abhängigkeit zur veränderten Übertragungsrate bei Client und Server in Mbps für Kyber und Berücksichtigung aller Sicherheitslevel sowie der hybriden Varianten

und SIKE überholt, welcher eine generelle Verzögerung aufgrund der geringen Effizienz in den Algorithmenoperationen aufweist. Abbildung 5.11b hebt diesen Vorgang noch einmal hervor und zeigt den vergrößerten Bereich zwischen 0 und 5 Prozent.

Im weiteren Verlauf sind die Werte zunächst weitestgehend konstant. Es kann bestätigt werden, dass erst ab etwa 8 bis 10 Prozent ein wirklicher Anstieg erkennbar ist und FrodoKEM bei 12 bis 15 Prozent, je nach Sicherheitslevel, einen großen Sprung macht. In den durchgeführten Experimenten kann ergänzend beobachtet werden, dass auch HQC, welcher ebenfalls eine größere Datenmenge aufweist, schneller ansteigt. Insgesamt bleibt der große Abstand von FrodoKEM jedoch bestehen. Die von der NIST zugesprochene hohe Gesamtperformanz der Kandidaten Kyber, Saber und NTRU, welche für *kyber*512 bereits durch Paquin, Stebila und Tamvada [196] überprüft wurde, kann hier für alle Sicherheitslevel bestätigt werden. Die Algorithmen weisen in der Regel erst ab etwa 14 bis 16 Prozent einen signifikanten Anstieg der Handshakezeit auf. Im für die Praxis besonders relevanten Bereich zwischen 0 und 10 ist aber mit Bezug auf den Median ein sehr gutes Ergebnis zu verzeichnen.

Der Bereich wurde auch für die schlechtesten 25 beziehungsweise 5 Prozent betrachtet, siehe A.10. Hier ist bereits bei 4 beziehungsweise 2 Prozent

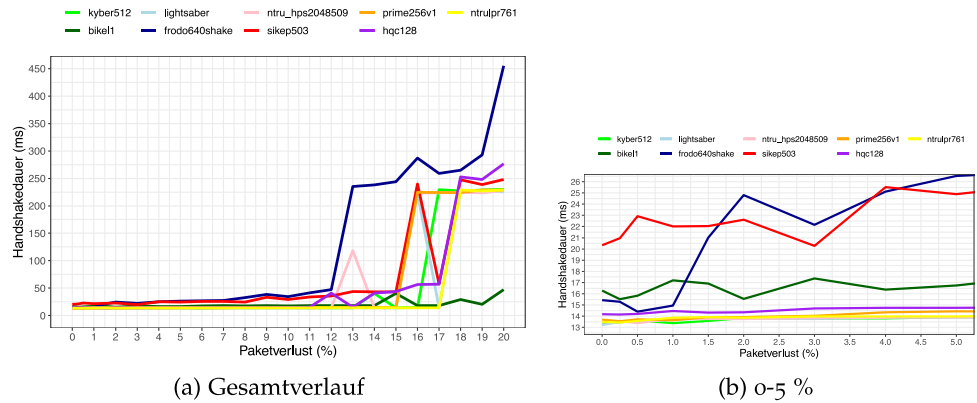

(a) Gesamtverlauf  (b) 0-5 %

**Abbildung 5.11:** Median der Handshakedauer in Abhängigkeit zur Paketverlustrate im Gesamtverlauf sowie Intervall von 0-5 % unter Berücksichtigung aller evaluierten Kandidaten und Alternativen für Sicherheitslevel 1

ein starker Anstieg der Zeiten zu verzeichnen und es fällt überraschenderweise auf, dass die Unterschiede zwischen allen Algorithmen vergleichsweise gering sind. Ein verringerter Unterschied innerhalb des 0.95-Quantils wurde jedoch auch durch die Vorgängerarbeit beschrieben.

Hybride Verfahren für Sicherheitslevel 1 zeigen im Vergleich zu PQ-Only Varianten keine signifikanten Unterschiede. Bei 0 bis 12 Prozent Paketverlust ist eine vergleichbare Geschwindigkeit aller Verfahren zu erwarten. Insbesondere im Bezug auf die Kandidaten kann bei Betrachtung der Handshakedauer bedenkenlos eine konservative, hybride Vorgehensweise gewählt werden. Bei Sicherheitslevel 3 ist hingegen ein deutlicher Unterschied erkennbar. Bereits bei keinem oder sehr geringem Verlust unterscheiden sich die Zeiten. Abbildung 5.12 zeigt den Bereich von 0 bis 4 Prozent. Der anfängliche Unterschied begründet sich durch die zusätzliche Verwendung des leicht ineffizienteren ECDH. Mit ansteigender Paketverlustrate wird der Anstieg der gemessenen Zeiten bei hybriden Verfahren schneller, während die anderen Verfahren gleichbleibend performen, siehe Abbildung A.11c. Zwischen 14 und 16 Prozent kommt es zu einer starken Verschlechterung, wobei nun auch PQ-Only Varianten betroffen sind. Abbildung A.11e macht deutlich, dass sich das beschriebene Verhalten für Sicherheitslevel 5 nochmals verstärkt.

Bei Betrachtung der Graphen bezüglich des Paketverlusts fällt auf, dass sich „Spitzen" bilden. Die Handshakedauer steigt sprunghaft an und fällt anschließend wieder ab. Beispielhaft sind unter anderem die Werteverläufe von *ntru_hps*2048509 sowie *SIKEp*503 in Abbildung 5.11a. Das Phänomen ist nicht gleichmäßig reproduzierbar und wird im folgenden Verlauf als „Ausreißer" tituliert. Eine detaillierte Betrachtung erfolgt in Abschnitt 5.3.

### 5.1.6  Korrupte Pakete

Vergleicht man die Graphen aus Abbildung 5.13, so fällt auf, dass zwischen den Verläufen bei verlorenen und fehlerhaften Paketen ein enger Zusam-menhang besteht. Das Verhalten der Algorithmen mit steigender Rate und die angenommenen Werte für die Dauer des Handshakes sind fast identisch, wenngleich die Graphen für korrupte Pakete um etwa 2 Prozent nach rechts entlang der x-Achse verschoben zu sein scheinen.

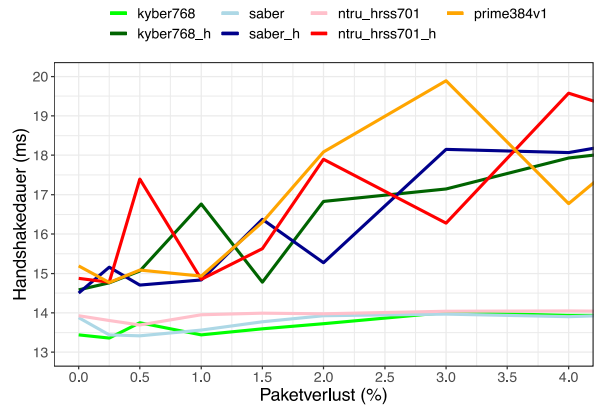

Abbildung 5.12: Handshakedauer in Millisekunden bei steigender Paketverlustrate von 0-4 % für hybride und PQ-Only Varianten bei Betrachtung des Median. Hybride Varianten werden innerhalb der Legende durch ein „h" gekennzeichnet. *Shake* wurde durch „s" ersetzt (z.B. $frodo640s$)

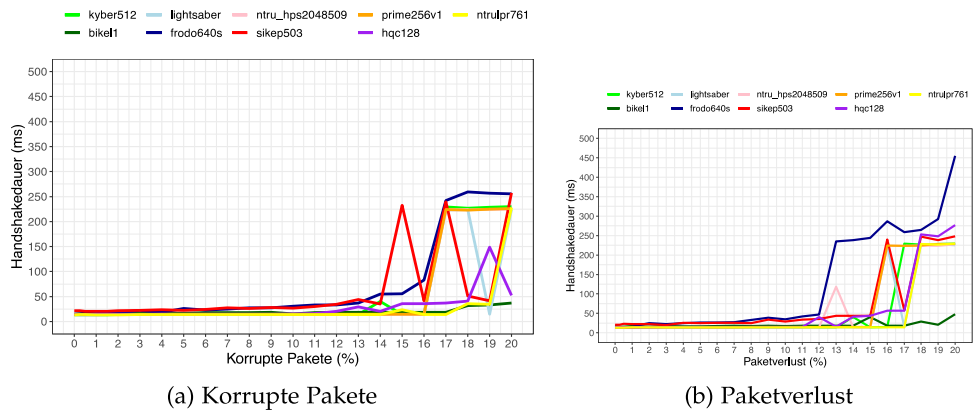

(a) Korrupte Pakete    (b) Paketverlust

Abbildung 5.13: Vergleich der Handshakedauer (Median) in Millisekunden bei steigender Anzahl korrupter und verlorener Pakete in Prozent. Berücksichtigt wurden alle evaluierten Kandidaten und Alternativen für Sicherheitslevel 1

Ähnliche Beobachtungen lassen sich auch für Sicherheitslevel 3 und 5 machen, siehe A.12. Um diesen Aspekt nachzuvollziehen, muss der Umgang mit fehlerhaften Paketen innerhalb des Netzwerks nachvollzogen werden:

- Da Schichten unterhalb von IP durch NetEm emuliert werden, können diese vernachlässigt werden.

- Tritt ein Fehler innerhalb des IP-Header auf, verwirft IP das Paket. Da das Protokoll zustandslos ist, ist der Vorgang auf dieser Ebene somit bereits abgeschlossen und für das darüber liegende Protokoll TCP ergibt sich eine äquivalente Situation zu einem verlorenen Paket. Tritt der Fehler hingegen an einer anderen Stelle auf und wird nicht durch IP bemerkt, wird das Segment an TCP übergeben.

- TCP überprüft die Integrität des Segments anhand einer Checksumme und verwirft das Paket, sofern Fehler angezeigt werden. Auch hier ist die Situation äquivalent zu einem verlorenen Paket.

- Da die TCP Checksumme vergleichsweise schwach ist [241], kann in seltenen Fällen ein korruptes Paket bei TLS eingehen. Befindet sich der Fehler innerhalb des TCP Headers, könnten die Daten im Optimalfall korrekt an TLS weitergereicht werden. andernfalls kommt es zu einem undefinierten Fehler. Bei einem Fehler innerhalb der TLS Daten bricht der Handshake, wie in RFC 8446 [213] beschrieben, mit einer Fehlermeldung ab.

In den meisten Fällen kommt es daher zu einer „Art von Paketverlust", welche den ähnlichen Verlauf der Graphen erklärt. Da nicht bei allen durch NetEm injizierten Bitfehlern ein Paket verworfen werden muss, ist der Anteil der tatsächlich wiederholt gesendeten Pakete jedoch geringer als bei direkter Angabe des Paketverlusts. Dies könnte die wahrgenommene Verschiebung des Graphen verursachen.

Die Beobachtungen aus Abschnitt 5.1.5 können daher unter Berücksichtigung der beschriebenen Unterschiede auch für die Veränderung für die Rate korrupter Pakete beobachtet werden. Insbesondere sind ebenfalls die beschriebenen Ausreißer zu beobachten.

## 5.2 PQC ALGORITHMEN

### 5.2.1 NIST Kandidaten

Die untersuchten Kandidaten Kyber, Saber und NTRU nutzen gitterbasierte Ansätze und weisen ähnliche Schlüssel- und Chiffratgrößen sowie Geschwindigkeiten bei Durchführung der Operationen auf. Dementsprechend fällt bei Berücksichtigung der untersuchten Netzwerkparameter auf, dass die Ergebnisse der Kandidaten grundsätzlich sehr nah beieinander liegen.

Bei gleichbleibend guter Netzwerkqualität betragen die Differenzen kaum mehr als eine Millisekunde und auch im Vergleich zur klassischen Variante mit ECDH ist kein signifikanter Unterschied erkennbar. Alle drei untersuchten KEM Kandidaten agieren sehr performant und können teils ECDH unterbieten: Sie benötigen im Durchschnitt weniger als 13.5 Millisekunden für die Durchführung des Handshakes bei Sicherheitsstufe 1 (*kyber*512 13.4 ms, *lightsaber* 13.2 ms, *ntru_hps*2048509 13.4 ms), während der Durchschnitt für die klassische Variante mit *prime*256v1 13.7 Millisekunden beträgt. Weiter fällt auf, dass die Zeiten der jeweiligen Algorithmenversionen entsprechend dem Sicherheitslevel kaum voneinander abweichen. Gleiches gilt für die hybriden Varianten.

Bei veränderter Latenz pro Paket sowie bei zusätzlichem Jitter sind kaum Unterschiede ersichtlich, wie die Abbildungen 5.5 sowie 5.6 deutlich machen. Teilweise kann der Verlauf der Graphen nicht mehr unterschieden werden, da sich diese gegenseitig verdecken. Bei verminderter Übertragungsrate können ähnliche Beobachtungen gemacht werden. Erst bei sehr geringen Raten kommt es zu Verzögerungen. Ab etwa 1.5 Mbps fällt der durchschnittliche Unterschied auf unter 30 Millisekunden. Lediglich im Bereich unterhalb von 1000 Kbps ist ein deutlicher Abfall der Zeiten im Vergleich zu ECDH ersichtlich und Kyber fällt leicht hinter seinen Konkurrenten Saber und NTRU zurück. Im Zweifel könnte im speziellen Fall ein Wechsel zu SIKE sinnvoll sein.

Bezüglich der Rate von Paketverlusten und fehlerhaften Paketen ist im herkömmlichen Bereich zwischen 0 und 10 Prozent bei niedrigem Sicherheitslevel und hoher Übertragungsrate weder ein Unterschied zwischen den Kandidaten noch zwischen den hybriden Varianten erkennbar. Sie halten dem Vergleich mit der klassischen Variante stand. Dies wird beispielsweise anhand der Grafik in 5.12 deutlich. Ein gravierender Anstieg der Handshakedauer bei sehr hohen Verlust- beziehungsweise Fehlerraten ist bei Kandidaten sowie ECDH gleichermaßen zu beobachten.

Laut NIST böten Kyber, Saber und NTRU eine sehr gute Balance zwischen Datenmengen und Effizienz der Operationen und die Gesamtperformanz sei exzellent. Dies kann anhand der Ergebnisse bestätigt werden.

### 5.2.2 NIST Alternativen

Die Gruppe der Alternativen weist eine größere Varianz bezüglich der theoretischen Ansätze und Implementierungen der Algorithmen auf. Schlüssel- und Chiffratgrößen sowie die Effizienz der Algorithmenoperationen unterscheiden sich teils deutlich. Dies macht sich auch in der Verteilung der gemessenen Handshakezeiten innerhalb der Ergebnisse deutlich.

**NTRU Prime** ist ebenso wie die untersuchten Kandidaten ein gitterbasiertes Verfahren und ähnelt in vielen Punkten dem Konkurrenten NTRU. Entsprechend unterscheiden sich die Ergebnisse zu den Kandidaten nur in wenigen Punkten. Unter anderem ist zwischen den durchschnittlichen Zeiten von NTRU Prime bei Variation der Sicherheitslevel kaum ein Unterschied erkennbar, wie Abbildung 5.2a zeigt. Gleiches gilt für eine hybride Durchführung sowie eine veränderte Latenz, Jitter oder Verlustrate. Auch bei veränderter Übertragungsrate größer als 2 Mbps besteht kein signifikanter Unterschied zu den Kandidaten oder der klassischen Variante, siehe Abbildung A.8. Lediglich unterhalb dieser Grenze vergrößert sich die Differenz. Einhergehend mit den Empfehlungen der NIST kann NTRU Prime ebenso wie die evaluierten Kandidaten Kyber, Saber und NTRU mit der klassische Variante mithalten und weist eine gute Performanz auf.

**Bike** und **HQC** sind beide codebasierte. Dennoch unterscheiden sich Datenmengen und Algorithmeneffizienz.

HQC Operationen sind deutlich effizienter durchführbar. Abbildung 5.1 zeigt, dass sich die Handshakedauer zwischen HQC und den zu evaluierten Kandidaten bei hinreichend hoher Verbindungsqualität kaum unterscheidet. Für BIKE ist hingegen ein Unterschied sichtbar. Auch bei Latenz und Jitter hat dies Auswirkungen auf die Ergebnisse wie die Abbildungen 5.5 und 5.6 zeigen.

Auf der anderen Seite fällt die Größe für öffentliche Schlüssel und Chiffrate bei BIKE generell deutlich geringer aus als bei HQC und kann teils mit den evaluierten Kandidaten mithalten. Die Abbildungen A.8 und A.9 zeigen beispielhaft, wie sich dies auf die Handshakedauer auswirkt. Mit sinkender Übertragungsrate ist aufgrund der geringeren zu übertragenden Datenmenge ein deutlicher Unterschied bei der Handshakedauer erkennbar und BIKE ist deutlich performanter als HQC. Im Vergleich zwischen Experimenten mit veränderter Clientrate und Experimenten mit veränderter Serverrate fällt auf, dass sich die gemessenen Zeiten bei HQC von Client- zu Serverrate um fast 90 Prozent vergrößern. Erst zwischen 15 und 20 Mbps Serverrate gleicht sich HQC den anderen Algorithmen an. Dies ist dem ungleichen Verhältnis zwischen den Größen von öffentlichen Schlüsseln und Chiffraten geschuldet, siehe Abbildung 2.2. Unterhalb einer Übertragungsrate von 15 bis 20 Mbps bei serverseitigen Nachrichten sollte HQC daher nicht genutzt werden. Bei Paketverlusten im typischen Bereich zwischen 0 und 10 bis maximal 12 Prozent ist hingegen eine vergleichbare Geschwindigkeit aller Verfahren zu erwarten und auch die Verwendung von HQC sowie hybriden Varianten bedenkenlos möglich. Erst bei extrem hohen Raten ist ein signifikanter Unterschied messbar. Gleiches gilt für fehlerhafte Pakete.

Zusammenfassend ist eine Verwendung von HQC und BIKE in vielen Fällen ohne größere Performanzeinschränkungen möglich. Abhängig vom angestrebten Sicherheitslevel und der vorherrschenden Netzwerkcharakteristik kann es jedoch insbesondere bei HQC zu Verzögerungen kommen, sodass eine generelle Verwendung, unabhängig vom Anwendungskontext, nicht empfohlen werden kann. Bei BIKE sollte die noch fehlende Variante für Sicherheitslevel 5 beachtet werden.

**SIKE** ist das einzige isogeniebasierte Verfahren und weist die Besonderheit auf, dass Schlüssel und Ciphertexte extrem kurz, die Operationen hingegen sehr ineffizient sind. Diese Eigenschaft spiegelt sich durchweg in den Messreihen wieder, siehe beispielsweise Abbildung 5.11. Bei Verbindungen mit hoher Qualität und wenigen Störungen ist die Dauer des Handshakes durchweg höher als bei performanteren Algorithmen wie den gitterbasierten Verfahren Saber und Kyber. Der Nachteil gleicht sich aus, sobald größere Einschränkungen auftreten und die Menge der gesendeten Daten an Bedeutung gewinnt. Bei reduzierter Übertragungsrate und Sicherheitslevel 1 empfiehlt sich, je nach betroffener Ausgangsrichtung der Daten, eine Nutzung von SIKE bei Raten unter 1.5 bis 2.5 Mbps. Bei Sicherheitslevel 5 ist unter 5 Mbps ein Vorteil erkennbar, siehe 5.14. Solch schlechte Übertragungsraten sind bei Funkübertragung sowie im Bereich IoT durchaus denkbar, siehe Abschnitt 4.2.3, sodass eine zusätzliche Standardisierung von SIKE durch die NIST möglicherweise lohnenswert wäre. Bei den beidseitig reduzierten Messreihen war jedoch bei 100 Kbps für keinen der verfügbaren Algorithmen eine erfolgreiche Durchführung möglich, da die konfigurierten Zeitschranken von TLS sowie TCP überschritten wurden. Hier würde auch ein Wechsel zu SIKE keine Besserung bringen.

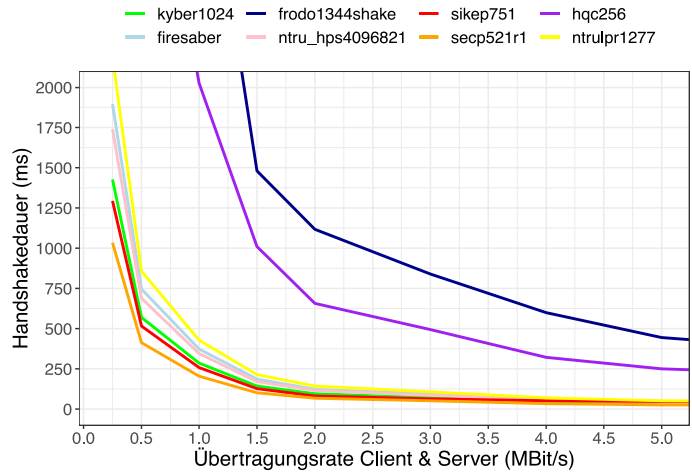

Abbildung 5.14: Handshakedauer (Median) in Millisekunden bei beidseitig veränderter Übertragungsrate zwischen 0 und 5 Mbps für alle evaluierten Kandidaten und Alternativen bezüglich Sicherheitslevel 5

Bei Paketverlust und fehlerhaften Paketen ist der Vorteil von SIKE gegenüber anderen PQC Verfahren weniger groß. Auch wenn Abbildung 5.11 zeigt, dass ab einer Paketverlustrate von über 2 Prozent, die benötigte Zeit des datenintensiven FrodoKEM über die von SIKE steigt (etwa 30- und 90-mal größere Schlüssel respektive Chiffratgröße als SIKE bei Sicherheitsstufe 1), so ist bei weiteren Algorithmen die Angleichung erst bei 12 bis 16 Prozent zu beobachten. [47, 65, 89, 196, 258] führen an, dass bei gängigen Client-Server-Verbindungen in der Regel Paketverlustraten von 0 bis maximal 5 Prozent zu erwarten seien. Laut [76, 256] sei eine Verlustrate von 10 Prozent sehr unwahrscheinlich. Unter herkömmlichen Bedingungen ist daher die Nutzung eines performanteren Algorithmus zu empfehlen.

Bezüglich der Sicherheitslevel sowie der hybriden Durchführung fällt auf, dass sich die Varianten von SIKE im Vergleich zu anderen KEM Algorithmen stark unterscheiden, siehe beispielsweise A.2. Bei hybrider Durchführung verschlechterte sich SIKE im Vergleich zu anderen Kandidaten und Alternativen.

**FrodoKEM** weist bei sehr guten Verbindungen einen erkennbaren, aber akzeptablen zeitlichen Unterschied zu anderen Kandidaten respektive Alternativen auf. Aufgrund der enormen Datenmengen ist jedoch beispielsweise bei etwa 2 Prozent Paketverlust oder 4 Prozent fehlerhaften Paketen ein Anstieg der benötigten Zeit zu erkennen. Dieser nimmt mit steigender Rate weiter zu. Abbildung 5.11 verdeutlicht den frühen Anstieg und das Ausmaß, gerade im Bereich der noch herkömmlichen Paketverlustraten von 1 bis 2 Prozent, siehe Abschnitt 4.2.3. Auch bei der Übertragungsrate spiegelt sich die Problematik in aller Deutlichkeit wieder. Bei typischen Werten von 30 bis 50 Mbps ist die Übertragungszeit drastisch erhöht. Bei sehr geringen Datenraten unter 5 Mbps scheint eine Nutzung undenkbar, bei 0.1 Mbps auf Clientseite ergibt der Median einen Wert von über 4.5 Sekunden.

FrodoKEM erweist sich daher schon bei geringen Qualitätseinbußen innerhalb des Netzes als unvorteilhaft. Zudem ist das Verhalten auch bei höheren Sicherheitsleveln zu beobachten. Oftmals vergrößern sich die Differenzen zu den Konkurrenten, siehe beispielsweise A.6d oder A.2d.

## 5.3  ausreisser

Insbesondere bei Veränderung des Anteils verlorener oder fehlerhafter Pakete kam es bei Betrachtung des Median ab höheren Prozentwerten zwischen 12 und 14 Prozent unabhängig vom gewählten Algorithmus und Sicherheitslevel zu Ausreißern. Beispielhaft sind unter anderem die Werteverläufe von *ntru_hps*2048509 sowie *sikep*503 in Abbildung 5.11. Die gemessenen Zeiten sind an bestimmten Stellen deutlich vergrößert, während darauf folgende Werte trotz steigendem Prozentsatz und somit trotz verminderter Netzwerkqualität deutlich abfallen. Zudem sind die Werte für arithmetisches Mittel und Standardabweichung im Vergleich zu anderen Netzwerkszenarien stark erhöht. Für das Verhalten können mehrere Gründe in Betracht gezogen werden:

- **Zufallskomponenten**: Die bei NetEm konfigurierte Wahrscheinlichkeit für das Auftreten des jeweiligen Phänomens, wie Paketverlust oder Reordering, wird auf jedes eingehende Paket einzeln und unabhängig voneinander angewendet. Durch die Zufallskomponente kann es zu Anomalien kommen. Die Annahme, dass die zufällige Verteilung der NetEm Parameter an dem Phänomen beteiligt ist, kann aufgrund der Tatsache bekräftigt werden, dass ausschließlich Experimente betroffen sind, welche über Wahrscheinlichkeiten definiert werden (Paketverlust, korrupte Pakete). Bei hoher Netzwerkqualität oder konstanten Verläufen wie Verringerung von Übertragungsraten oder Erhöhung der Latenz treten keine Ausreißer auf.

- **Parallele Berechnungen**: Auch, wenn der genutzte Rechner vom Netzwerk getrennt war und unnötige Prozesse gestoppt wurden, könnte die Leistung durch parallele Berechnungen beeinflusst worden sein. Bei der Durchführung wurden für ein Netzwerkszenario (CSV-Datei) und für eine Netzwerkkonfiguration (Zeile) jeweils alle Algorithmen durchlaufen. Pro Algorithmus erfolgten 200 Messungen, bevor die nächste Konfiguration (Zeile) getestet wurde. Das heißt, es könnte rechnerbedingt zu Problemen innerhalb des Intervalls der 200 Messungen gekommen sein. Würden die Ausreißer gänzlich unabhängig von den Messreihen, rein aufgrund aufwändiger paralleler Berechnungen entstehen, so müssten in jeder Messreihe derartige Phänomene auftreten. Bei weitgehend störfreien Messreihen oder Experimenten mit kon-stanten Veränderungen der Parameter konnten jedoch keine derartigen Ausreißer festgestellt werden. Allerdings sind die Wahrscheinlichkeiten für das Auftreten eines Ausreißers unter der oben beschriebenen Annahme unterschiedlich verteilt. Bei längeren Handshake Zeiten kann es eher zu Überschneidung mit parallelen Prozessen kommen. Zu erwarten wäre in dem Fall, dass der Verzögerungsprozess mitunter über zwei Messreihen hinweg angehalten hat. Aufgrund des beschriebenen Ablaufs müssten innerhalb eines Experiments bei gleicher Netzwerkkonfiguration zwei aufeinander folgende Algorithmen betroffen sein. Dies kann tatsächlich beobachtet werden. Beispielsweise ist bei einer Paketverlustrate von 15 Prozent der Median von *sikep*503 stark vergrößert. Auf SIKE folgte die Messung von *bikel*1 als hybride Variante (*p256_bikel*1). Dessen Median ist ebenfalls vergrößert, wenngleich nicht in dem extremen Maß wie SIKE. Darauf folgende Messreihen sind nicht mehr betroffen. Bei einer Paketverlustrate von 19 Prozent ist eine ähnliches Phänomen zu beobachten. Der Median für *kyber*768 ist extrem groß und auch für den darauf folgenden Algorithmus *kyber*90*s*768 lässt sich eine Abweichung feststellen. Diese Beobachtungen können jedoch nicht in allen Fällen gemacht werden.

- **Beteiligte Netzwerk-Kontrollmechanismen**: Wie bereits in vorherigen Abschnitten angeführt, können verschiedene Kontrollmechanismen auf den zwischenliegenden Netzwerkprotokollen greifen, sofern die Verbindungsqualität niedrig ist. Während IP und TLS weniger in den Prozess eingebunden sind, regelt TCP die Segmentierung, die Flusskontrolle sowie die Fehlerbehandlung und sorgt bei Verzögerung oder Verlust von Segmenten für entsprechende Maßnahmen. Diese könnten zu Unregelmäßigkeiten bei den Übertragungszeiten führen. Mechanismen zur Flusskontrolle beziehungsweise -steuerung schätzen unter anderem mittels der Bestätigungsnachrichten (*ACK*) des Empfängers die maximalen Kapazitäten der Verbindung ab. Anschließend kann eine Bündelung einzelner Nachrichtenpakete oder eine Zeitanpassung bis zu einem erneuten Senden eines Segments bei Ausbleiben der Bestätigung erfolgen. Um trotz der unrealistischen, kontinuierlichen Veränderung der NetEm Netzwerkkonfigurationen ein möglichst realistisches Verhalten zu erzeugen und eine gewisse Nachvollziehbarkeit zu gewährleisten, wurden linuxeigene sowie einige TCP Mechanismen zur Netzwerkauslastung während der Versuche ausgesetzt. Zum einen wurden mittels *ethtool* die Segmentierungs- und Optimierungsmechanismen TCP Segmentation Offload (TSO), Generic Segmentation Offload (GSO) und Generic Receive Offload (GRO) für Linux Schnittstellen ausgeschaltet:

```
# Turn off optimizations
# that dent realism.
ip netns exec ${CLIENT_NS} \
    ethtool -K ${CLIENT_VETH} gso off gro off tso off

ip netns exec ${SERVER_NS} \
    ethtool -K ${SERVER_VETH} gso off gro off tso off
```

Für tiefergehende Informationen siehe [253]. Zum anderen wurde Nagle's Algorithmus sowie jegliche verfügbare Komprimierung für die eröffneten TCP-Sockets deaktiviert:

```
SSL_CTX_set_options(ssl_ctx, SSL_OP_NO_COMPRESSION);
BIO_set_conn_mode(conn, BIO_SOCK_NODELAY);
```

Trotzdem ist ein Einfluss der verbleibenden TCP Mechanismen möglich.

Sikeridis, Kampanakis und Devetsikiotis [234] integrieren unter anderem PQC Kandidaten für KEM und Authentifizierung innerhalb von TLS 1.3 und berichten von einem Zusammenhang zwischen der Handshakedauer und der Größe des TCP Congestion Window. Das Congestion Window bestimmt, wie viele TCP Segmente gesendet werden können, bevor ein *ACK* erforderlich ist, siehe auch Abschnitt 2.4.1. Bei geringer initialer Fenstergröße muss insbesondere bei großen zusendenden Datenmengen zunächst mehrfach auf eine Empfangsbestätigung gewartet werden, was den Handshake verzögert. Sikeridis, Kampanakis und Devetsikiotis [234] zeigen in Experimenten, dass durch Vergrößerung des Parameters *initcwnd* eine deutliche Verringerung der Handshakedauer erzielt werden könne. Da die Größe des Congestion Window generell variabel gehandelt wird und sich insbesondere auch am Anteil der nicht bestätigten respektive verlorenen Pakete orientiert, kann eine automatische Änderung als Ursache für die Ausreißer in Betracht gezogen werden. Folgender Ablauf wäre denkbar:

Es seien 20 Segmente zu senden und für die Größe *cwndSize* des Congestion Windows gilt zunächst *cwndSize* = 4. Nachdem Senden der ersten 4 Segmente und erfolgreicher Bestätigung verbleiben 12. Ohne Störung sind 5 *ACK*-Nachrichten des Empfängers erforderlich. Bei jedem sechsten Segment muss zunächst die Bestätigung abgewartet werden. Ist *cwndSize* = 5, sind lediglich 4 *ACK*-Nachrichten erforderlich. Der Prozess spart 0.5 RTTs und kann schneller durchgeführt werden. Bei Paketverlusten ist hingegen bei der vergrößerten *cwndSize* eine ausbleibende Bestätigung für 5 anstatt 4 Segmente zu verkraften. Zudem ist bei gegebener Verlustrate bei großem *cwndSize* die Wahrscheinlichkeit erhöht, dass mindestens ein Paket innerhalb des Fensters verloren geht. Entsprechend müssen mehr Segmente erneut gesendet werden. Daher ist ein geeigneter Trade-Off zwischen *cwndSize* und Netzwerkqualität erforderlich.

In der Praxis wird die Größe des Congestion Window dynamisch angepasst. Das Verhalten ist in RFC 5681 [48] definiert. Mittels des Slow-Start-Algorithmus wird zunächst ein kleiner Wert *cwndSize* angenommen und im Verlauf fortlaufend vergrößert, bis das Netz beziehungsweise der Empfänger überlastet ist und ein Verlust auftritt. Die anfängliche Steigerung ist anhand der verringerten Handshakedauer bei den ersten Messergebnissen der Versuchsreihen zu erkennen. Nach dem Verlust und dem folglich ausbleibenden *ACK* halbiert der Sender *cwndSize* und passt es mit zunehmend erfolgreich versendeten Segmenten wieder nach oben an. Der Prozess wird stetig fortgeführt.

Folglich ergibt sich eine Verzögerung aufgrund des erneuten Sendens der unbestätigten Pakete sowie aufgrund des zunächst verringerten *cwndSize*. Bei mehreren ausbleibenden Bestätigungen wird *cwndSize* zudem wiederholt halbiert. Die Verzögerung von Nachrichtenpaketen kann somit nicht unabhängig voneinander betrachtet werden, wenngleich für jeden Handshake ein neuer TCP Port eröffnet wird.

Der Einfluss dieser Flusssteuerung wurde mehrfach in wissenschaftlichen Kontext diskutiert [197, 232, 255]. Insbesondere im Bereich Mobilfunk und *Low Latency*, da hier neue Anforderungen beziehungsweise Prioritäten in den Fokus rücken [2, 3, 218, 266].

- Ein weiterer Aspekt sind Fehler beim Entkapseln beziehungsweise Entschlüsseln der Chiffrate. Jeder Algorithmus hat eine individuelle Fehlerrate, welche aus Sicherheits- und Performanzgründen möglichst gering gehalten werden sollte. In seltenen Fällen kann die Entkapselung trotz korrekter Übertragung des Chiffrats einen falschen Schlüssel ergeben. Sollte eine Entschlüsselung nicht erfolgreich durchgeführt werden, könnte es aufgrund notwendiger Wiederholungen eventuell zu Verzögerungen kommen. Innerhalb des Experiments sind jedoch andere Auswirkungen zu erwarten. Weitere Inhalte innerhalb des *ServerHello* sowie des *Finished* wie beispielsweise das Zertifikat sind bereits mit dem vom Server anhand des *key_shares* des Clients erzeugten Schlüssel verschlüsselt. Bei falscher Entkapselung auf Clientseite würde ein fehlerhafter Schlüssel somit die weitere Bearbeitung des *ServerHello* verhindern und zu einem Abbruch des Handshakes führen. Folglich würde keine gültige Zeit für den Testdurchlauf erfasst und keine Verzögerung registriert.

Auch eine Kombination der Ursachen ist denkbar und weitere, bis dahin unbekannte Faktoren könnten beteiligt sein. Um das Verhalten genauer zu analysieren und beispielsweise den Einfluss des TCP RTO auf den Ablauf nachzuvollziehen, sollten weitere Messungen mit größerem Stichprobenumfang durchgeführt werden. Parallel sollte der Netzwerkverkehr, beispielsweise mit *tcpdump* oder *Wireshark* aufgezeichnet werden und zudem könnten die Experimente mit unterschiedlichen Werten für *initcwnd* wiederholt werden.

# 6

## FAZIT & AUSBLICK

Mit der voranschreitenden Entwicklung der Quantencomputer steigt auch die Bedrohung im Bezug auf die Anwendungsfelder klassischer kryptografischer Verfahren wie RSA, ECDSA, ECDH und DSA. Ebenso steigt der Bedarf an quantensicheren und praktikablen Alternativen, die Anforderungen bezüglich geeigneter Algorithmen sind jedoch vielseitig und komplex. Die notwendige Umstellung betrifft einen Großteil der digitalen Infrastruktur, bestehend aus einer sehr heterogenen Menge von Software- und Hardwarekomponenten.

Auch, wenn das Verfahren der NIST hinsichtlich der Standardisierung geeigneter PQC Algorithmen weiter vorangetrieben wird und sich eine Vielzahl von Forschungsarbeiten mit der Migration und Evaluierung beschäftigen, müssen die Algorithmen noch besser verstanden und insbesondere innerhalb der diversen Anwendungsszenarien evaluiert werden. Die Ergebnisse dieser Arbeit sollen zu einem besseren Verständnis bezüglich der Performanz der Algorithmen im Kontext von TLS 1.3 verhelfen und somit einen Teil zur Bewertung und Einordnung der Algorithmen beitragen.

Zunächst konnten die Einschätzungen von Paquin, Stebila und Tamvada [196] bestätigt werden. Bei Netzwerkverbindungen mit schlechter Qualität vergrößerte sich die Handshakedauer, wobei datenintensive Algorithmen wie FrodoKEM in größerem Maß betroffen waren. Zudem wurde deutlich, dass viele Kandidaten und Alternativen dem Vergleich mit der klassischen Variante unter herkömmlichen Bedingungen standhalten können. Dies galt im besonderen Maß für die evaluierten Kandidaten. Bei Übertragungsraten über 2 Mbps sowie weniger als 10 Prozent verlorenen oder fehlerhaften Paketen waren sie sogar performanter als die vergleichbare ECDH Variante. Somit kann die Bewertung der NIST bezüglich einer hohen Gesamtperformanz der Kandidaten Kyber, Saber und NTRU bestätigt werden.

Bemerkenswert ist, dass die gitterbasierte Alternative NTRU Prime annähernd durchgängig gleich gute Ergebnisse erzielte wie die Kandidaten. HQC und insbesondere FrodoKEM zeigten hingegen diverse Schwachstellen aufgrund ihrer großen Datenmengen. Insbesondere FrodoKEM war bereits bei herkömmlichen Netzwerkkonfigurationen, wie beispielsweise Übertragungsraten unter 50 Mbps, anfällig. BIKE und Sike fielen bei besserer Verbindungsqualität aufgrund ihrer ineffizienten Operationen negativ auf. Sike besticht jedoch durch die geringen Schlüssel- und Chiffratgrößen, welche in speziellen Anwendungsfällen, insbesondere bei Übertragungsraten von weniger als 2 Mbps, durchaus nützlich sein könnten. Möglicherweise wäre eine in Abhängigkeit zur zugrundeliegenden Netzwerkcharakteristik getroffene Auswahl des TLS KEM beziehungsweise ein entsprechender Wechsel zu Sike sinnvoll.

Zudem sollte Sike weiter analysiert und eine zukünftige Standardisierung nicht frühzeitig ausgeschlossen werden.

Mit Blick auf die zusätzlich untersuchten Unterschiede zwischen Sicherheitsleveln und hybriden sowie PQ-Only Varianten war kein einheitliches Muster für alle Algorithmen erkennbar. Die Unterschiede zwischen den jeweiligen Varianten eines Algorithmus waren bei den Kandidaten sowie bei NTRU Prime für fast alle Experimente sehr gering. Auch bei höheren Paketverlustraten, steigender Latenz oder geringer Übertragungsrate kann daher ohne signifikante Verzögerungen eine eher konservative Variante gewählt werden. Bei Alternativen, vornehmlich bei Sike und FrodoKEM, konnten hingegen größere Verzögerungen bei höheren Sicherheitsleveln oder hybriden Varianten beobachtet werden. Dies wurde insbesondere bei geringer Übertragungsrate oder steigender Latenz deutlich. Sofern Alternativen in Betracht gezogen werden, sollte daher bei der Auswahl des Algorithmus das angestrebte Sicherheitslevel berücksichtigt werden.

Bei Veränderung der Übertragungsraten wurde ersichtlich, dass vom Server ausgehende Verbindungen eine größere Bedeutung besitzen als vom Client ausgehende Verbindungen. Dies ist auf die unterschiedlichen Verhältnisse zwischen Schlüssel- und Chiffratgrößen pro Algorithmus zurückzuführen.

Des Weiteren fiel auf, dass der Werteverlauf, insbesondere bei steigender Wahrscheinlichkeit bezüglich verlorener oder fehlerhafter Pakete, nicht durchweg gleichmäßig war. Die Ursachen konnten nicht endgültig geklärt werden. Möglicherweise sind TCP Mechanismen zur effizienten und fehlerfreien Übertragung der Nachrichten für dieses Verhalten mit verantwortlich. Beispielsweise bestimmt die Größe des *Congestion Windows* die Anzahl der Pakete, welche mittels TCP ohne Bestätigung des Empfängers parallel gesendet werden können. Sie richtet sich insbesondere nach der Paketverlustrate und beeinflusst maßgeblich den tatsächlichen Durchsatz der Verbindung und somit die Performanz der Algorithmen. Die Zusammenhänge sollten besser untersucht werden. Eventuell könnte eine optimierte TCP Konfigurationen, je nach Netzwerkcharakteristik, ausfindig gemacht werden, welche die Nachteile der neuartigen PQC Algorithmen ausgleichen könnte. Zudem führten Paquin, Stebila und Tamvada [196] an, dass auch die MTU entscheidenden Einfluss auf den Durchsatz habe. Auch die Wahl anderer Netzwerkprotokolle wie UDP, QUIC oder IPv6 könnte sich merklich auf die Performanz des Handshakes auswirken.

Letztlich zeigte sich, dass die Performanz des TLS Handshakes nicht alleine von der Wahl des PQC Algorithmus abhängt. Weitere Faktoren auf Netzwerkebene spielen eine ebenso wichtige Rolle und eine Optimierung oder Anpassung dieser könnte folglich mögliche Nachteile von PQC eliminieren beziehungsweise diese obsolet erscheinen lassen.

An dieser Stelle werden jedoch auch die Grenzen der Emulation sichtbar:

1. Alle signifikanten äußeren Einflüsse müssen frühzeitig ausfindig gemacht und in die Emulation integriert werden, um ein für die Realität

exemplarisches Szenario zu schaffen. Werden wichtige Faktoren vergessen, kann die Aussagekraft der Ergebnisse beeinträchtigt werden.

2. Es existieren Unmengen an Netzwerkszenarien beziehungsweise Kombinationsmöglichkeiten von Netzwerkparametern und Algorithmenvarianten, welche bei Durchführung und Analyse berücksichtigt werden müssen, um die Gesamtheit der in der Realität möglichen Verbindungen zu berücksichtigen.

3. Nicht alle möglichen Kombinationen spiegeln die Realität beziehungsweise typische TLS Szenarien wieder. Beispielsweise verläuft die Veränderung der Netzwerkparameter wie Latenz oder Paketverlust selten durchgehend konstant. Atypische Konstellationen können jedoch irreführende Reaktionen der beteiligten Netzwerkschichten hervorrufen

Zusammenfassend stellt sich die Frage, inwieweit die Emulation beziehungsweise die erzielten Ergebnisse die Realität abbilden.

In Zukunft wären weitere Experimente in diesem Bereich, teils unter Nutzung des erweiterten Frameworks aus dieser Arbeit, sinnvoll, um die Ergebnisse zu überprüfen und weitere Bereiche zu ergänzen:

- In weiteren Untersuchungen könnte die Auswahl der Netzwerkparameter analysiert und Experimente mit weiteren Szenarien durchgeführt werden. Zudem könnte eine Untersuchung der aufgetretenen Ausreißer sinnvoll sein. Unter anderem ist eine Vergrößerung der Anzahl der Handshakes beziehungsweise Messungen für eine bessere Genauigkeit denkbar. Auf Netzwerkebene könnten Segmente mit geeigneten Softwarewerkzeugen wie *tcpdump* oder *wireshark* nachverfolgt werden.

- Wie in [196] wären als Vergleich Experimente mit realen Servern mit unterschiedlichen Distanzen möglich. Die Experimente könnten erweitert werden, indem verschiedene Übertragungsmedien und Endgeräte eingebunden werden. Alternativ wäre eine komplexere Emulation beispielsweise mit *Mininet* oder *NetMirage* denkbar. Bei beiden Vorgehensweisen werden jedoch Nachvollziehbarkeit und dezidierte Analyse erschwert.

- Mit Blick auf die aktuelle OpenSLL Version wäre die Erweiterung mit OpenSSL 3.0.0, siehe Abschnitt 4.1.4, zu erwähnen. Möglicherweise ist in Zukunft auch eine Implementierung mit Classic McEliece verfügbar oder es kommt zu Änderungen bei anderen Algorithmen.

- Wie bereits angeführt, könnten die beteiligten Netzwerkprotokolle ersetzt werden, um deren Einfluss und Möglichkeiten zur Optimierung zu überprüfen. Denkbar wäre beispielsweise eine Nutzung von QUIC, SPDY, UDP oder IPv6. Bei UDP könnte entsprechend auf DTLS zurückgegriffen werden.

- Neben der Schlüsselvereinbarung soll auch die Authentifizierung mittels digitaler Signaturen und Zertifikate innerhalb des TLS Handshakes in Zukunft quantensicher erfolgen. Die Algorithmen wurden in den Kapiteln 2 und 3 beschrieben. Das Framework könnte entsprechend erweitert werden, sodass Experimente mit unterschiedlichen Signaturalgorithmen möglich sind. Dabei ist insbesondere auch die Kombination qunatensicherer KEMs sowie Signaturen wünschenswert. Interessant wäre die Verwendung von mehreren Zertifikaten und unterschiedlichen Algorithmen innerhalb der Zertifikatsketten.

- Neben der klassischen Authentifizierung mittels digitaler Signaturen könnte auch *authenticated TLS* ohne digitale Signaturen nach [152, 224] untersucht werden. Dieser Ansatz ist insbesondere interessant, da nur ein neuartiger Algorithmus verwendet werden muss, sodass die Gefahr durch *Downgrade Attacken* reduziert wird und weniger Komponenten umgestellt werden müssen.

Teil II

ANHANG

# A

GRAFISCHE DARSTELLUNG DER MESSERGEBNISSE

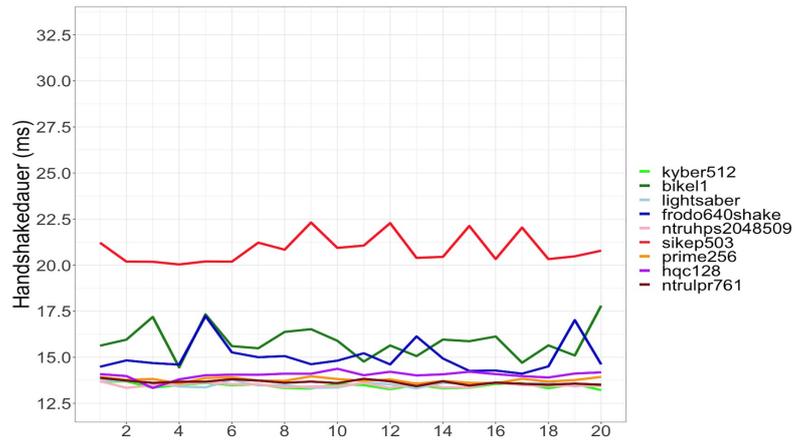

(a) Median

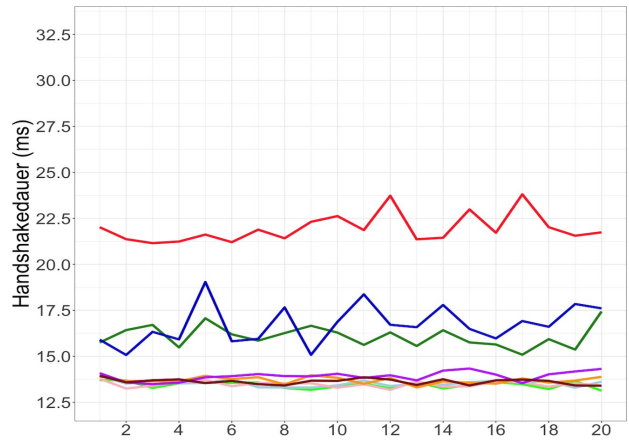

(b) Arithmetisches Mittel

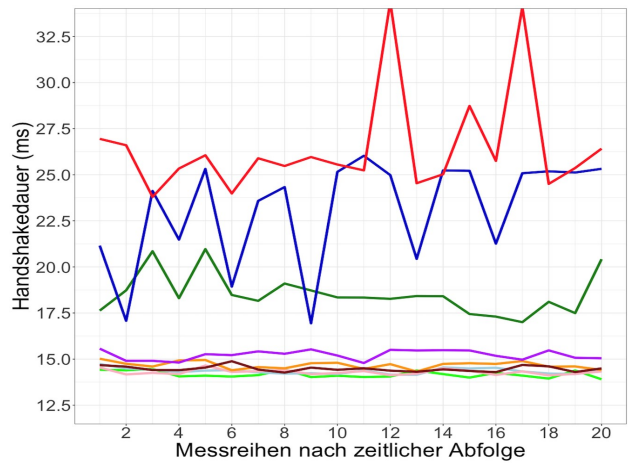

(c) 0.95-Quantil

Abbildung A.1: Handshakedauer von 20 Messreihen mit konstanten Netzwerkparametern bei Median, arithmetischem Mittel und 0.95-Quantil

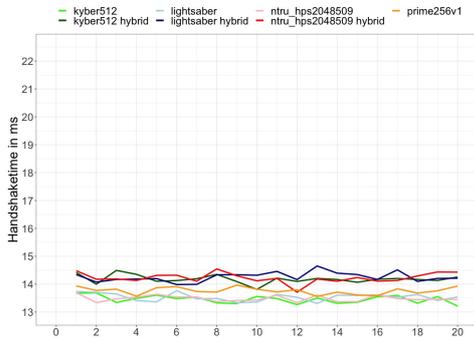
(a) Kandidaten Sicherheitslevel 1

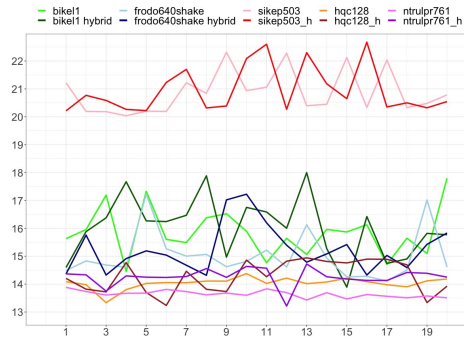
(b) Alternativen Sicherheitslevel 1

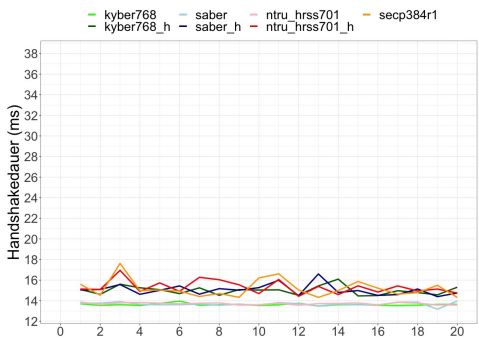
(c) Kandidaten Sicherheitslevel 3

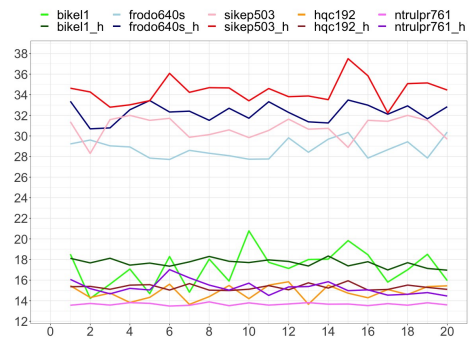
(d) Alternativen Sicherheitslevel 3

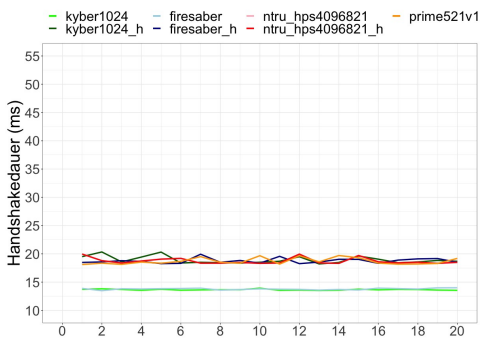
(e) Kandidaten Sicherheitslevel 5

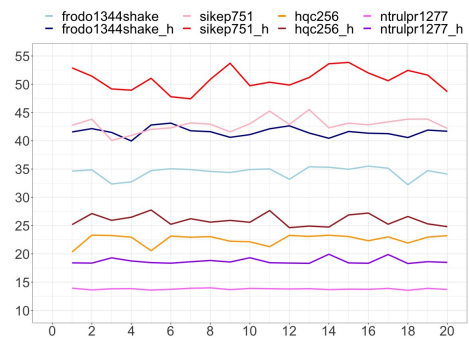
(f) Alternativen Sicherheitslevel 5

Abbildung A.2: Handshakedauer für hybride und PQ-Only Varianten in Millisekunden auf der y-Achse in Abhängigkeit zur Nummerierung der durchgeführten Messung bei unveränderten Netzwerkparametern und Betrachtung des Median bei Kandidaten und Alternativen. Hybride Varianten werden innerhalb der Legende durch ein „h" gekennzeichnet

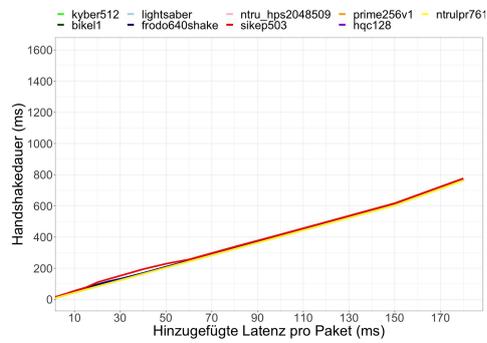
(a) Median Sicherheitslevel 1

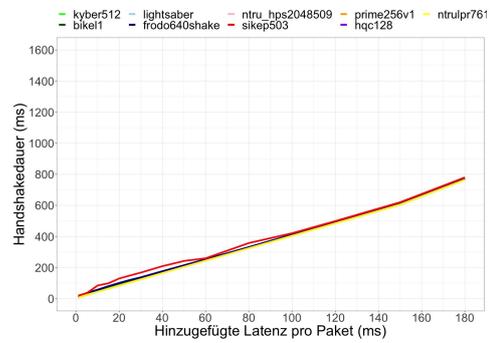
(b) 0.95-Quantil Sicherheitslevel 1

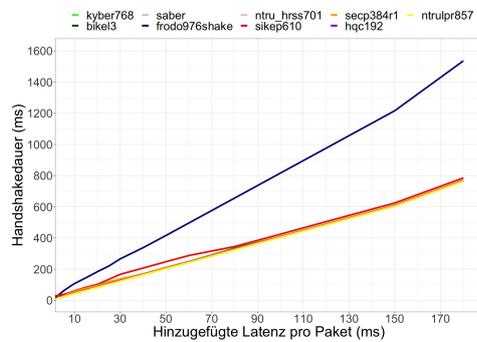
(c) Median Sicherheitslevel 3

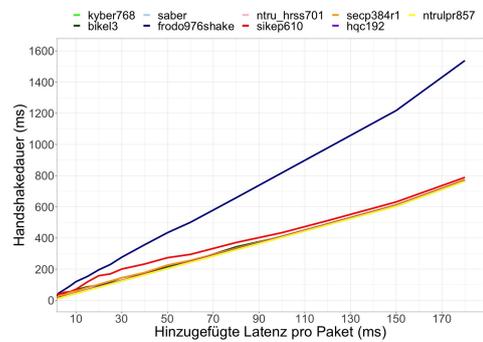
(d) 0.95-Quantil Sicherheitslevel 3

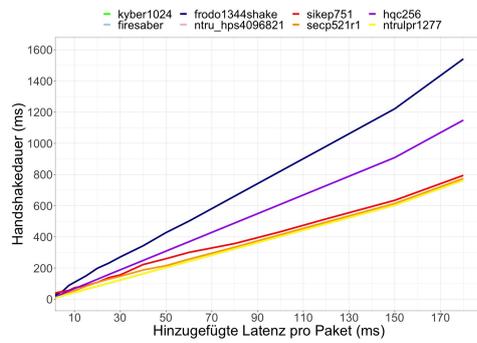
(e) Median Sicherheitslevel 5

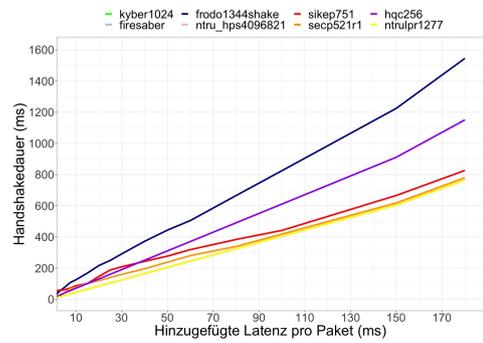
(f) 0.95-Quantil Sicherheitslevel 5

Abbildung A.3: Handshakedauer für alle Kandidaten und Alternativen sowie alle Sicherheitslevel in Abhängigkeit zum Jitter und einer Latenz von 20 ms. Betrachtung des Medians (links) und des 0.95-Quantils (rechts).

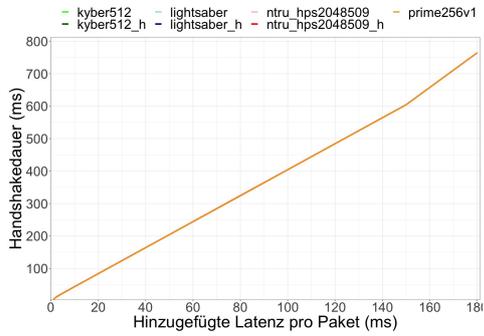
(a) Kandidaten Sicherheitslevel 1

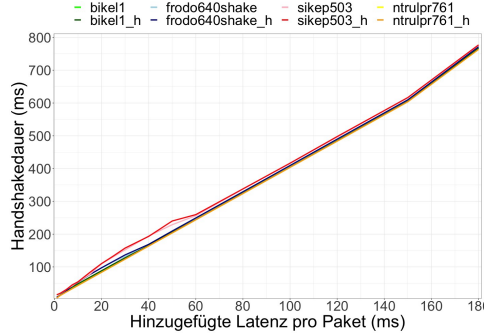
(b) Alternativen Sicherheitslevel 1

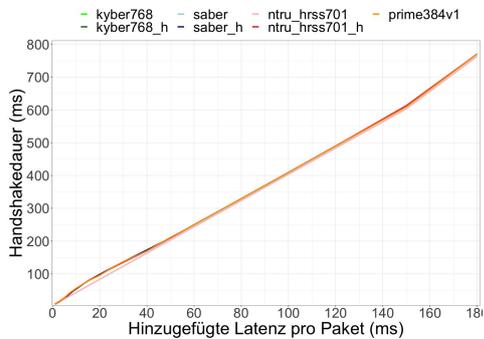
(c) Kandidaten Sicherheitslevel 3

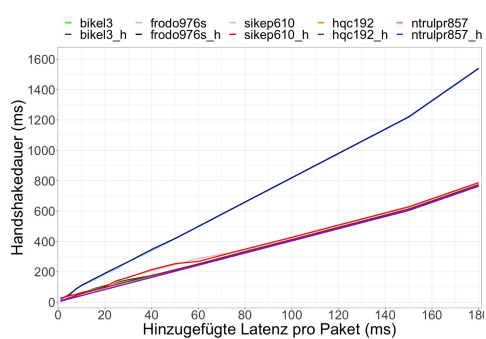
(d) Alternativen Sicherheitslevel 3

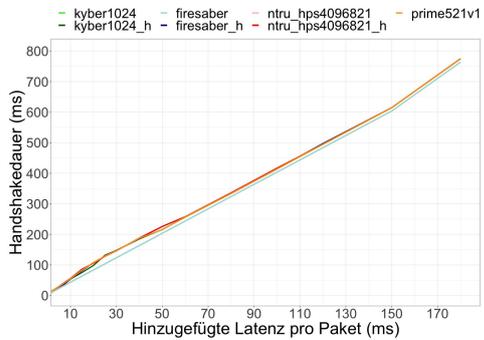
(e) Kandidaten Sicherheitslevel 5

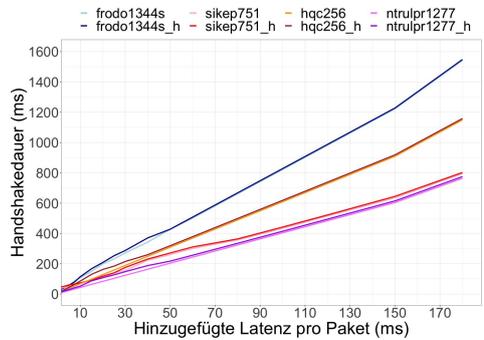
(f) Alternativen Sicherheitslevel 5

Abbildung A.4: Handshakedauer für hybride und PQ-Only Varianten in Abhängigkeit zur pro Paket hinzugefügten Latenz, jeweils in Millisekunden bei Betrachtung des Median. Hybride Varianten werden innerhalb der Legende durch ein „h" gekennzeichnet. *frodo*640*shake* und alle weiteren Varianten wurde mit *frodo*640*s* abgekürzt.

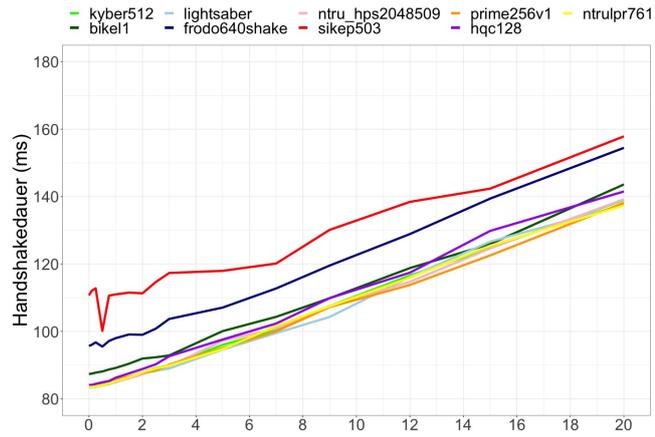

(a) Median

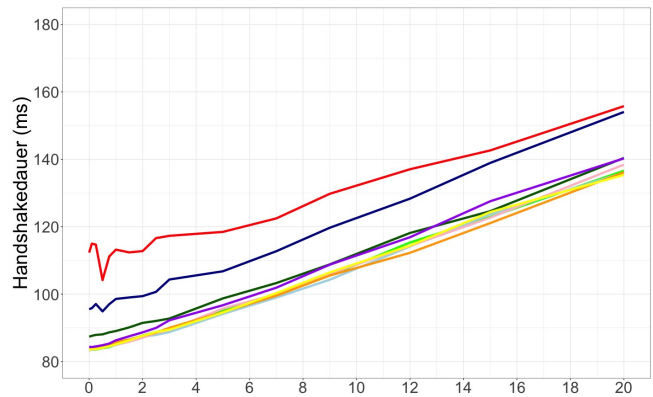

(b) Arithmetisches Mittel

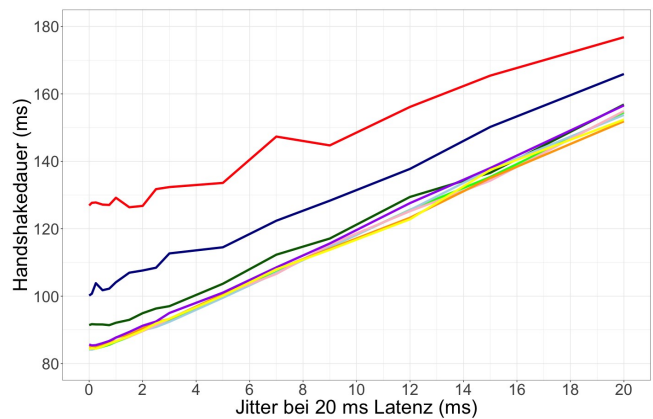

(c) 0.95-Quantil

Abbildung A.5: Handshakedauer bei 20 ms Latenz und steigender Varianz (Jitter) für Median, arithmetisches Mittel und 0.95-Quantil. Abbildung aller Kandidaten und Alternativen für Sicherheitslevel 1

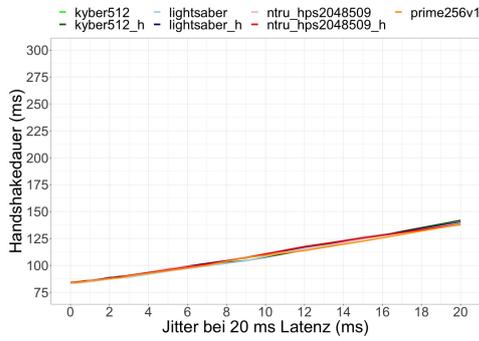
(a) Kandidaten Sicherheitslevel 1

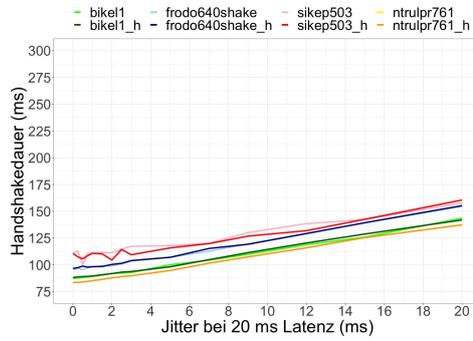
(b) Alternativen Sicherheitslevel 1

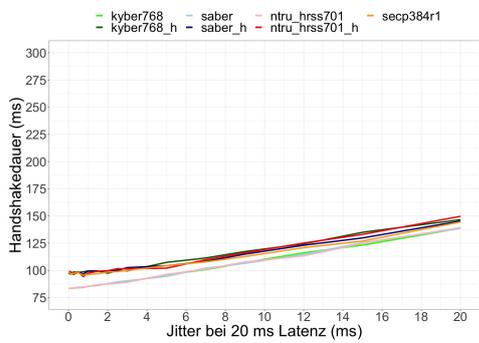
(c) Kandidaten Sicherheitslevel 3

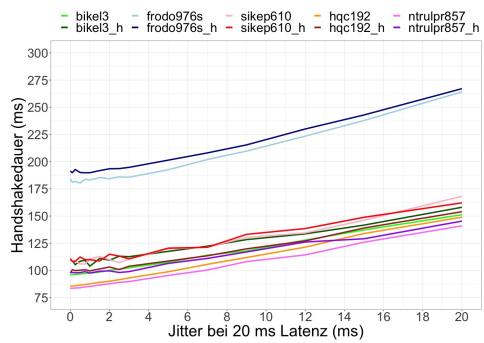
(d) Alternativen Sicherheitslevel 3

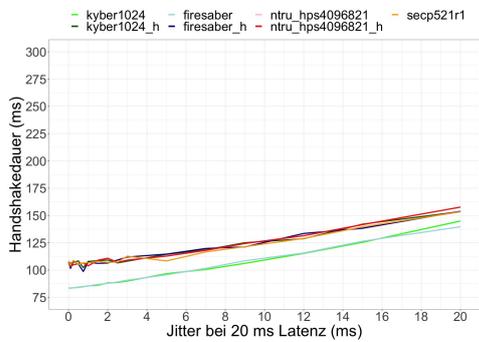
(e) Kandidaten Sicherheitslevel 5

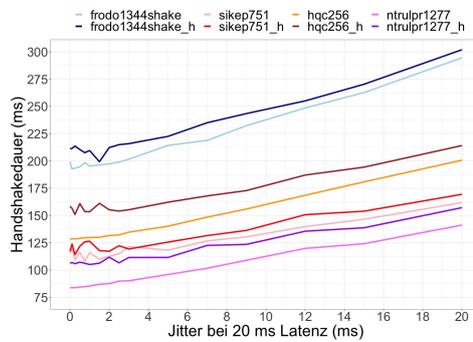
(f) Alternativen Sicherheitslevel 5

Abbildung A.6: Handshakedauer bei steigendem Jitter und einer Latenz von 20 ms für hybride und PQ-Only Varianten, jeweils in Millisekunden bei Betrachtung des Median. Hybride Varianten werden innerhalb der Legende durch ein „h" gekennzeichnet. *shake* wurde durch „s" ersetzt (z.B. *frodo*640*s*)

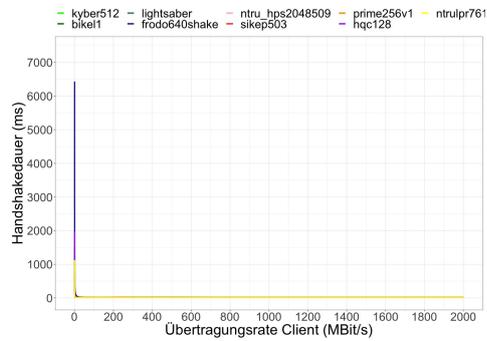
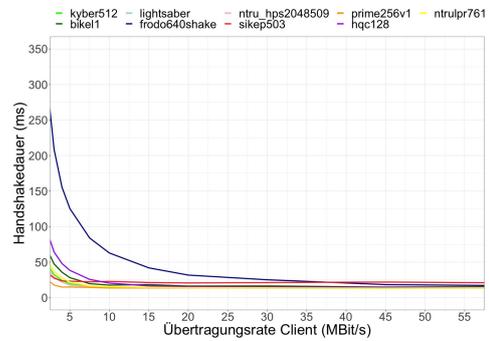

(a) Gesamtverlauf Client　　　　　(b) 5-55 Mbps Client

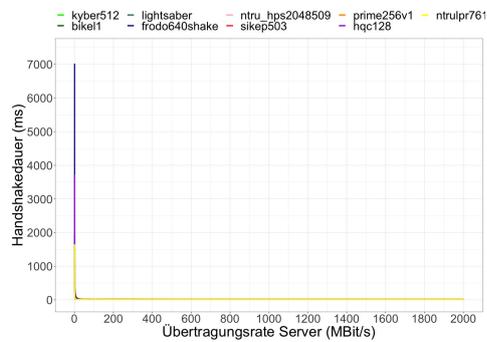
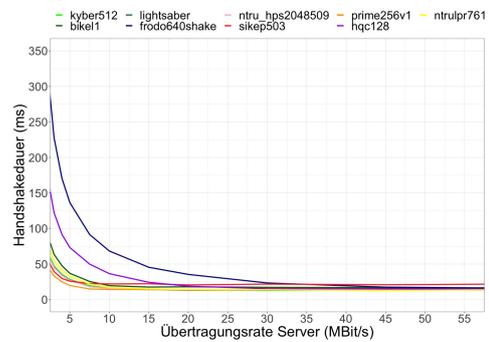

(c) Gesamtverlauf Server　　　　　(d) 5-55 Mbps Server

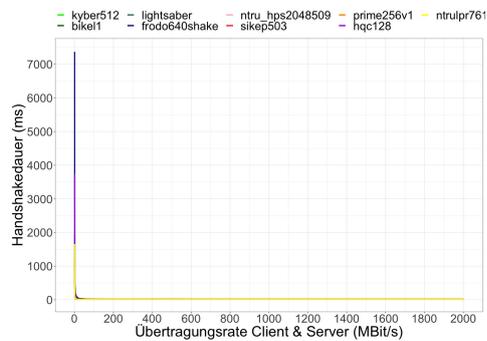
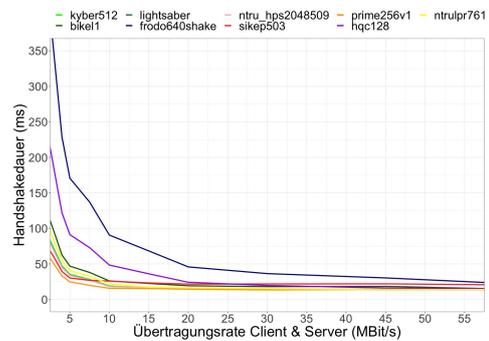

(e) Gesamtverlauf Beide　　　　　(f) 5-55 Mbps Beide

Abbildung A.7: Handshakedauer in Millisekunden bei veränderter Übertragungsrate in Mbps für Client, Server, sowie beidseitig. Betrachtung des Median bei allen evaluierten Kandidaten und Alternativen

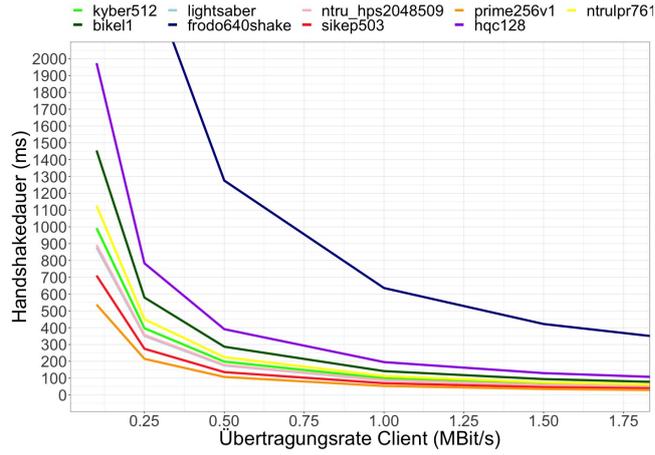

(a) Client

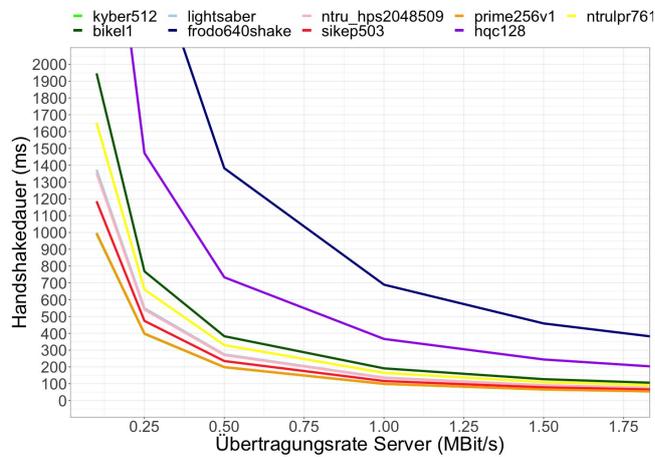

(b) Server

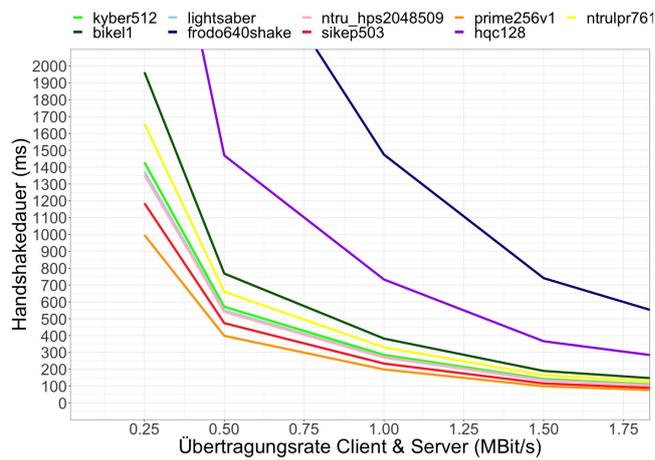

(c) Beidseitig

Abbildung A.8: Handshakedauer in Millisekunden bei veränderter Übertragungsrate im Bereich 0 bis 2 Mbps für Client, Server, sowie beidseitig. Betrachtung des Median bei allen evaluierten Kandidaten und Alternativen

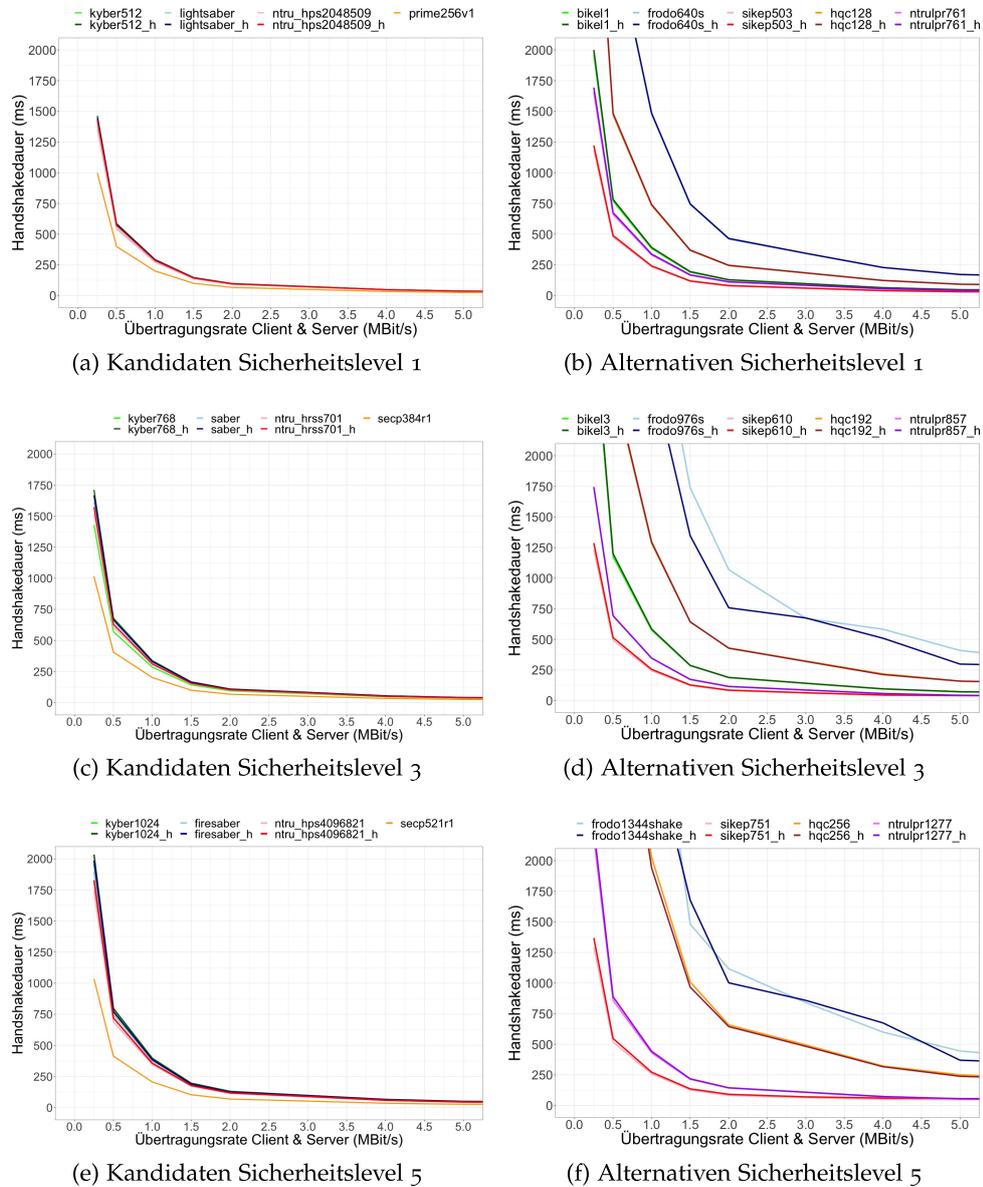

(a) Kandidaten Sicherheitslevel 1 (b) Alternativen Sicherheitslevel 1

(c) Kandidaten Sicherheitslevel 3 (d) Alternativen Sicherheitslevel 3

(e) Kandidaten Sicherheitslevel 5 (f) Alternativen Sicherheitslevel 5

Abbildung A.9: Handshakedauer bei beidseitig veränderter Übertragungsrate und einer Latenz von 20 ms für hybride und PQ-Only Varianten, jeweils in Millisekunden bei Betrachtung des Median. Hybride Varianten werden innerhalb der Legende durch ein „h" gekennzeichnet. *shake* wurde durch „s" ersetzt (z.B. *frodo640s*)

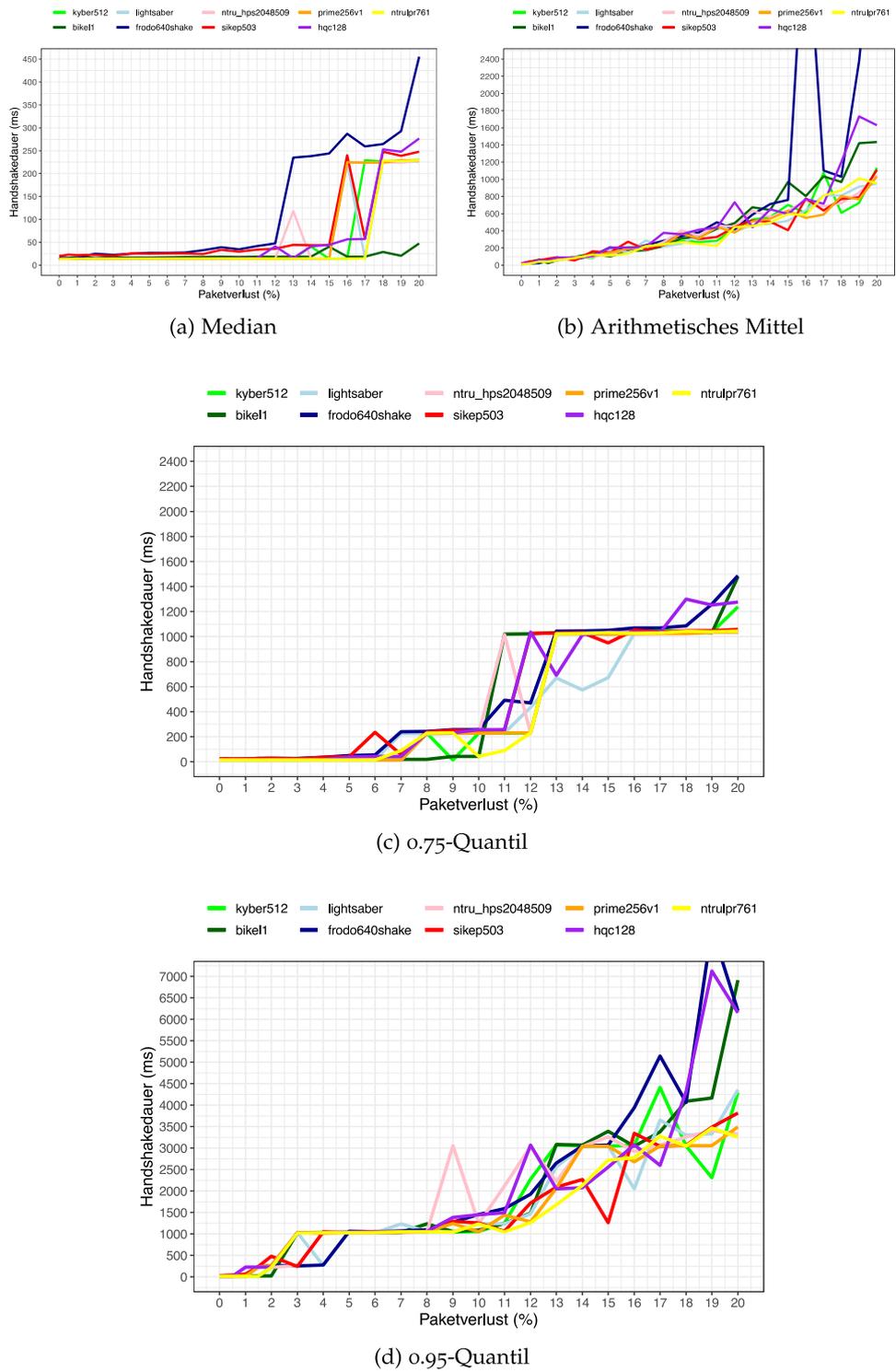

(a) Median

(b) Arithmetisches Mittel

(c) 0.75-Quantil

(d) 0.95-Quantil

Abbildung A.10: Handshakedauer in Millisekunden bei steigender Paketverlustrate in Prozent. Betrachtung von Median, arithmetischem Mittel, 0.75-Quantil und 0.95-Quantil für alle evaluierten Kandidaten und Alternativen mit Sicherheitslevel 1

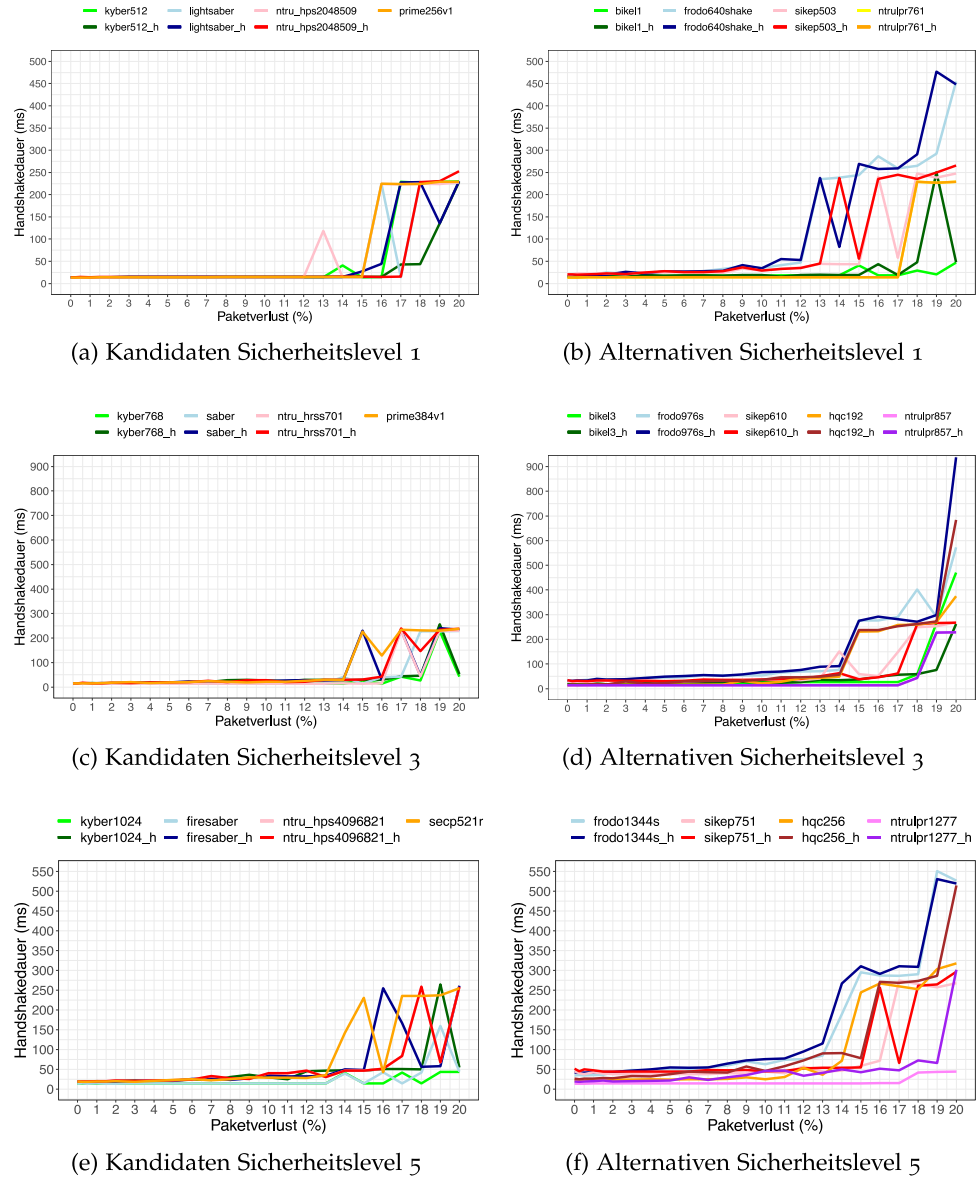

(a) Kandidaten Sicherheitslevel 1 (b) Alternativen Sicherheitslevel 1

(c) Kandidaten Sicherheitslevel 3 (d) Alternativen Sicherheitslevel 3

(e) Kandidaten Sicherheitslevel 5 (f) Alternativen Sicherheitslevel 5

Abbildung A.11: Handshakedauer in Millisekunden bei steigender Paketverlustrate in Prozent für hybride und PQ-Only Varianten bei Betrachtung des Median. Hybride Varianten werden innerhalb der Legende durch ein „h" gekennzeichnet. *shake* wurde durch „s" ersetzt (z.B. $frodo640s$)

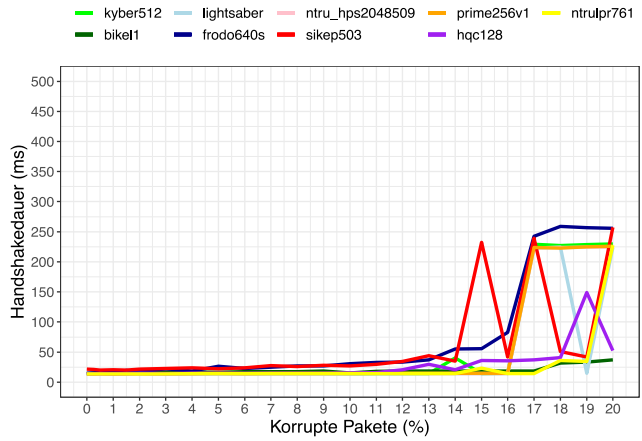
(a) Sicherheitslevel 1

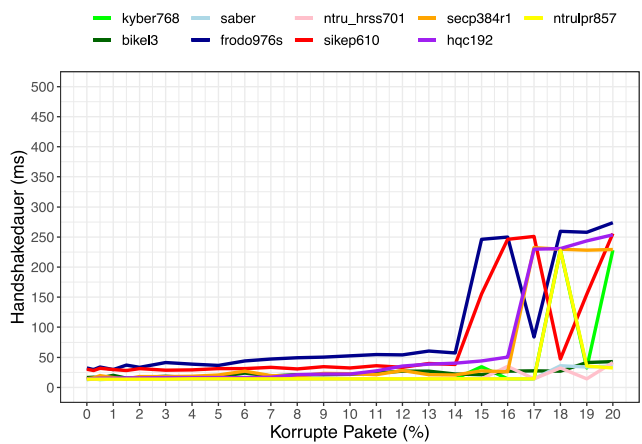
(b) Sicherheitslevel 3

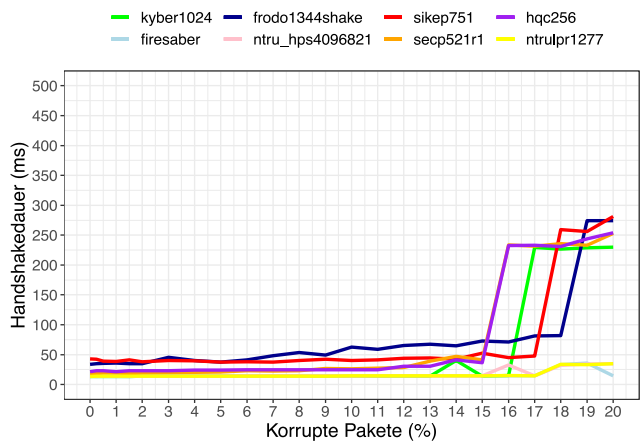
(c) Sicherheitslevel 5

Abbildung A.12: Handshakedauer in Millisekunden bei veränderter Übertragungsrate im Bereich 0 bis 2 Mbps für Client, Server, sowie beidseitig. Betrachtung des Median bei allen evaluierten Kandidaten und Alternativen

Teil III

LITERATUR

# LITERATUR